%% file: main.tex
\RequirePackage{lineno}
\documentclass[aps,prd,twocolumn,showpacs,superscriptaddress,nofootinbib,floatfix,letterpaper]{revtex4-1}
\pdfoutput=1
\usepackage{placeins}
\usepackage{multirow}
\usepackage[utf8]{inputenc}
\usepackage{color}
\usepackage[caption=false]{subfig}
\usepackage{xspace}
\usepackage{graphicx}
\usepackage[pdftex,bookmarks,hidelinks]{hyperref}
\usepackage{amsmath}

\graphicspath{ {graphics/} }

\newcommand{\numu}{\ensuremath{\nu_\mu}\xspace}
\newcommand{\nue}{\ensuremath{\nu_e}\xspace}
\newcommand{\nutau}{\ensuremath{\nu_\tau}\xspace}
\newcommand{\anumu}{\ensuremath{\bar\nu_\mu}\xspace}
\newcommand{\anue}{\ensuremath{\bar\nu_e}\xspace}

\newcommand{\dm}[1]{\ensuremath{\Delta m^2_{#1}}\xspace} 
\newcommand{\sinst}[1]{\ensuremath{\sin^2\theta_{#1}}\xspace} 
\newcommand{\sinstt}[1]{\ensuremath{\sin^22\theta_{#1}}\xspace}  
\newcommand{\deltacp}{\ensuremath{\delta_{\rm CP}}\xspace}   

\def\bracketbar{\hbox{\kern-8pt\raise1pt%
    \hbox{{\tiny(}{\lower1.5pt\hbox{\bf--}}{\tiny)}}}}
\newcommand{\dchisq}{\ensuremath{\Delta\chi^{2}}\xspace}
\newcommand{\dchisqcrit}{\ensuremath{\Delta\chi^{2}_{c}}\xspace}
\newcommand{\dchisqMO}{\ensuremath{\Delta\chi^{2}_{\mathrm{MO}}}\xspace}
\newcommand{\dchisqCPV}{\ensuremath{\Delta\chi^{2}_{\mathrm{CPV}}}\xspace}
\newcommand{\dchisqFC}{\ensuremath{\Delta\chi^{2}_{\mathrm{FC}}}\xspace}

\begin{document}
\title{Low exposure long-baseline neutrino oscillation sensitivity of the DUNE experiment}
\date{\today}
\input{sections/author_list}
\begin{abstract}
The Deep Underground Neutrino Experiment (DUNE) will produce world-leading neutrino oscillation measurements over the lifetime of the experiment. In this work, we explore DUNE's sensitivity to observe charge-parity violation (CPV) in the neutrino sector, and to resolve the mass ordering, for exposures of up to 100 kiloton-megawatt-years (kt-MW-yr). The analysis includes detailed uncertainties on the flux prediction, the neutrino interaction model, and detector effects. We demonstrate that DUNE will be able to unambiguously resolve the neutrino mass ordering at a 3$\sigma$ (5$\sigma$) level, with a 66 (100) kt-MW-yr far detector exposure, and has the ability to make strong statements at significantly shorter exposures depending on the true value of other oscillation parameters. We also show that DUNE has the potential to make a robust measurement of CPV at a 3$\sigma$ level with a 100 kt-MW-yr exposure for the maximally CP-violating values $\deltacp = \pm\pi/2$. Additionally, the dependence of DUNE's sensitivity on the exposure taken in neutrino-enhanced and antineutrino-enhanced running is discussed. An equal fraction of exposure taken in each beam mode is found to be close to optimal when considered over the entire space of interest.
\end{abstract}

\maketitle

\input{sections/introduction}

\input{sections/analysis_framework}

\input{sections/run_plan_opt}

\input{sections/cp_sens}
\input{sections/mh_sens}
\input{sections/conclusion}
\input{sections/acknowledgements}
\FloatBarrier
\bibliography{tdr-citedb}

\appendix*
\input{sections/fc_appendix}

\end{document}

%% file: sections/author_list.tex
\newcommand{\Abilene}{Abilene Christian University, Abilene, TX 79601, USA}
\newcommand{\Albanysuny}{University of Albany, SUNY, Albany, NY 12222, USA}
\newcommand{\Amsterdam}{University of Amsterdam, NL-1098 XG Amsterdam, The Netherlands}
\newcommand{\Antalya}{Antalya Bilim University, 07190 D{\"o}{\c{s}}emealt{\i}/Antalya, Turkey}
\newcommand{\Antananarivo}{University of Antananarivo, Antananarivo 101, Madagascar}
\newcommand{\AntonioNarino}{Universidad Antonio Nari{\~n}o, Bogot{\'a}, Colombia}
\newcommand{\Argonne}{Argonne National Laboratory, Argonne, IL 60439, USA}
\newcommand{\Arizona}{University of Arizona, Tucson, AZ 85721, USA}
\newcommand{\Asuncion}{Universidad Nacional de Asunci{\'o}n, San Lorenzo, Paraguay}
\newcommand{\Athens}{University of Athens, Zografou GR 157 84, Greece}
\newcommand{\Atlantico}{Universidad del Atl{\'a}ntico, Barranquilla, Atl{\'a}ntico, Colombia}
\newcommand{\Augustana}{Augustana University, Sioux Falls, SD 57197, USA}
\newcommand{\Banaras}{Banaras Hindu University, Varanasi - 221 005, India}
\newcommand{\Basel}{University of Basel, CH-4056 Basel, Switzerland}
\newcommand{\Bern}{University of Bern, CH-3012 Bern, Switzerland}
\newcommand{\Beykent}{Beykent University, Istanbul, Turkey}
\newcommand{\Birmingham}{University of Birmingham, Birmingham B15 2TT, United Kingdom}
\newcommand{\BolognaUniversity}{Universit{\`a} del Bologna, 40127 Bologna, Italy}
\newcommand{\Boston}{Boston University, Boston, MA 02215, USA}
\newcommand{\Bristol}{University of Bristol, Bristol BS8 1TL, United Kingdom}
\newcommand{\Brookhaven}{Brookhaven National Laboratory, Upton, NY 11973, USA}
\newcommand{\Bucharest}{University of Bucharest, Bucharest, Romania}
\newcommand{\CBPF}{Centro Brasileiro de Pesquisas F\'isicas, Rio de Janeiro, RJ 22290-180, Brazil}
\newcommand{\CEASaclay}{IRFU, CEA, Universit{\'e} Paris-Saclay, F-91191 Gif-sur-Yvette, France}
\newcommand{\CERN}{CERN, The European Organization for Nuclear Research, 1211 Meyrin, Switzerland}
\newcommand{\CIEMAT}{CIEMAT, Centro de Investigaciones Energ{\'e}ticas, Medioambientales y Tecnol{\'o}gicas, E-28040 Madrid, Spain}
\newcommand{\CUSB}{Central University of South Bihar, Gaya, 824236, India }
\newcommand{\CalBerkeley}{University of California Berkeley, Berkeley, CA 94720, USA}
\newcommand{\CalDavis}{University of California Davis, Davis, CA 95616, USA}
\newcommand{\CalIrvine}{University of California Irvine, Irvine, CA 92697, USA}
\newcommand{\CalLosangeles}{University of California Los Angeles, Los Angeles, CA 90095, USA}
\newcommand{\CalRiverside}{University of California Riverside, Riverside CA 92521, USA}
\newcommand{\CalSantabarbara}{University of California Santa Barbara, Santa Barbara, California 93106 USA}
\newcommand{\Caltech}{California Institute of Technology, Pasadena, CA 91125, USA}
\newcommand{\Cambridge}{University of Cambridge, Cambridge CB3 0HE, United Kingdom}
\newcommand{\Campinas}{Universidade Estadual de Campinas, Campinas - SP, 13083-970, Brazil}
\newcommand{\CataniaUniversitadi}{Universit{\`a} di Catania, 2 - 95131 Catania, Italy}
\newcommand{\Catolica}{Universidad Cat{\'o}lica del Norte, Antofagasta, Chile}
\newcommand{\Charles}{Institute of Particle and Nuclear Physics of the Faculty of Mathematics and Physics of the Charles University, 180 00 Prague 8, Czech Republic }
\newcommand{\Chicago}{University of Chicago, Chicago, IL 60637, USA}
\newcommand{\ChungAng}{Chung-Ang University, Seoul 06974, South Korea}
\newcommand{\Cincinnati}{University of Cincinnati, Cincinnati, OH 45221, USA}
\newcommand{\Cinvestav}{Centro de Investigaci{\'o}n y de Estudios Avanzados del Instituto Polit{\'e}cnico Nacional (Cinvestav), Mexico City, Mexico}
\newcommand{\Colima}{Universidad de Colima, Colima, Mexico}
\newcommand{\ColoradoBoulder}{University of Colorado Boulder, Boulder, CO 80309, USA}
\newcommand{\ColoradoState}{Colorado State University, Fort Collins, CO 80523, USA}
\newcommand{\Columbia}{Columbia University, New York, NY 10027, USA}
\newcommand{\CzechAcademyofSciences}{Institute of Physics, Czech Academy of Sciences, 182 00 Prague 8, Czech Republic}
\newcommand{\CzechTechnical}{Czech Technical University, 115 19 Prague 1, Czech Republic}
\newcommand{\DakotaState}{Dakota State University, Madison, SD 57042, USA}
\newcommand{\Dallas}{University of Dallas, Irving, TX 75062-4736, USA}
\newcommand{\DannecyleVieux}{Laboratoire d{\textquoteright}Annecy de Physique des Particules, Univ. Grenoble Alpes, Univ. Savoie Mont Blanc, CNRS, LAPP-IN2P3, 74000 Annecy, France}
\newcommand{\Daresbury}{Daresbury Laboratory, Cheshire WA4 4AD, United Kingdom}
\newcommand{\Drexel}{Drexel University, Philadelphia, PA 19104, USA}
\newcommand{\Duke}{Duke University, Durham, NC 27708, USA}
\newcommand{\Durham}{Durham University, Durham DH1 3LE, United Kingdom}
\newcommand{\EIA}{Universidad EIA, Envigado, Antioquia, Colombia}
\newcommand{\ETH}{ETH Zurich, Zurich, Switzerland}
\newcommand{\Edinburgh}{University of Edinburgh, Edinburgh EH8 9YL, United Kingdom}
\newcommand{\FCULport}{Faculdade de Ci{\^e}ncias da Universidade de Lisboa - FCUL, 1749-016 Lisboa, Portugal}
\newcommand{\FederaldeAlfenas}{Universidade Federal de Alfenas, Po{\c{c}}os de Caldas - MG, 37715-400, Brazil}
\newcommand{\FederaldeGoias}{Universidade Federal de Goias, Goiania, GO 74690-900, Brazil}
\newcommand{\FederaldeSaoCarlos}{Universidade Federal de S{\~a}o Carlos, Araras - SP, 13604-900, Brazil}
\newcommand{\FederaldoABC}{Universidade Federal do ABC, Santo Andr{\'e} - SP, 09210-580, Brazil}
\newcommand{\FederaldoRio}{Universidade Federal do Rio de Janeiro,  Rio de Janeiro - RJ, 21941-901, Brazil}
\newcommand{\Fermi}{Fermi National Accelerator Laboratory, Batavia, IL 60510, USA}
\newcommand{\Ferrarauniv}{University of Ferrara, Ferrara, Italy}
\newcommand{\Florida}{University of Florida, Gainesville, FL 32611-8440, USA}
\newcommand{\Fluminense}{Fluminense Federal University, 9 Icara{\'\i} Niter{\'o}i - RJ, 24220-900, Brazil }
\newcommand{\Genova}{Universit{\`a} degli Studi di Genova, Genova, Italy}
\newcommand{\Georgian}{Georgian Technical University, Tbilisi, Georgia}
\newcommand{\GranSasso}{Gran Sasso Science Institute, L'Aquila, Italy}
\newcommand{\GranSassoLab}{Laboratori Nazionali del Gran Sasso, L'Aquila AQ, Italy}
\newcommand{\Granada}{University of Granada {\&} CAFPE, 18002 Granada, Spain}
\newcommand{\Grenoble}{University Grenoble Alpes, CNRS, Grenoble INP, LPSC-IN2P3, 38000 Grenoble, France}
\newcommand{\Guanajuato}{Universidad de Guanajuato, Guanajuato, C.P. 37000, Mexico}
\newcommand{\Harish}{Harish-Chandra Research Institute, Jhunsi, Allahabad 211 019, India}
\newcommand{\Harvard}{Harvard University, Cambridge, MA 02138, USA}
\newcommand{\Hawaii}{University of Hawaii, Honolulu, HI 96822, USA}
\newcommand{\Houston}{University of Houston, Houston, TX 77204, USA}
\newcommand{\Hyderabad}{University of  Hyderabad, Gachibowli, Hyderabad - 500 046, India}
\newcommand{\IFAE}{Institut de F{\'\i}sica d{\textquoteright}Altes Energies (IFAE){\textemdash}Barcelona Institute of Science and Technology (BIST), Barcelona, Spain}
\newcommand{\IFIC}{Instituto de F{\'\i}sica Corpuscular, CSIC and Universitat de Val{\`e}ncia, 46980 Paterna, Valencia, Spain}
\newcommand{\IGFAE}{Instituto Galego de F{\'\i}sica de Altas Enerx{\'\i}as (IGFAE), Universidade de Santiago de Compostela, E-15782, Galicia, Spain}
\newcommand{\INFNBologna}{Istituto Nazionale di Fisica Nucleare Sezione di Bologna, 40127 Bologna BO, Italy}
\newcommand{\INFNCatania}{Istituto Nazionale di Fisica Nucleare Sezione di Catania, I-95123 Catania, Italy}
\newcommand{\INFNFerrara}{Istituto Nazionale di Fisica Nucleare Sezione di Ferrara, I-44122 Ferrara, Italy}
\newcommand{\INFNGenova}{Istituto Nazionale di Fisica Nucleare Sezione di Genova, 16146 Genova GE, Italy}
\newcommand{\INFNLecce}{Istituto Nazionale di Fisica Nucleare Sezione di Lecce, 73100 - Lecce, Italy}
\newcommand{\INFNMilanBicocca}{Istituto Nazionale di Fisica Nucleare Sezione di Milano Bicocca, 3 - I-20126 Milano, Italy}
\newcommand{\INFNMilano}{Istituto Nazionale di Fisica Nucleare Sezione di Milano, 20133 Milano, Italy}
\newcommand{\INFNNapoli}{Istituto Nazionale di Fisica Nucleare Sezione di Napoli, I-80126 Napoli, Italy}
\newcommand{\INFNPadova}{Istituto Nazionale di Fisica Nucleare Sezione di Padova, 35131 Padova, Italy}
\newcommand{\INFNPavia}{Istituto Nazionale di Fisica Nucleare Sezione di Pavia,  I-27100 Pavia, Italy}
\newcommand{\INFNSud}{Istituto Nazionale di Fisica Nucleare Laboratori Nazionali del Sud, 95123 Catania, Italy}
\newcommand{\INR}{Institute for Nuclear Research of the Russian Academy of Sciences, Moscow 117312, Russia}
\newcommand{\IPLyon}{Institut de Physique des 2 Infinis de Lyon, 69622 Villeurbanne, France}
\newcommand{\IPM}{Institute for Research in Fundamental Sciences, Tehran, Iran}
\newcommand{\ISTlisboa}{Instituto Superior T{\'e}cnico - IST, Universidade de Lisboa, Portugal}
\newcommand{\Idaho}{Idaho State University, Pocatello, ID 83209, USA}
\newcommand{\Illinoisinstitute}{Illinois Institute of Technology, Chicago, IL 60616, USA}
\newcommand{\Imperial}{Imperial College of Science Technology and Medicine, London SW7 2BZ, United Kingdom}
\newcommand{\IndGuwahati}{Indian Institute of Technology Guwahati, Guwahati, 781 039, India}
\newcommand{\IndHyderabad}{Indian Institute of Technology Hyderabad, Hyderabad, 502285, India}
\newcommand{\Indiana}{Indiana University, Bloomington, IN 47405, USA}
\newcommand{\Ingenieria}{Universidad Nacional de Ingenier{\'\i}a, Lima 25, Per{\'u}}
\newcommand{\Insubria }{University of Insubria, Via Ravasi, 2, 21100 Varese VA, Italy}
\newcommand{\Iowa}{University of Iowa, Iowa City, IA 52242, USA}
\newcommand{\IowaState}{Iowa State University, Ames, Iowa 50011, USA}
\newcommand{\Iwate}{Iwate University, Morioka, Iwate 020-8551, Japan}
\newcommand{\JINR}{Joint Institute for Nuclear Research, Dzhelepov Laboratory of Nuclear Problems 6 Joliot-Curie, Dubna, Moscow Region, 141980 RU }
\newcommand{\Jammu}{University of Jammu, Jammu-180006, India}
\newcommand{\Jawaharlal}{Jawaharlal Nehru University, New Delhi 110067, India}
\newcommand{\Jeonbuk}{Jeonbuk National University, Jeonrabuk-do 54896, South Korea}
\newcommand{\Jyvaskyla}{University of Jyvaskyla, FI-40014, Finland}
\newcommand{\KEK}{High Energy Accelerator Research Organization (KEK), Ibaraki, 305-0801, Japan}
\newcommand{\KISTI}{Korea Institute of Science and Technology Information, Daejeon, 34141, South Korea}
\newcommand{\KL}{K L University, Vaddeswaram, Andhra Pradesh 522502, India}
\newcommand{\Kansasstate}{Kansas State University, Manhattan, KS 66506, USA}
\newcommand{\Kavli}{Kavli Institute for the Physics and Mathematics of the Universe, Kashiwa, Chiba 277-8583, Japan}
\newcommand{\Kure}{National Institute of Technology, Kure College, Hiroshima, 737-8506, Japan}
\newcommand{\Kyiv}{Taras Shevchenko National University of Kyiv, 01601 Kyiv, Ukraine}
\newcommand{\LIP}{Laborat{\'o}rio de Instrumenta{\c{c}}{\~a}o e F{\'\i}sica Experimental de Part{\'\i}culas, 1649-003 Lisboa and 3004-516 Coimbra, Portugal}
\newcommand{\Lancaster}{Lancaster University, Lancaster LA1 4YB, United Kingdom}
\newcommand{\LawrenceBerkeley}{Lawrence Berkeley National Laboratory, Berkeley, CA 94720, USA}
\newcommand{\Liverpool}{University of Liverpool, L69 7ZE, Liverpool, United Kingdom}
\newcommand{\LosAlmos}{Los Alamos National Laboratory, Los Alamos, NM 87545, USA}
\newcommand{\Louisanastate}{Louisiana State University, Baton Rouge, LA 70803, USA}
\newcommand{\Lucknow}{University of Lucknow, Uttar Pradesh 226007, India}
\newcommand{\Madrid}{Madrid Autonoma University and IFT UAM/CSIC, 28049 Madrid, Spain}
\newcommand{\Manchester}{University of Manchester, Manchester M13 9PL, United Kingdom}
\newcommand{\Massinsttech}{Massachusetts Institute of Technology, Cambridge, MA 02139, USA}
\newcommand{\Maxplanck}{Max-Planck-Institut, Munich, 80805, Germany}
\newcommand{\Medellin}{University of Medell{\'\i}n, Medell{\'\i}n, 050026 Colombia }
\newcommand{\Michigan}{University of Michigan, Ann Arbor, MI 48109, USA}
\newcommand{\Michiganstate}{Michigan State University, East Lansing, MI 48824, USA}
\newcommand{\MilanoBicocca}{Universit{\`a} del Milano-Bicocca, 20126 Milano, Italy}
\newcommand{\MilanoUniv}{Universit{\`a} degli Studi di Milano, I-20133 Milano, Italy}
\newcommand{\Minnduluth}{University of Minnesota Duluth, Duluth, MN 55812, USA}
\newcommand{\Minntwin}{University of Minnesota Twin Cities, Minneapolis, MN 55455, USA}
\newcommand{\Mississippi}{University of Mississippi, University, MS 38677 USA}
\newcommand{\Newmexico}{University of New Mexico, Albuquerque, NM 87131, USA}
\newcommand{\Niewodniczanski}{H. Niewodnicza{\'n}ski Institute of Nuclear Physics, Polish Academy of Sciences, Cracow, Poland}
\newcommand{\Nikhef}{Nikhef National Institute of Subatomic Physics, 1098 XG Amsterdam, Netherlands}
\newcommand{\Northdakota}{University of North Dakota, Grand Forks, ND 58202-8357, USA}
\newcommand{\Northernillinois}{Northern Illinois University, DeKalb, IL 60115, USA}
\newcommand{\Northwestern}{Northwestern University, Evanston, Il 60208, USA}
\newcommand{\NotreDame}{University of Notre Dame, Notre Dame, IN 46556, USA}
\newcommand{\Occidental}{Occidental College, Los Angeles, CA  90041}
\newcommand{\Ohiostate}{Ohio State University, Columbus, OH 43210, USA}
\newcommand{\OregonState}{Oregon State University, Corvallis, OR 97331, USA}
\newcommand{\Oxford}{University of Oxford, Oxford, OX1 3RH, United Kingdom}
\newcommand{\PacificNorthwest}{Pacific Northwest National Laboratory, Richland, WA 99352, USA}
\newcommand{\Padova}{Universt{\`a} degli Studi di Padova, I-35131 Padova, Italy}
\newcommand{\Panjab}{Panjab University, Chandigarh, 160014 U.T., India}
\newcommand{\Parissaclay}{Universit{\'e} Paris-Saclay, CNRS/IN2P3, IJCLab, 91405 Orsay, France}
\newcommand{\Parisuniversite}{Universit{\'e} de Paris, CNRS, Astroparticule et Cosmologie, F-75006, Paris, France}
\newcommand{\Pavia}{Universit{\`a} degli Studi di Pavia, 27100 Pavia PV, Italy}
\newcommand{\Penn}{University of Pennsylvania, Philadelphia, PA 19104, USA}
\newcommand{\PennState}{Pennsylvania State University, University Park, PA 16802, USA}
\newcommand{\PhysicalResearchLaboratory}{Physical Research Laboratory, Ahmedabad 380 009, India}
\newcommand{\Pisa}{Universit{\`a} di Pisa, I-56127 Pisa, Italy}
\newcommand{\Pitt}{University of Pittsburgh, Pittsburgh, PA 15260, USA}
\newcommand{\Pontificia}{Pontificia Universidad Cat{\'o}lica del Per{\'u}, Lima, Per{\'u}}
\newcommand{\PuertoRico}{University of Puerto Rico, Mayaguez 00681, Puerto Rico, USA}
\newcommand{\Punjab}{Punjab Agricultural University, Ludhiana 141004, India}
\newcommand{\QMUL}{Queen Mary University of London, London E1 4NS, United Kingdom }
\newcommand{\Radboud}{Radboud University, NL-6525 AJ Nijmegen, Netherlands}
\newcommand{\Rochester}{University of Rochester, Rochester, NY 14627, USA}
\newcommand{\Royalholloway}{Royal Holloway College London, TW20 0EX, United Kingdom}
\newcommand{\Rutgers}{Rutgers University, Piscataway, NJ, 08854, USA}
\newcommand{\Rutherford}{STFC Rutherford Appleton Laboratory, Didcot OX11 0QX, United Kingdom}
\newcommand{\SLAC}{SLAC National Accelerator Laboratory, Menlo Park, CA 94025, USA}
\newcommand{\SURF}{Sanford Underground Research Facility, Lead, SD, 57754, USA}
\newcommand{\Salento}{Universit{\`a} del Salento, 73100 Lecce, Italy}
\newcommand{\Sanjosestate}{San Jose State University, San Jos{\'e}, CA 95192-0106, USA}
\newcommand{\SergioArboleda}{Universidad Sergio Arboleda, 11022 Bogot{\'a}, Colombia}
\newcommand{\Sheffield}{University of Sheffield, Sheffield S3 7RH, United Kingdom}
\newcommand{\SouthDakotaSchool}{South Dakota School of Mines and Technology, Rapid City, SD 57701, USA}
\newcommand{\SouthDakotaState}{South Dakota State University, Brookings, SD 57007, USA}
\newcommand{\Southcarolina}{University of South Carolina, Columbia, SC 29208, USA}
\newcommand{\SouthernMethodist}{Southern Methodist University, Dallas, TX 75275, USA}
\newcommand{\StonyBrook}{Stony Brook University, SUNY, Stony Brook, NY 11794, USA}
\newcommand{\Sunyatsen}{Sun Yat-Sen University, Guangzhou, 510275}
\newcommand{\Sussex}{University of Sussex, Brighton, BN1 9RH, United Kingdom}
\newcommand{\Syracuse}{Syracuse University, Syracuse, NY 13244, USA}
\newcommand{\Tecnologica }{Universidade Tecnol{\'o}gica Federal do Paran{\'a}, Curitiba, Brazil}
\newcommand{\TexasAMcollege}{Texas A{\&}M University, College Station, Texas 77840}
\newcommand{\TexasAMcorpuscristi}{Texas A{\&}M University - Corpus Christi, Corpus Christi, TX 78412, USA}
\newcommand{\TexasArlington}{University of Texas at Arlington, Arlington, TX 76019, USA}
\newcommand{\Texasaustin}{University of Texas at Austin, Austin, TX 78712, USA}
\newcommand{\Toronto}{University of Toronto, Toronto, Ontario M5S 1A1, Canada}
\newcommand{\Tufts}{Tufts University, Medford, MA 02155, USA}
\newcommand{\UNIST}{Ulsan National Institute of Science and Technology, Ulsan 689-798, South Korea}
\newcommand{\Unifesp}{Universidade Federal de S{\~a}o Paulo, 09913-030, S{\~a}o Paulo, Brazil}
\newcommand{\UniversityCollegeLondon}{University College London, London, WC1E 6BT, United Kingdom}
\newcommand{\ValleyCity}{Valley City State University, Valley City, ND 58072, USA}
\newcommand{\VariableEnergy}{Variable Energy Cyclotron Centre, 700 064 West Bengal, India}
\newcommand{\VirginiaTech}{Virginia Tech, Blacksburg, VA 24060, USA}
\newcommand{\Warsaw}{University of Warsaw, 02-093 Warsaw, Poland}
\newcommand{\Warwick}{University of Warwick, Coventry CV4 7AL, United Kingdom}
\newcommand{\Wellesley}{Wellesley College, Wellesley, MA 02481, USA}
\newcommand{\Wichita}{Wichita State University, Wichita, KS 67260, USA}
\newcommand{\WilliamMary}{William and Mary, Williamsburg, VA 23187, USA}
\newcommand{\Wisconsin}{University of Wisconsin Madison, Madison, WI 53706, USA}
\newcommand{\Yale}{Yale University, New Haven, CT 06520, USA}
\newcommand{\Yerevan}{Yerevan Institute for Theoretical Physics and Modeling, Yerevan 0036, Armenia}
\newcommand{\York}{York University, Toronto M3J 1P3, Canada}
\affiliation{\Abilene}
\affiliation{\Albanysuny}
\affiliation{\Amsterdam}
\affiliation{\Antalya}
\affiliation{\Antananarivo}
\affiliation{\AntonioNarino}
\affiliation{\Argonne}
\affiliation{\Arizona}
\affiliation{\Asuncion}
\affiliation{\Athens}
\affiliation{\Atlantico}
\affiliation{\Augustana}
\affiliation{\Banaras}
\affiliation{\Basel}
\affiliation{\Bern}
\affiliation{\Beykent}
\affiliation{\Birmingham}
\affiliation{\BolognaUniversity}
\affiliation{\Boston}
\affiliation{\Bristol}
\affiliation{\Brookhaven}
\affiliation{\Bucharest}
\affiliation{\CBPF}
\affiliation{\CEASaclay}
\affiliation{\CERN}
\affiliation{\CIEMAT}
\affiliation{\CUSB}
\affiliation{\CalBerkeley}
\affiliation{\CalDavis}
\affiliation{\CalIrvine}
\affiliation{\CalLosangeles}
\affiliation{\CalRiverside}
\affiliation{\CalSantabarbara}
\affiliation{\Caltech}
\affiliation{\Cambridge}
\affiliation{\Campinas}
\affiliation{\CataniaUniversitadi}
\affiliation{\Catolica}
\affiliation{\Charles}
\affiliation{\Chicago}
\affiliation{\ChungAng}
\affiliation{\Cincinnati}
\affiliation{\Cinvestav}
\affiliation{\Colima}
\affiliation{\ColoradoBoulder}
\affiliation{\ColoradoState}
\affiliation{\Columbia}
\affiliation{\CzechAcademyofSciences}
\affiliation{\CzechTechnical}
\affiliation{\DakotaState}
\affiliation{\Dallas}
\affiliation{\DannecyleVieux}
\affiliation{\Daresbury}
\affiliation{\Drexel}
\affiliation{\Duke}
\affiliation{\Durham}
\affiliation{\EIA}
\affiliation{\ETH}
\affiliation{\Edinburgh}
\affiliation{\FCULport}
\affiliation{\FederaldeAlfenas}
\affiliation{\FederaldeGoias}
\affiliation{\FederaldeSaoCarlos}
\affiliation{\FederaldoABC}
\affiliation{\FederaldoRio}
\affiliation{\Fermi}
\affiliation{\Ferrarauniv}
\affiliation{\Florida}
\affiliation{\Fluminense}
\affiliation{\Genova}
\affiliation{\Georgian}
\affiliation{\GranSasso}
\affiliation{\GranSassoLab}
\affiliation{\Granada}
\affiliation{\Grenoble}
\affiliation{\Guanajuato}
\affiliation{\Harish}
\affiliation{\Harvard}
\affiliation{\Hawaii}
\affiliation{\Houston}
\affiliation{\Hyderabad}
\affiliation{\IFAE}
\affiliation{\IFIC}
\affiliation{\IGFAE}
\affiliation{\INFNBologna}
\affiliation{\INFNCatania}
\affiliation{\INFNFerrara}
\affiliation{\INFNGenova}
\affiliation{\INFNLecce}
\affiliation{\INFNMilanBicocca}
\affiliation{\INFNMilano}
\affiliation{\INFNNapoli}
\affiliation{\INFNPadova}
\affiliation{\INFNPavia}
\affiliation{\INFNSud}
\affiliation{\INR}
\affiliation{\IPLyon}
\affiliation{\IPM}
\affiliation{\ISTlisboa}
\affiliation{\Idaho}
\affiliation{\Illinoisinstitute}
\affiliation{\Imperial}
\affiliation{\IndGuwahati}
\affiliation{\IndHyderabad}
\affiliation{\Indiana}
\affiliation{\Ingenieria}
\affiliation{\Insubria }
\affiliation{\Iowa}
\affiliation{\IowaState}
\affiliation{\Iwate}
\affiliation{\JINR}
\affiliation{\Jammu}
\affiliation{\Jawaharlal}
\affiliation{\Jeonbuk}
\affiliation{\Jyvaskyla}
\affiliation{\KEK}
\affiliation{\KISTI}
\affiliation{\KL}
\affiliation{\Kansasstate}
\affiliation{\Kavli}
\affiliation{\Kure}
\affiliation{\Kyiv}
\affiliation{\LIP}
\affiliation{\Lancaster}
\affiliation{\LawrenceBerkeley}
\affiliation{\Liverpool}
\affiliation{\LosAlmos}
\affiliation{\Louisanastate}
\affiliation{\Lucknow}
\affiliation{\Madrid}
\affiliation{\Manchester}
\affiliation{\Massinsttech}
\affiliation{\Maxplanck}
\affiliation{\Medellin}
\affiliation{\Michigan}
\affiliation{\Michiganstate}
\affiliation{\MilanoBicocca}
\affiliation{\MilanoUniv}
\affiliation{\Minnduluth}
\affiliation{\Minntwin}
\affiliation{\Mississippi}
\affiliation{\Newmexico}
\affiliation{\Niewodniczanski}
\affiliation{\Nikhef}
\affiliation{\Northdakota}
\affiliation{\Northernillinois}
\affiliation{\Northwestern}
\affiliation{\NotreDame}
\affiliation{\Occidental}
\affiliation{\Ohiostate}
\affiliation{\OregonState}
\affiliation{\Oxford}
\affiliation{\PacificNorthwest}
\affiliation{\Padova}
\affiliation{\Panjab}
\affiliation{\Parissaclay}
\affiliation{\Parisuniversite}
\affiliation{\Pavia}
\affiliation{\Penn}
\affiliation{\PennState}
\affiliation{\PhysicalResearchLaboratory}
\affiliation{\Pisa}
\affiliation{\Pitt}
\affiliation{\Pontificia}
\affiliation{\PuertoRico}
\affiliation{\Punjab}
\affiliation{\QMUL}
\affiliation{\Radboud}
\affiliation{\Rochester}
\affiliation{\Royalholloway}
\affiliation{\Rutgers}
\affiliation{\Rutherford}
\affiliation{\SLAC}
\affiliation{\SURF}
\affiliation{\Salento}
\affiliation{\Sanjosestate}
\affiliation{\SergioArboleda}
\affiliation{\Sheffield}
\affiliation{\SouthDakotaSchool}
\affiliation{\SouthDakotaState}
\affiliation{\Southcarolina}
\affiliation{\SouthernMethodist}
\affiliation{\StonyBrook}
\affiliation{\Sunyatsen}
\affiliation{\Sussex}
\affiliation{\Syracuse}
\affiliation{\Tecnologica }
\affiliation{\TexasAMcollege}
\affiliation{\TexasAMcorpuscristi}
\affiliation{\TexasArlington}
\affiliation{\Texasaustin}
\affiliation{\Toronto}
\affiliation{\Tufts}
\affiliation{\UNIST}
\affiliation{\Unifesp}
\affiliation{\UniversityCollegeLondon}
\affiliation{\ValleyCity}
\affiliation{\VariableEnergy}
\affiliation{\VirginiaTech}
\affiliation{\Warsaw}
\affiliation{\Warwick}
\affiliation{\Wellesley}
\affiliation{\Wichita}
\affiliation{\WilliamMary}
\affiliation{\Wisconsin}
\affiliation{\Yale}
\affiliation{\Yerevan}
\affiliation{\York}
\author{A.~Abed Abud} \affiliation{\Liverpool}\affiliation{\CERN}
\author{B.~Abi} \affiliation{\Oxford}
\author{R.~Acciarri} \affiliation{\Fermi}
\author{M.~A.~Acero} \affiliation{\Atlantico}
\author{M.~R.~Adames} \affiliation{\Tecnologica }
\author{G.~Adamov} \affiliation{\Georgian}
\author{D.~Adams} \affiliation{\Brookhaven}
\author{M.~Adinolfi} \affiliation{\Bristol}
\author{A.~Aduszkiewicz} \affiliation{\Houston}
\author{J.~Aguilar} \affiliation{\LawrenceBerkeley}
% ...
\collaboration{The DUNE Collaboration}
\noaffiliation

%% file: sections/introduction.tex
\section{Introduction}
\label{sec:intro}

The Deep Underground Neutrino Experiment (DUNE)~\cite{Abi:2020wmh} is a next-generation long-baseline neutrino oscillation experiment which will utilize high-intensity \numu and \anumu beams with peak neutrino energies of $\approx$2.5~GeV over a 1285 km baseline to carry out a detailed study of neutrino mixing. Some of DUNE's key scientific goals are the definitive determination of the neutrino mass ordering, the definitive observation of charge-parity symmetry violation (CPV) for more than 50\% of possible true values of the charge-parity violating phase, \deltacp, and the precise measurement of other three-neutrino oscillation parameters.
These measurements will help guide theory in understanding if there are new symmetries in the neutrino sector and whether there is a relationship between the generational structure of quarks and leptons~\cite{Qian:2015waa}. Observation of CPV in neutrinos would be an important step in understanding the origin of the baryon asymmetry of the universe~\cite{Fukugita:1986hr, Davidson:2008bu}. DUNE has a rich physics program beyond the three-neutrino oscillation accelerator neutrino program described here. These include beyond standard model searches~\cite{Abi:2020kei}, supernova neutrino detection~\cite{Abi:2020lpk}, and solar neutrino detection~\cite{Capozzi:2018dat}. Additional physics possibilities with DUNE are discussed in Refs.~\cite{Abi:2020evt} and~\cite{AbedAbud:2021hpb}.

Neutrino oscillation experiments have so far measured five of the three-neutrino mixing parameters~\cite{Capozzi:2017ipn,deSalas:2020pgw,Esteban:2020cvm}: the three mixing angles $\theta_{12}$, $\theta_{23}$, and $\theta_{13}$; and the two squared-mass differences $\Delta m^{2}_{21}$ and $|\Delta m^{2}_{31}|$, where $\Delta m^2_{ij} = m^2_{i} - m^{2}_{j}$ is the difference between the squares of the neutrino masses.
The neutrino mass ordering (the sign of $\Delta m^{2}_{31}$) is not currently known, though recent results show a weak preference for the normal ordering (NO), where $\Delta m^{2}_{31} > 0$, over the inverted ordering (IO)~\cite{Abe:2021gky,PhysRevD.97.072001,PhysRevLett.123.151803}.
The value of \deltacp is not well known, though neutrino oscillation data are beginning to provide some information on its value~\cite{Abe:2019vii,Abe:2021gky}.

The oscillation probability of $\;\nu^{\bracketbar}_\mu \rightarrow \nu^{\bracketbar}_e$ through matter in the standard three-flavor model and a constant matter density approximation can be written as~\cite{Nunokawa:2007qh}:
\begin{linenomath*}
  \begin{equation}
    \begin{aligned}
      P(\;\nu^{\bracketbar}_\mu \rightarrow \nu^{\bracketbar}_e) & \simeq \sin^2 \theta_{23} \sin^2 2 \theta_{13} 
      \frac{ \sin^2(\Delta_{31} - aL)}{(\Delta_{31}-aL)^2} \Delta_{31}^2 \\
      & + \sin 2 \theta_{23} \sin 2 \theta_{13} \sin 2 \theta_{12}\frac{ \sin(\Delta_{31} - aL)}{(\Delta_{31}-aL)} \Delta_{31} \\
      &\times \frac{\sin(aL)}{(aL)} \Delta_{21} \cos (\Delta_{31} \pm \deltacp) & \\
      & + \cos^2 \theta_{23} \sin^2 2 \theta_{12} \frac {\sin^2(aL)}{(aL)^2} \Delta_{21}^2,
    \end{aligned}
    \label{eqn:appprob}
  \end{equation}
\end{linenomath*}
where
\begin{linenomath*}
  \begin{equation*}
    a = \pm \frac{G_{\mathrm{F}}N_e}{\sqrt{2}} \approx \pm\frac{1}{3500~\mathrm{km}}\left(\frac{\rho}{3.0~\mathrm{g/cm}^{3}}\right),
  \end{equation*}
\end{linenomath*}
$aL \approx 0.367$, $G_{\mathrm{F}}$ is the Fermi constant, $N_e$ is the number density of electrons in the Earth's crust, $\Delta_{ij} = 1.267 \Delta m^2_{ij} L/E_\nu$, $\Delta m^2_{ij}$ is in eV$^{2}$, $L$ is the baseline in km, and $E_\nu$ is the neutrino energy in GeV.
Both \deltacp and $a$ terms are positive (negative) for $\numu \to \nue$ ($\anumu \to \anue$) oscillations. The matter effect asymmetry encapsulated in the $a$ terms arises from the presence of electrons and absence of positrons in the Earth's crust~\cite{Wolfenstein:1977ue,Mikheev:1986gs}. In the analysis described here, the oscillation probabilities are calculated exactly~\cite{Kopp:2006wp}.

DUNE has published sensitivity estimates~\cite{Abi:2020qib} to CPV and the neutrino mass ordering, as well as other oscillation parameters, for large exposures of up to 1104 kiloton-megawatt-years (kt-MW-yr), which show the ultimate sensitivity of the experiment. Sophisticated studies with a detailed treatment of systematic uncertainties were carried out only at large exposures. In this work, DUNE's sensitivity at low exposures is explored further, with a detailed systematics treatment, including an investigation into how the run plan may be optimized to enhance sensitivity to CPV and/or mass ordering. It is shown that DUNE will produce world-leading results at relatively short exposures, which highlights the need for a high-performance near detector complex from the beginning of the experiment. 

The DUNE far detector (FD) will ultimately consist of four modules, each with a 17 kt total mass. The neutrino beam-line has an initial design intensity of 1.2 MW, with a planned upgrade to 2.4 MW. We assume a combined yearly Fermilab accelerator and neutrino beam-line uptime of 57\%~\cite{Abi:2020evt}, and include this in our calculation of calendar years in this work. As the FD deployment schedule and beam power scenarios are both subject to change, the results shown in this work are consistently given in terms of exposure in units of kt-MW-yr, which is agnostic to the exact staging scenario, but can easily be expressed in terms of experiment years for any desired scenario. For example, with two FD modules, assuming a fiducial mass of 10 kt and a beam intensity of 1.2 MW, exposure would accumulate at a rate of 24 kt-MW-yr per calendar year, although a ramp up in beam power is expected before reaching the design intensity in early running. In this work, the single-phase horizontal drift technology is assumed for all FD modules (see Section~\ref{sec:fd}), which is a necessary simplification, but alternative technologies which may have slightly different performance are under investigation for some FD modules.

The analysis framework is described in Section~\ref{sec:analysis_framework}, including a description of the flux, neutrino interaction and detector models and associated uncertainties. A study on the dependence of the sensitivity to CPV and mass ordering on the fraction of data collected in neutrino-enhanced or antineutrino-enhanced running is given in Section~\ref{sec:run_plan_opt}. A detailed study on the CPV and mass ordering sensitivities at low exposures are described in Sections~\ref{sec:cp_sens} and~\ref{sec:mh_sens}, respectively. Finally, conclusions are presented in Section~\ref{sec:conclude}.

%% file: sections/analysis_framework.tex
\section{Analysis framework}
\label{sec:analysis_framework}
This work uses the flux, neutrino interaction and detector model described in detail in Ref.~\cite{Abi:2020qib}, implemented in the CAFAna framework~\cite{CAFAna}. This section provides an overview of the key analysis features. Further details on all aspects can be found in Ref.~\cite{Abi:2020qib}.

\subsection{Neutrino flux}
DUNE will operate with two different beam modes, which depend on the polarity of the electromagnetic horns used to focus secondary particles produced after protons from the primary beam line interact in the target. Forward horn current (FHC) corresponds to neutrino-enhanced running, and reverse horn current (RHC) corresponds to antineutrino-enhanced running. In both FHC and RHC there are significant contributions from neutrinos with energies between 0.5--6 GeV, with a flux peak at $\approx$2.5~GeV. The neutrino flux prediction is generated with G4LBNF~\cite{Aliaga:2016oaz,Abi:2020evt}, using the LBNF optimized beam design~\cite{Abi:2020evt}. Flux uncertainties are due to the production rates and kinematic distributions of hadrons produced in the target and the parameters of the beam line, such as horn currents and horn and target positioning (``focusing uncertainties'')~\cite{Abi:2020evt}. They are evaluated using current measurements of hadron production and estimates of alignment tolerances, giving flux uncertainties of approximately 8\% at the first oscillation maximum, which are highly correlated across energy bins and neutrino flavors. A flux covariance as a function of neutrino energy, beam mode, detector, and neutrino species is generated with a ``toy throw'' approach, which is built using variations (``throws'') of the systematics propagated through the full beam line simulation. To reduce the number of parameters used in the fit, the covariance matrix is diagonalized, and each principal component is treated as an uncorrelated nuisance parameter. Only the first $\approx$30 principal components (out of 108) were found to have a significant effect in the analysis and were included. The shapes of the unoscillated fluxes at the ND and FD are similar, and the differences between them are understood at the percent level.

\subsection{Neutrino interaction model}
The interaction model used is based on GENIE v2.12.10~\cite{Andreopoulos:2009rq,Andreopoulos:2015wxa}, although the combination of models used is much closer to some of the physics tunes available with GENIE v3.00.06, including a number of uncertainties beyond those provided by either GENIE version. These are motivated by data, although the available (anti)neutrino data taken on argon targets is sparse, leading to an uncertainty model that relies in a number of places on light target (mostly hydrocarbon) data. Variations in the cross sections are implemented either using GENIE reweighting parameters, or with {\em ad hoc} weights of events designed to parameterize uncertainties or cross-section corrections currently not implemented within GENIE. The latter were developed using alternative generators or GENIE configurations, or custom weightings using the NUISANCE package~\cite{Stowell:2016jfr}. Further details about the uncertainties used can be found in Ref.~\cite{Abi:2020qib} (Section 3).

The nuclear model which describes the initial state of nucleons in the nucleus is the Bodek-Ritchie global Fermi gas model~\cite{BodekRitchie}, which includes empirical modifications to the nucleon momentum distribution to account for short-range correlation effects. The quasi-elastic model uses the Llewelyn-Smith formalism~\cite{llewelyn-smith} with a simple dipole axial form factor, and BBBA05 vector form factors~\cite{bbba05}. Nuclear screening effects and uncertainties are included based on the T2K 2017/8 parameterization~\cite{Abe:2018wpn} of the Valencia group's~\cite{nieves1,nieves2} Random Phase Approximation model. The Valencia model of the multi-nucleon, 2p2h, contribution to the cross section~\cite{nieves1,nieves2} is used, as described in Ref.~\cite{Schwehr:2016pvn}. Both MINERvA~\cite{Rodrigues:2015hik} and NOvA~\cite{NOvA:2018gge} have shown that this model underpredicts observed event rates on carbon at relevant neutrino energies for DUNE. Modifications to the model are constructed to produce agreement with MINERvA CC-inclusive data~\cite{Rodrigues:2015hik}, which are used in the analysis to introduce additional uncertainties on the 2p2h contribution, with energy dependent uncertainties, and extra freedom between neutrinos and antineutrinos. Uncertainties are added on scaling the 2p2h prediction from carbon to argon on electron-scattering measurements of short-range correlated (SRC) pairs taken on multiple targets~\cite{Colle:2015ena}, separately for neutrinos and antineutrinos. GENIE uses a modified version of the Rein-Sehgal (R-S) model for pion production~\cite{Rein:1980wg}. A data-driven modification to the GENIE model is included based on reanalyzed neutrino--deuterium bubble chamber data~\cite{Wilkinson:2014yfa,Rodrigues:2016xjj}. The Deep Inelastic Scattering (DIS) model implemented in GENIE uses the Bodek-Yang parametrization~\cite{Bodek:2002ps}, and GRV98 parton distribution functions~\cite{Gluck:1998xa}. Hadronization is described by the AKGY model~\cite{Yang:2009zx}, which uses the KNO scaling model~\cite{Koba:1972ng} for invariant masses $W \leq 2.3$ GeV and PYTHIA6~\cite{Sjostrand:2006za} for invariant masses $W \geq 3$ GeV, with a smooth transition between the two models for intermediate invariant masses. Additional uncertainties developed by the NOvA Collaboration~\cite{nova_2018} to describe their resonance to DIS transition region data are also included. The final state interaction model and uncertainties available in GENIE are retained~\cite{Dytman:2011zz,Dytman:2015taa,intranuke_2009}.

The cross sections include terms proportional to the lepton mass, which are significant at low energies where quasielastic processes dominate. Some of the form factors in these terms have significant uncertainties in the nuclear environment. Separate (and anticorrelated) uncertainties on the cross section ratio $\sigma_\mu/\sigma_e$ for neutrinos and antineutrinos are adopted from Ref.~\cite{Day:2012gb}. Additionally, some $\nu_e$ charged-current (CC) interactions occur at four-momentum transfers where $\nu_\mu$ CC interactions are kinematically forbidden, and so cannot be constrained by $\nu_\mu$ cross-section measurements. To reflect this, a 100\% uncertainty is applied in the phase space present for $\nu_e$ but absent for $\nu_\mu$.

\subsection{Near detector simulation and reconstruction}
The near detector (ND) hall will be located 574 m downstream of the proton target and $\approx$60~m underground. The reference design for the DUNE ND system is fully described in Ref.~\cite{AbedAbud:2021hpb}, and consists of a liquid argon (LAr) time projection chamber (TPC) referred to as ND-LAr, a magnetized high-pressure gaseous argon TPC (ND-GAr), and an on-axis beam monitor (SAND). Additionally, ND-LAr and ND-GAr are designed to move perpendicular to the beam axis in order to take data at various off-axis angles (the DUNE-PRISM technique). ND-LAr is a modular detector based on the ArgonCube design~\cite{argoncube_loi, Dwyer:2018phu, arclight}, with a total active LAr volume of $105$~m$^{3}$ (a LAr mass of 147 tons). ND-GAr is implemented in this analysis as a cylindrical TPC filled with a 90/10 mixture of argon and CH$_{4}$ at 10 bar, surrounded by a granular, high-performance electromagnetic calorimeter (ECal). ND-GAr sits immediately downstream of the LAr cryostat and serves as a muon spectrometer for ND-LAr~\cite{AbedAbud:2021hpb}.

Neutrino interactions are simulated in the active volume of ND-LAr. The propagation of neutrino interaction products through the ND-LAr and ND-GAr detector volumes is simulated using a Geant4-based program~\cite{Agostinelli:2002hh}. As pattern recognition and reconstruction software has not yet been fully developed for the ND, this analysis uses a parameterized reconstruction based on the Geant4 simulated energy deposits in active detector volumes.

Only CC-inclusive interactions originating in the LAr are considered in this analysis, with a fiducial volume (FV) which excludes 50 cm from the sides and upstream edge, and 150 cm from the downstream edge of the active region, containing a total fiducial mass of $\approx$50~t. Most muons with kinetic energies greater than 1 GeV exit ND-LAr. Energetic forward-going muons pass into ND-GAr, where their momentum and charge are reconstructed by curvature. Muon energy is reconstructed by range for tracks that stop in the LAr, and the charge cannot be determined event-by-event. Events with muons that exit the LAr active volume and do not match to a track in ND-GAr are rejected, as the muon momentum is not well reconstructed. For FHC beam running, the wrong-sign background is small and the charge is assumed to be negative for all LAr-contained muons. For RHC beam running, a Michel electron is required at the end of these stopped tracks to suppress the wrong-sign $\mu^-$ by a factor of four.

All generated muons and charged pions are evaluated as potential muon candidates. Tracks are classified as muons if their length is at least 1 m, and their mean energy deposit is less than 3 MeV/cm. The minimum length requirement imposes an effective threshold on the true muon kinetic energy of about 200 MeV, but greatly suppresses potential neutral current (NC) backgrounds with low-energy, non-interacting charged pions. Charged-current events are required to have exactly one muon candidate, and if the charge is reconstructed by curvature, it must be of the appropriate sign. Hadronic energy in the ND is reconstructed by summing all charge deposits in the LAr active volume that are not associated with the muon. To remove events where the hadronic energy is badly reconstructed due to charged particles exiting the detector, a veto region is defined as the outer 30 cm of the active volume on all sides, and events with more than 30 MeV total energy deposited in the veto region are rejected. Only a fraction of neutron kinetic energy is typically observed (24\% on average with large fluctuations), resulting in poor energy reconstruction of events with energetic neutrons. The reconstructed neutrino energy, $E_{\nu}^{\mathrm{rec}} = E_{\mu}^{\mathrm{rec}} + E_{\mathrm{had}}^{\mathrm{rec}}$, is the sum of the reconstructed hadronic energy, $E_{\mathrm{had}}^{\mathrm{rec}}$, and the reconstructed muon energy, $E_{\mu}^{\mathrm{rec}}$. The reconstructed inelasticity, $y_{\mathrm{rec}} = 1 - E_{\mu}^{\mathrm{rec}}/E_{\nu}^{\mathrm{rec}}$, is the fraction of the neutrino energy that is carried by hadrons.

\subsection{Far detector simulation and reconstruction}
\label{sec:fd}
The DUNE FD design consists of four separate LArTPC detector modules, each with a total LAr mass of 17 kt, installed $\approx$1.5~km underground at the Sanford Underground Research Facility (SURF)~\cite{Abi:2018dnh}. The technologies to be deployed for the four modules and their order of construction are still under investigation, so in this analysis, only the single-phase design with a horizontal drift~\cite{Abi:2020loh} is used. In this design, signals from drift electrons in the 13.3 $\times$ 12.0 $\times$ 57.5 m$^{3}$ active volume are read out by $\approx$5~mm spaced wires in anode readout planes. Scintillation light produced at the time of the neutrino interaction is detected and used to reconstruct the start time of the electron drift. We have developed a full simulation chain, which generates neutrino events in a Geant4 model of the FD geometry and simulates the electronics readout. We have developed a reconstruction package to calculate efficiencies and reconstructed neutrino energy estimators for the four CC-inclusive FD samples used in the analysis (\numu-like FHC, \nue-like FHC, \anumu-like RHC and \anue-like RHC).

The electronics response to the ionization electrons and scintillation light is simulated in the wire planes and photon detectors, respectively. Algorithms are applied to remove the impact of the LArTPC electric field and electronics response from the raw detector signal to identify hits, and to cluster hits that may be grouped together due to proximity in time and space. Clusters from different wire planes are matched to form high-level objects such as tracks and showers using the Pandora toolkit~\cite{Marshall:2015rfa,Acciarri:2017hat}. Event classification is carried out through image recognition techniques using a convolutional neural network~\cite{cvn_paper} which classifies events as $\nu^{\bracketbar}_{\mu}$-CC, $\nu^{\bracketbar}_{e}$-CC, $\nu^{\bracketbar}_{\tau}$-CC, and NC. The $\nue^{\bracketbar}$ and $\numu^{\bracketbar}$ efficiencies in both beam modes exceed 90\% in the flux peak.

The neutrino energy for $\,\nu^{\bracketbar}_{\mu}$-CC ($\,\nu^{\bracketbar}_{e}$-CC) events is estimated by the sum of the energy of the longest reconstructed track (highest energy reconstructed electromagnetic shower) and the hadronic energy. For both event types, the hadronic energy is estimated from the charge of reconstructed hits that are not in the primary track or shower, and corrections are applied to each hit charge for recombination and electron lifetime effects. For $\nu^{\bracketbar}_{\mu}$-CC events, the energy of the longest track is estimated by range if the track is contained or by multiple Coulomb scattering if it is exiting. For 0.5--4 GeV neutrino energies, the observed neutrino energy resolution is 15--20\%. The muon energy resolution is 4\% for contained tracks and 18\% for exiting tracks. The electron energy resolution is approximately $4\% \oplus 9\%/\sqrt{E}$. The hadronic energy resolution is 34\%.

\subsection{Detector systematics}
Detector effects impact the event selection efficiency as well as the reconstruction of neutrino energy and inelasticity (the variables used in the oscillation fits). The main sources of detector systematic uncertainties are limitations of the expected calibration and modeling of particles in the detector. Important differences between the ND and FD LArTPC design, size, detector environment, and calibration strategy, lead to uncertainties that do not fully correlate between the two detectors. The degree of correlation is under active study, but in this analysis they are treated as being completely uncorrelated. Detailed simulations of detector effects are under development. In this analysis, uncertainties on the energy scale, energy resolution, particle responses, and detector acceptance are included to encapsulate these effects. The absolute scale uncertainties shift the reconstructed energy distributions, while the resolution uncertainties narrow or broaden them.

An uncertainty on the overall energy scale is included in the analysis presented here, as well as particle energy scale and resolution uncertainties that are separate and uncorrelated between four particle classes: muons, charged hadrons, neutrons, and electromagnetic showers. In the ND, muons reconstructed by range in LAr and by curvature in the ND-GAr are treated separately and assigned uncorrelated uncertainties. For each class of particle, uncertainties on the energy scale are introduced as a function of the reconstructed particle energy, $E$, with a constant term, a term proportional to $\sqrt{E}$, and a term proportional to $1/\sqrt{E}$. A 10\% uncertainty on the energy resolution is also included, and treated as uncorrelated between the four particle classes. The parameters produce a shift to the kinematic variables in an event, as opposed to simply assigning a weight to each simulated event. The scale of the uncertainties is motivated by what has been achieved in recent experiments, including calorimetric based approaches (NOvA, MINERvA) and LArTPCs (LArIAT, MicroBooNE, ArgoNeut).

In addition to impacting energy reconstruction, the E-field model also affects the definition of the FD FV, which is sensitive to electron drift. An additional 1\% uncertainty is therefore included on the total fiducial mass, which is conservatively treated as uncorrelated between the $\nu^{\bracketbar}_{\mu}$ and $\nu^{\bracketbar}_{e}$ samples due to the potential distortion caused by large electromagnetic showers in the electron sample.

The FD is sufficiently large that acceptance is not expected to vary significantly as a function of event kinematics. However, the ND acceptance does vary as a function of both muon and hadronic kinematics due to various containment criteria. Uncertainties are evaluated on the muon and hadron acceptance at the ND based on the change in the acceptance as a function of muon kinematics and true hadronic energy.

\subsection{Sensitivity Methods}
Systematics are implemented in the analysis using one-dimensional response functions for each analysis bin, and oscillation weights are calculated exactly, in fine (50 MeV) bins of true neutrino energy. For a given set of inputs --- flux, oscillation parameters, cross sections, detector energy response matrices, and detector efficiency --- an expected event rate can be produced. Minimization is performed using the {\sc minuit}~\cite{James:1994vla} package.

\begin{table}[htbp]
    \centering
    \begin{tabular}{lcc}
      \hline
      Parameter &    Central value & Relative uncertainty \\
      \hline\hline
      $\theta_{12}$ & 0.5903 & 2.3\% \\ 
      $\theta_{23}$ (NO) & 0.866  & 4.1\% \\ 
      $\theta_{23}$ (IO) & 0.869  & 4.0\% \\
      $\theta_{13}$ (NO) & 0.150  & 1.5\% \\ 
      $\theta_{13}$ (IO) & 0.151  & 1.5\% \\
      $\Delta m^2_{21}$ & 7.39$\times10^{-5}$~eV$^2$ & 2.8\% \\
      $\Delta m^2_{32}$ (NO) & 2.451$\times10^{-3}$~eV$^2$ &  1.3\% \\
      $\Delta m^2_{32}$ (IO) & -2.512$\times10^{-3}$~eV$^2$ &  1.3\% \\
      $\rho$ & 2.848 g cm$^{-3}$ & 2\% \\
      \deltacp (NO) & -2.53 (rad.) & -- \\
      \deltacp (IO) & -1.33 (rad.) & -- \\
      \hline
    \end{tabular}
    \caption{Central value and relative uncertainty of neutrino oscillation parameters from a global fit~\cite{Esteban:2018azc,nufitweb} to neutrino oscillation data. The matter density is taken from Ref.~\cite{Roe:2017zdw}. Because the probability distributions are somewhat non-Gaussian (particularly for $\theta_{23}$), the relative uncertainty is computed using 1/6 of the 3$\sigma$ allowed range from the fit, rather than 1/2 of the 1$\sigma$ range. For $\theta_{23}$, $\theta_{13}$, and $\Delta m^2_{31}$, the best-fit values and uncertainties depend on whether NO or IO is assumed. The best fit for \deltacp is used as a test point in the analysis, but no uncertainty is assigned.}
    \label{tab:oscpar_nufit}
\end{table}

Oscillation sensitivities are obtained by simultaneously fitting the \numu-like FHC, \nue-like FHC, \anumu-like RHC and \anue-like RHC FD spectra along with the $\nu_{\mu}$ FHC and $\bar{\nu}_{\mu}$ RHC samples from the ND. In the studies, all oscillation parameters shown in Table~\ref{tab:oscpar_nufit} are allowed to vary. Gaussian penalty terms (taken from Table~\ref{tab:oscpar_nufit}) are applied to $\theta_{12}$, \dm{21}, and the matter density, $\rho$, of the Earth along the DUNE baseline~\cite{Roe:2017zdw}. Some studies presented in this work include a Gaussian penalty term on $\theta_{13}$ (also taken from NuFIT 4.0, given in Table~\ref{tab:oscpar_nufit}), which is precisely measured by experiments sensitive to reactor antineutrino disappearance~\cite{Abrahao:2020ztg,Adey:2018zwh,Bak:2018ydk}. The remaining parameters, \sinst{23}, $\Delta m^{2}_{32}$, and \deltacp are allowed to vary freely, with no penalty terms. The penalty terms are treated as uncorrelated with each other, and uncorrelated with other parameters.

Flux, cross-section, and FD detector parameters are allowed to vary in the fit, but are constrained by a penalty term corresponding to the prior uncertainty. ND detector uncertainties are included via a covariance matrix based on the shape difference between ND prediction and the ``data'' (which comes from the simulation in this sensitivity study). The covariance matrix is constructed with a throwing technique. For each ``throw'', all ND energy scale, resolution, and acceptance parameters are simultaneously thrown according to their respective uncertainties, and the modified prediction is produced by varying the relevant quantities away from the nominal prediction according to the thrown parameter values. The bin-to-bin covariance is determined by comparing the resulting spectra with the nominal prediction, in the same binning as is used in the oscillation sensitivity analysis.

The compatibility of a particular oscillation hypothesis with both ND and FD data is evaluated using the standard Poisson log-likelihood ratio~\cite{Tanabashi:2018oca}:
\begin{linenomath*}
  \begin{equation}
    \begin{aligned}
      \chi^2(\vec{\vartheta}, \vec{x}) &= -2\log\mathcal{L}(\vec{\vartheta}, \vec{x}) \\
      &= 2\sum_i^{N_{\rm bins}}\left[ M_i(\vec{\vartheta}, \vec{x})-D_i+D_i\ln\left({D_i\over M_i(\vec{\vartheta}, \vec{x})}\right) \right] \\
      &+ \sum_{j}^{N_{\mathrm{systs}}}\left[ \frac{\Delta x_{j}}{\sigma_{j}} \right]^{2} \\
      &+ \sum^{N^{\mathrm{ND}}_{\mathrm{bins}}}_{k}\sum^{N^{\mathrm{ND}}_{\mathrm{bins}}}_{l} \left(M_k(\vec{x})-D_k \right) V^{-1}_{kl}\left(M_l(\vec{x})-D_l \right),
    \end{aligned}
    \label{eq:chisq}
  \end{equation}
\end{linenomath*}
where $\vec{\vartheta}$ and $\vec{x}$ are the vector of oscillation parameter and nuisance parameter values, respectively; $M_i(\vec{\vartheta}, \vec{x})$ and $D_{i}$ are the MC expectation and fake data in the $i$th reconstructed bin (summed over all selected samples), with the oscillation parameters neglected for the ND; $\Delta x_{j}$ and $\sigma_{j}$ are the difference between the nominal and current value, and the prior uncertainty on the $j$th nuisance parameter; and $V_{kl}$ is the covariance matrix between ND bins described previously. To protect against false minima, all fits are repeated starting at four different \deltacp values (-$\pi$, -$\pi$/2, 0, $\pi$/2), in both mass orderings, and in both \sinst{23} octants, and the lowest obtained $\chi^{2}$ value is taken as the true minimum.

\begin{table}
  \centering
  \begin{tabular}{lcc}
    \hline
    Parameter & Prior & Range\\ \hline\hline
    $\sin^{2}\theta_{23}$ & Uniform & [0.4; 0.6] \\
    $|\Delta m^{2}_{32}|$ ($\times 10^{-3}$ eV$^{2}$) & Uniform & $|[2.3;2.7]|$ \\
    \deltacp ($\pi$) & Uniform & [-1;1] \\
    $\theta_{13}$ (NO) & Gaussian & 0.1503 $\pm$ 0.0023 (rad.)  \\
    $\theta_{13}$ (IO) & Gaussian & 0.1510 $\pm$ 0.0023 (rad.)  \\
    \hline
  \end{tabular}
  \caption{Treatment of the oscillation parameters for the simulated data set studies. The value and uncertainty for $\theta_{13}$ in both NO and IO used in the analysis come from NuFIT 4.0~\cite{Esteban:2018azc,nufitweb}.}
  \label{table:OA_throw}
\end{table}
Two approaches are used for the sensitivity studies presented in this work. Asimov studies~\cite{Cowan:2010js} are carried out (in Section~\ref{sec:run_plan_opt}) in which the fake (Asimov) dataset is the same as the nominal MC. In these, the true value of all systematic uncertainties and oscillation parameters are set to their nominal value (see Table~\ref{tab:oscpar_nufit}) except the parameters of interest, which are set to a test point. Then a fit is carried out in which all parameters can vary, constrained by their prior uncertainty where applicable. For the smallest exposures investigated with an Asimov study in this work, all samples have at least 100 events, satisfying the Gaussian approximation inherent in the Asimov method. Toy throw studies are performed (in Sections~\ref{sec:cp_sens} and~\ref{sec:mh_sens}) in which an ensemble of systematic, oscillation parameter and statistical throws are made. Systematic throws are made according to their prior Gaussian uncertainties, oscillation parameters are randomly chosen as described in Table~\ref{table:OA_throw}, and Poisson fluctuations are then applied to all analysis bins, based on the mean event count for each bin after the systematic adjustments have been applied. For each throw in the ensemble, the test statistic is minimized, and the best-fit value of all parameters is determined. The expected resolution for parameters of interest are then determined from the spread in the distribution of their post-fit values.

Asimov studies are computationally efficient, and for Gaussian parameters and uncertainties, give a good sense of the median sensitivity of an experiment. Toy throwing studies are computationally expensive, fully explore the parameter space, and make fewer assumptions about the behavior of parameters and uncertainties.

\subsection{Near and far detector samples and statistics}
In this work, the sensitivity as a function of FD exposure is explored and results are reported in terms of kt-MW-yr, which does not assume any specific FD or beam intensity staging scenario. However, the ND used in this analysis (ND-LAr with a dowstream muon spectrometer) is assumed not to be staged, and as such the ND sample size corresponding to a particular FD exposure will vary based on the staging scenario. The nominal staging scenario from Ref.~\cite{Abi:2020qib} is therefore retained for the purpose of normalizing the ND samples at each FD exposure. In that scenario, a 7 year exposure corresponds to 336 kt-MW-yr at the FD, and 480 t-MW-yr at the ND, summed over both beam modes. The ND statistics used in this analysis are scaled assuming this ratio throughout, using the same fraction of exposure in each beam mode as used at the FD. The ND samples used in this analysis are relatively quickly systematics limited in both beam modes, and so these approximations are unlikely to have a significant impact on the results.

\begin{figure*}
  \centering
  \subfloat[FHC]{\includegraphics[width=0.8\linewidth]{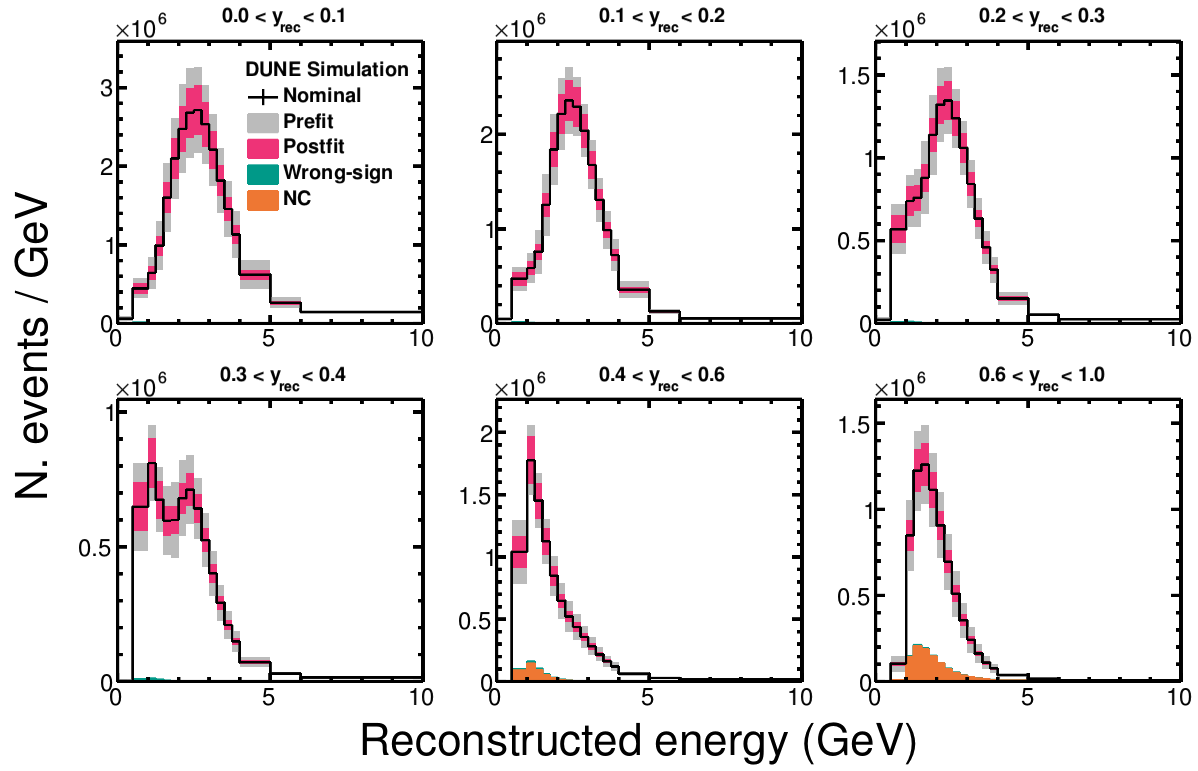}}\\
  \subfloat[RHC]{\includegraphics[width=0.8\linewidth]{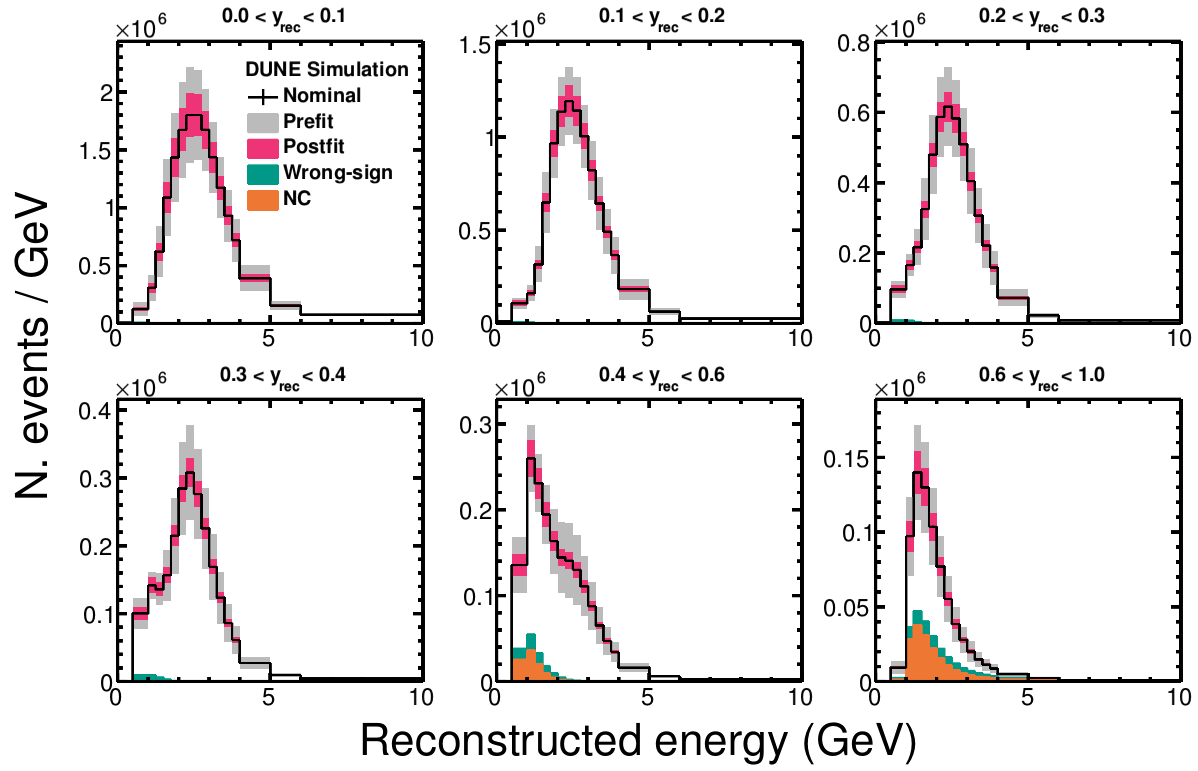}}
  \caption{ND samples in both FHC and RHC, shown in the reconstructed neutrino energy and reconstructed inelasticity binning ($y_{\mathrm{rec}}$) used in the analysis, for a 105 t-MW-yr exposure (equivalent to a 100 kt-MW-yr exposure at the FD), with an equal split between FHC and RHC. The size of the systematic uncertainty bands from all of the flux, cross-section and ND detector systematics used in the analysis are shown, as well as the postfit uncertainty bands obtained by performing an Asimov fit to the ND data. NC backgrounds and wrong-sign contributions to the total event rate are also shown. Statistical uncertainties are too small to be visible on this plot scale.}
 \label{fig:nd_samples}
\end{figure*}
The oscillation analysis presented here includes two CC-inclusive samples originating in the ND-LAr FV, an FHC $\nu_{\mu}$ and an RHC $\bar{\nu}_{\mu}$ sample. These samples are both binned in two dimensions, as a function of reconstructed neutrino energy and inelasticity, $y_{\mathrm{rec}} = 1 - E^{\mathrm{rec}}_{\mu}/E^{\mathrm{rec}}_{\nu}$. The sample distributions for both FHC and RHC are shown in Figure~\ref{fig:nd_samples} for an exposure of 105 t-MW-yr, corresponding to 100 kt-MW-yr at the far detector with the assumptions stated above. The size of the systematic uncertainty bands from all of the flux, cross-section and ND detector systematics used in the analysis and described above are shown, as well as the postfit uncertainty bands obtained by performing an Asimov fit to the ND data. It is clear that even after a relatively small exposure of 105 t-MW-yr, the ND samples are very high statistics, and are systematics limited in the binning used in the analysis. Backgrounds in the $\nu^{\bracketbar}_{\mu}$-CC samples are also shown in Figure~\ref{fig:nd_samples}. NC backgrounds are predominantly from NC $\pi^{\pm}$ production where the pion leaves a long track and does not shower. Wrong-sign contamination in the beam is a background where the charge of the muon is not reconstructed, which particularly affects low reconstructed neutrino energies in RHC. The wrong-sign background is also larger at high reconstructed inelasticity, $y_{\mathrm{rec}}$, due to the kinematics of neutrino and antineutrino scattering.

\begin{figure}[htbp]
 \subfloat[FHC]{\includegraphics[width=0.8\linewidth]{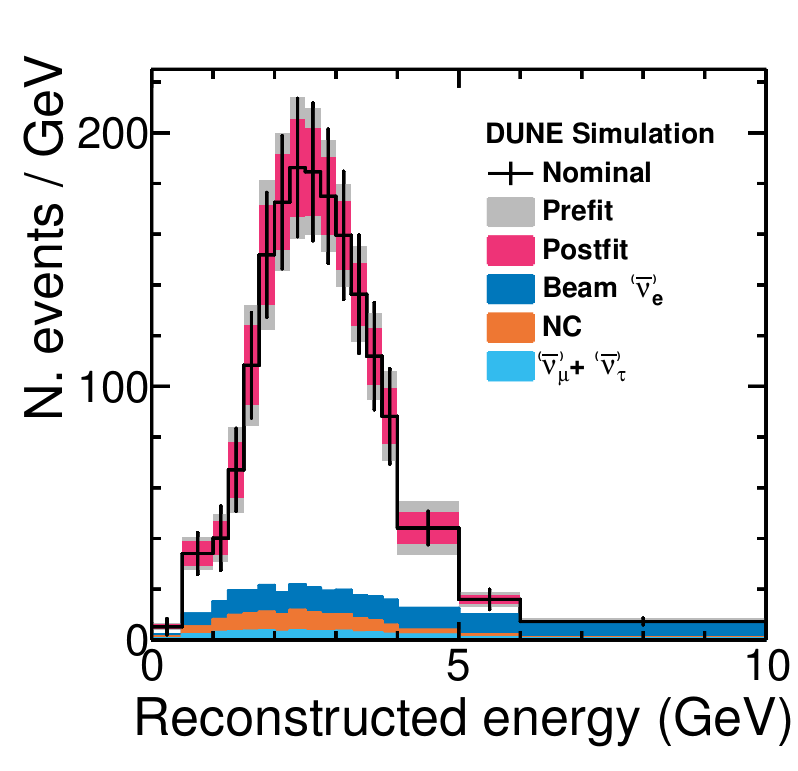}}\\
 \subfloat[RHC]{\includegraphics[width=0.8\linewidth]{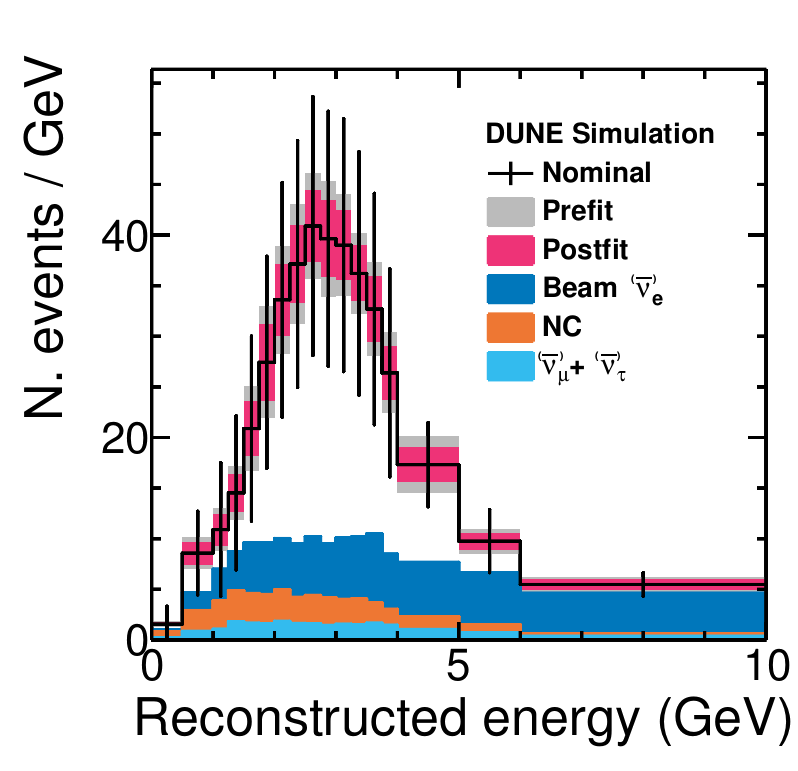}}
 \caption{Reconstructed energy distribution of selected CC $\;\nue^{\bracketbar}$-like events in the FD, for a 50 kt-MW-yr exposure in both FHC and RHC beam modes, for a total 100 kt-MW-yr exposure. The plots are shown for NO, all other oscillation parameters are set to their NuFIT 4.0 best-fit values (see Table~\ref{tab:oscpar_nufit}). The size of the systematic uncertainty bands from all of the flux, cross-section and FD detector systematics used in the analysis are shown, as well as the postfit uncertainty bands with parameters constrained by ND data. Backgrounds are also shown, the largest contribution comes from intrinsic $\nue^{\bracketbar}$ contamination in the beam, although NC and other flavors, $\numu^{\bracketbar} + \nutau^{\bracketbar}$, also contribute.}
 \label{fig:appspectra}
\end{figure}
The expected FD FHC \nue and RHC \anue samples are shown in Figure~\ref{fig:appspectra} for a 100 kt-MW-yr total FD exposure, split equally between FHC and RHC beam modes. The systematic uncertainty bands with and without the ND constraint applied are shown, as well as the background contributions. There are contributions from both \nue and \anue in both beam modes. The NC, intrinsic beam $\nue^{\bracketbar}$, and wrong flavor contamination is also shown; the largest background comes from the intrinsic $\;\nue^{\bracketbar}$ beam contribution in both modes. After a 50 kt-MW-yr exposure in FHC, the \nue sample statistical uncertainty is close to the systematic uncertainty before the ND constraint, although is still clearly statistics limited when the ND constraint is applied. The \anue sample is still strongly statistics limited after 50 kt-MW-yr exposure in RHC. The difference is largely due to the difference in the \nue and \anue cross sections. 

\begin{figure}[htbp]
  \subfloat[FHC]{\includegraphics[width=0.8\linewidth]{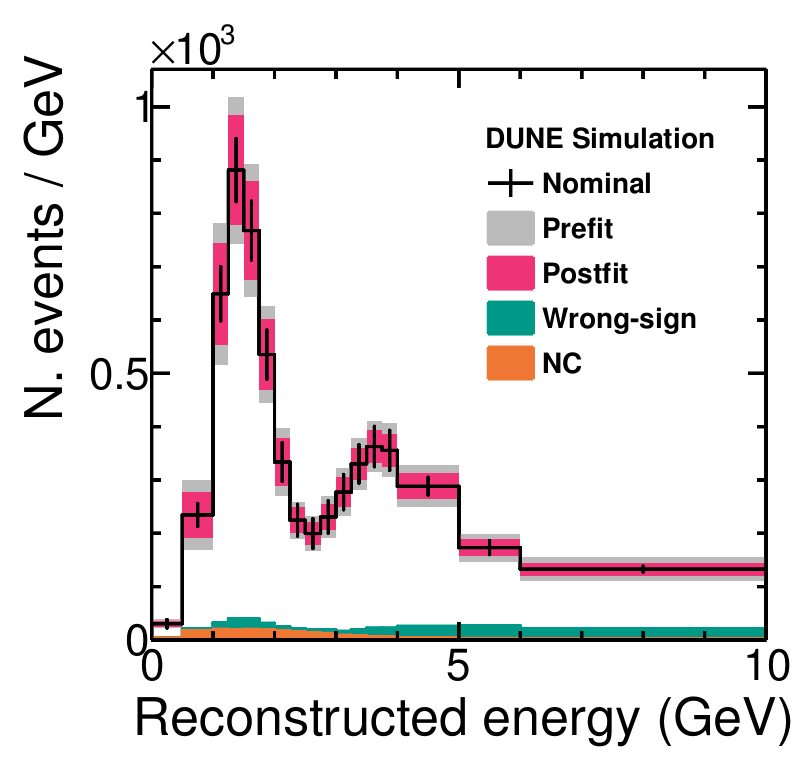}}\\
  \subfloat[RHC]{\includegraphics[width=0.8\linewidth]{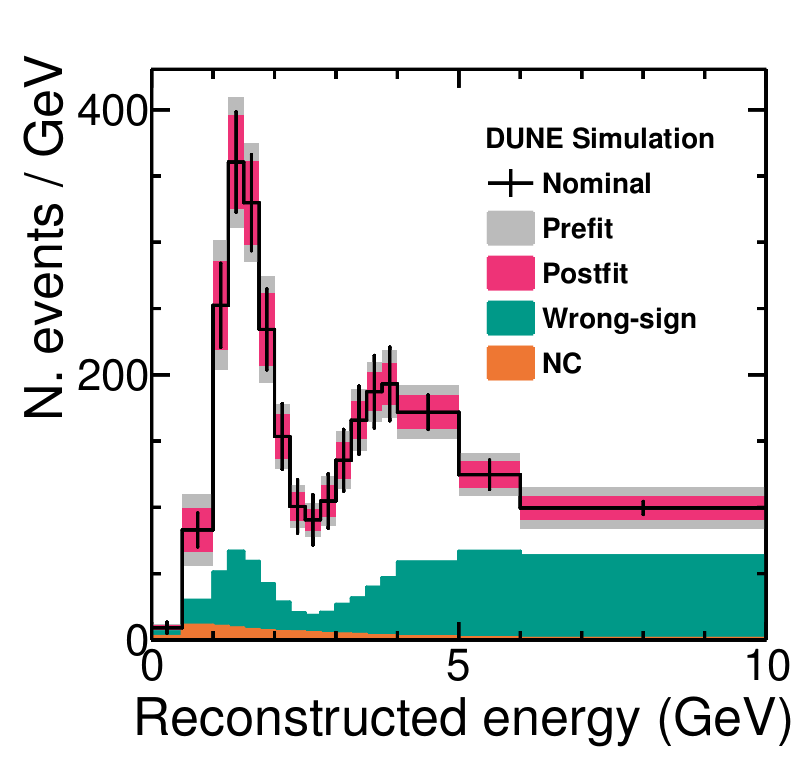}}
\caption{Reconstructed energy distribution of selected CC $\;\nu^{\bracketbar}_{\mu}$-like events in the FD, for 50 kt-MW-yr exposure in both FHC and RHC beam modes, for a total 100 kt-MW-yr exposure. The plots are shown for NO, all other oscillation parameters are set to their NuFIT 4.0 best-fit values (see Table~\ref{tab:oscpar_nufit}). The size of the systematic uncertainty bands from all of the flux, cross-section and ND detector systematics used in the analysis are shown, as well as the postfit uncertainty bands with parameters constrained by ND data. NC backgrounds and wrong-sign contributions to the event rate are also shown.}
\label{fig:disspectra}
\end{figure}
The expected FD FHC \numu and RHC \anumu samples are shown in Figure~\ref{fig:disspectra} for a 100 kt-MW-yr total FD exposure, split equally between FHC and RHC beam modes. The systematic uncertainty bands with and without the ND constraint applied are shown, as well as the background contributions. Although the wrong-sign \numu contribution to the RHC \anumu sample is shown separately, it still provides useful information for constraining the oscillation parameters and is included in the analysis. The statistics are much higher than in Figure~\ref{fig:appspectra}; the statistical uncertainty on the \numu FHC sample is smaller than the systematic uncertainty band for some regions of phase space, even after the ND constraint is applied, although the statistical uncertainty is larger than the constrained systematic uncertainty in the ``dip'' region, around 2.5 GeV, which is likely to have the most impact on the analysis. The statistical uncertainty on the \anumu RHC sample is larger, again due to the smaller \anumu (than \numu) cross section and lower fluxes in RHC running. The statistical uncertainty around the 2.5 GeV dip region is significantly larger than the systematic uncertainty band, although as for the FHC \numu sample, the statistical uncertainty is smaller than the systematics for some regions of the parameter space.

Events with a reconstructed neutrino energy of less than 0.5 GeV (which are shown in Figures~\ref{fig:nd_samples}, ~\ref{fig:appspectra} and~\ref{fig:disspectra}) or a reconstructed neutrino energy greater than 10 GeV are not included in the analysis for any of the FD or ND samples.

%% file: sections/run_plan_opt.tex
\section{Run plan optimization}
\label{sec:run_plan_opt}
In previous DUNE sensitivity studies~\cite{Abi:2020qib}, equal running times in FHC and RHC were assumed, based on early sensitivity estimates for different scenarios. In this section, the dependence of the median CPV and mass ordering significances are studied, for different fractions of time spent in each beam mode, using the full analysis framework described in Section~\ref{sec:analysis_framework}.
\begin{figure*}[htbp]
  \centering
  \subfloat[NO, with $\theta_{13}$-penalty] {\includegraphics[width=0.4\linewidth]{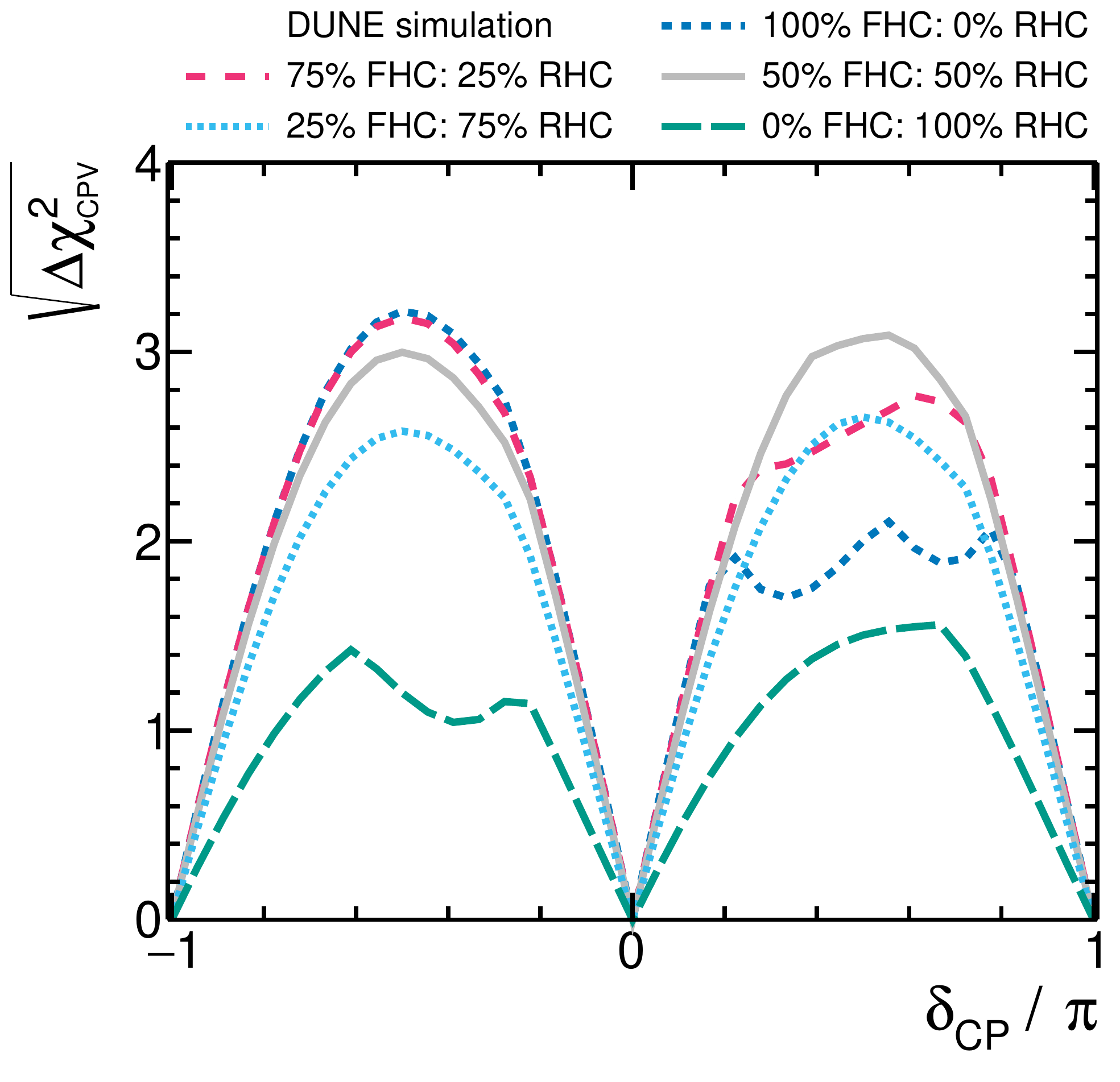}}
  \subfloat[IO, with $\theta_{13}$-penalty] {\includegraphics[width=0.4\linewidth]{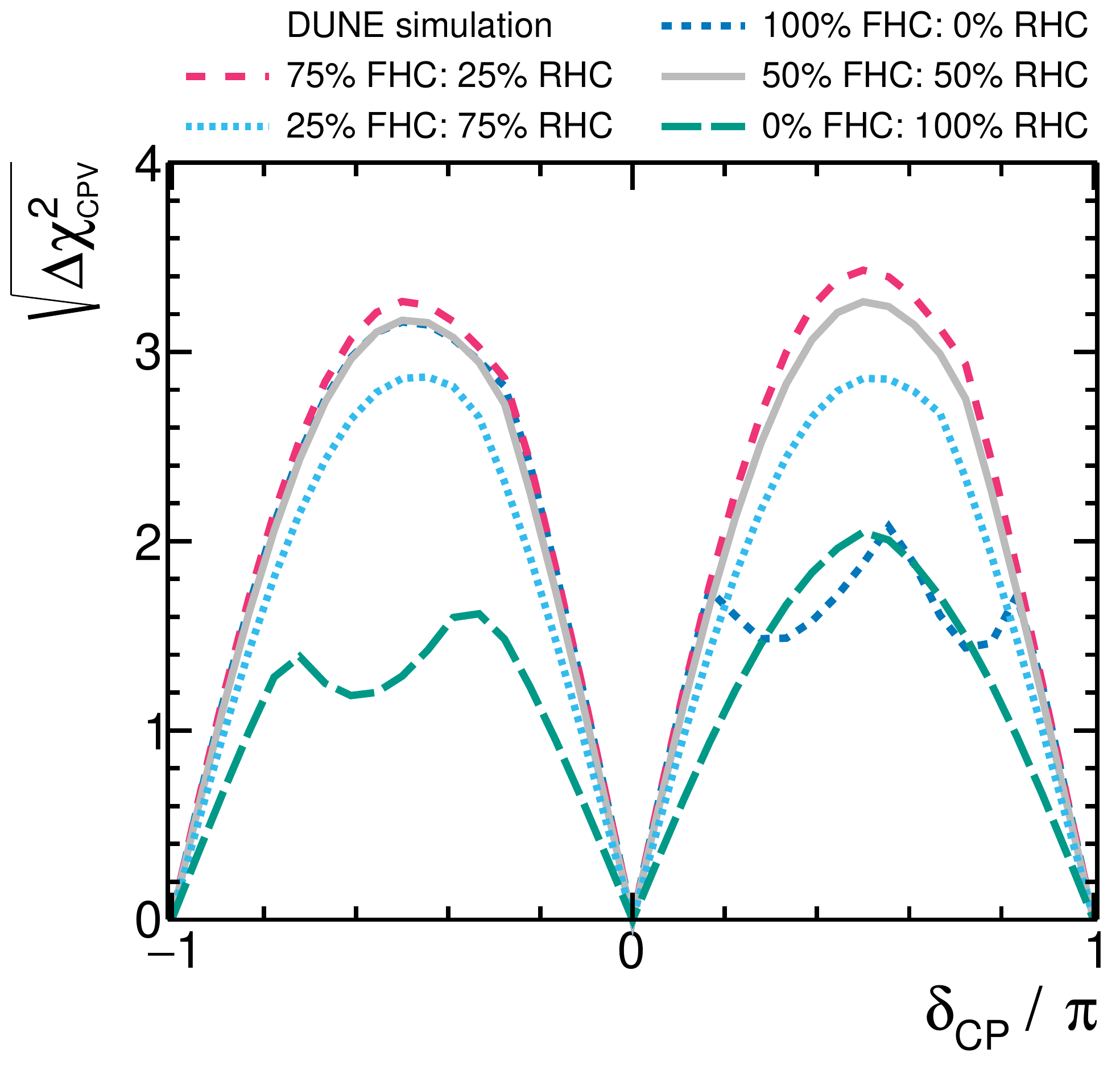}}\\
  \subfloat[NO, no $\theta_{13}$-penalty]   {\includegraphics[width=0.4\linewidth]{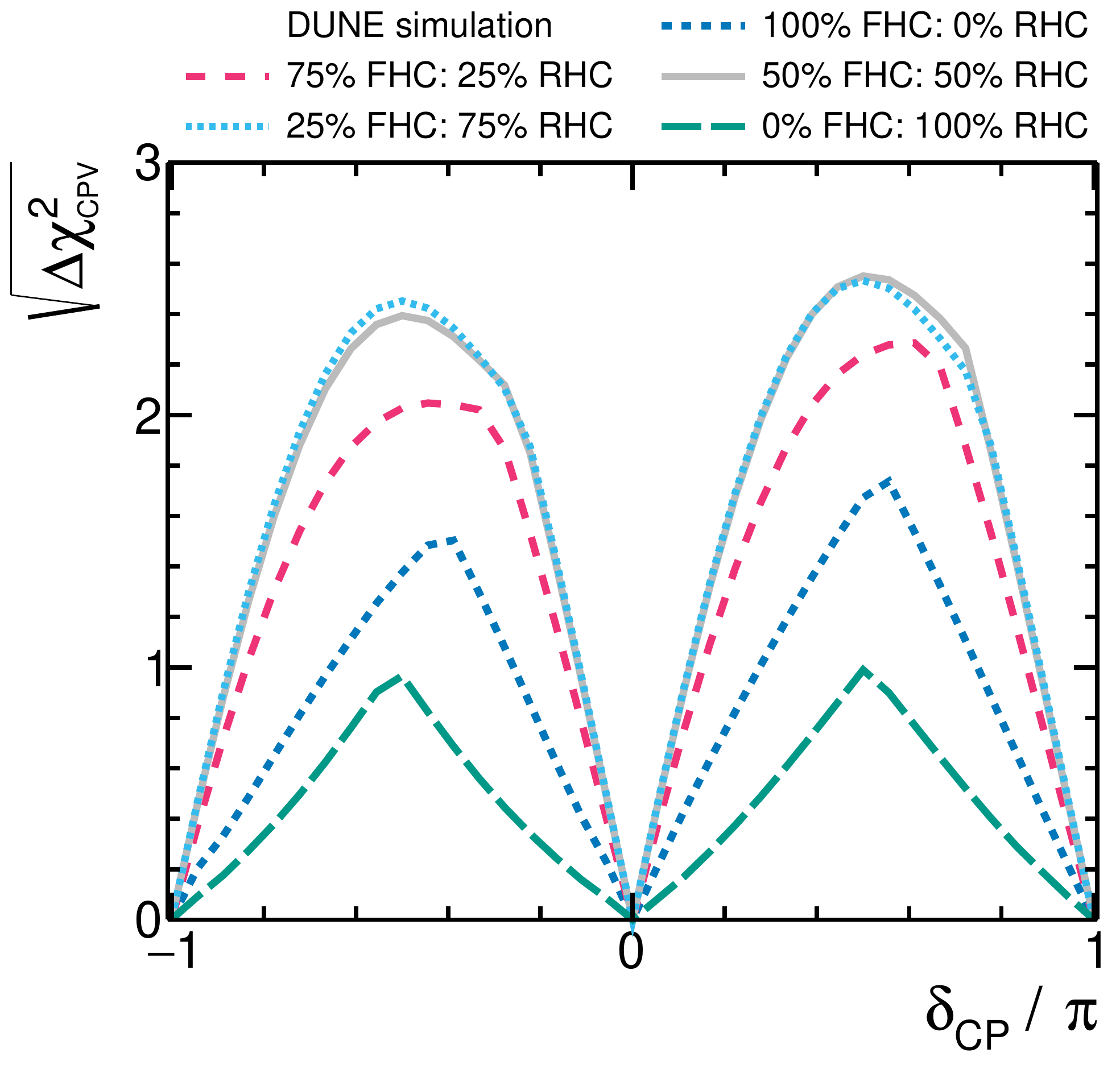}}
  \subfloat[IO, no $\theta_{13}$-penalty]   {\includegraphics[width=0.4\linewidth]{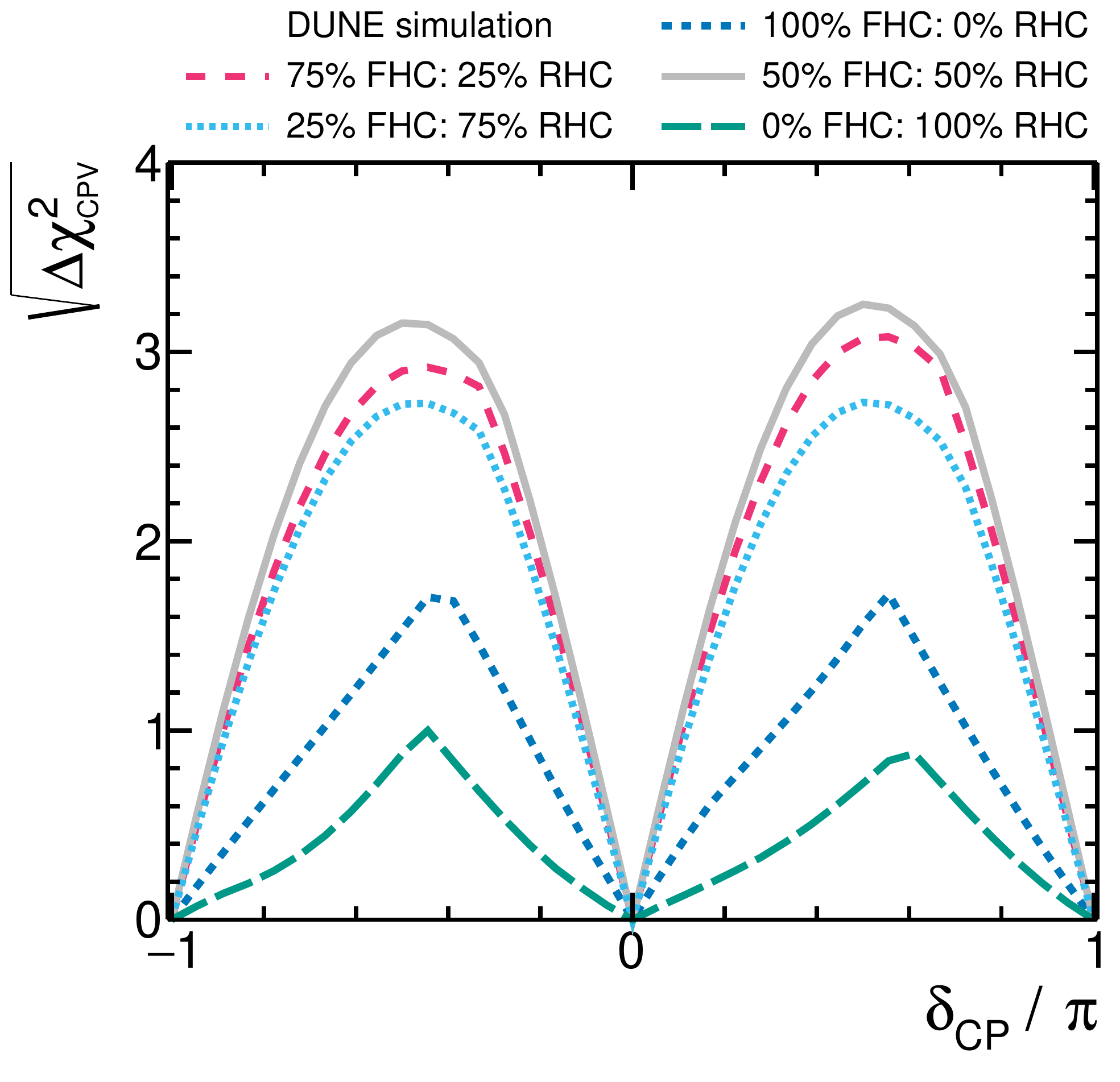}}
  \caption{The Asimov CPV sensitivity as a function of the true value of \deltacp, for a total exposure of 100 kt-MW-yr with different fractions of FHC and RHC running, with and without a $\theta_{13}$ penalty applied in the fit. Results are shown for both true normal and inverted ordering, with the true oscillation parameter values set to the NuFit 4.0 best fit point in each ordering (see Table~\ref{tab:oscpar_nufit}).}
  \label{fig:run_opt_cpv}
\end{figure*}

Figure~\ref{fig:run_opt_cpv} shows DUNE's Asimov sensitivity to CPV for a total 100 kt-MW-yr far detector exposure, with different fractions of FHC and RHC running, at the NuFIT 4.0 best fit value in both NO and IO (see Table~\ref{tab:oscpar_nufit}), shown with and without a penalty on $\theta_{13}$ applied. For each point tested, all oscillation and nuisance parameters are allowed to vary, and three fits are carried out, two where \deltacp is set to the CP-conserving values $\deltacp = 0$ and $\deltacp = \pm\pi$, the minimum of which is the CP-conserving best-fit value, and another where \deltacp is allowed to vary. The difference in the best-fit $\chi^{2}$ values is calculated:
\begin{linenomath*}
  \begin{equation}
    \dchisqCPV = \min\left\{\chi^{2}_{\deltacp = 0},\chi^{2}_{\deltacp = \pm\pi}\right\} - \chi^{2}_{\mathrm{CPV}},
    \label{eq:cpv_chi2}
  \end{equation}
\end{linenomath*}
\noindent and the square root of the difference is used as the figure of merit on the y-axis in Figure~\ref{fig:run_opt_cpv}. There are some caveats associated with this figure of merit, which are discussed in Section~\ref{sec:cp_sens}. A 100 kt-MW-yr exposure is shown as it was identified in Ref~\cite{Abi:2020qib} as the exposure at which DUNE's median CPV significance exceeds 3$\sigma$ at $\deltacp = \pm\pi/2$, an important milestone in DUNE's physics program (with equal beam mode running). 

Figure~\ref{fig:run_opt_cpv} shows that when the reactor constraint on $\theta_{13}$ is included, the sensitivity to CPV can be increased in some regions of \deltacp parameter space with more FHC than RHC running. However, this degrades the sensitivity in other regions, most notably for $\deltacp > 0$ regardless of the true mass ordering. This is due to a degeneracy between \deltacp and the octant of $\sinstt{23}$ because both parameters impact the rate of \nue appearance. The degeneracy is resolved by including antineutrino data; the octant of $\sinstt{23}$ affects the rate of \nue and \anue appearance in the same way, but the effect of \deltacp is reversed for antineutrinos.

For regions of phase space where the octant degeneracy does not affect the result (e.g., $\sin^{2}\theta_{23} \approx 0.5$), there is no degradation in the sensitivity, and enhanced FHC running increases the sensitivity for all values of \deltacp. Increasing the fraction of RHC decreases the sensitivity for the entire \deltacp range when the reactor $\theta_{13}$ constraint is included, relative to equal beam mode running. This is due to the lower statistics of the \anue sample (see Figure~\ref{fig:appspectra}) because of the reduced antineutrino flux and cross section. For short exposures, DUNE will not have a competitive independent measurement of $\theta_{13}$, so the main analysis will include the reactor $\theta_{13}$ constraint. Nonetheless, it is instructive to look at the results without the penalty applied. In this case, the sensitivity is severely degraded (as expected) for 100\% running in either beam mode.

\begin{figure*}[htbp]
  \centering
  \subfloat[NO, with $\theta_{13}$-penalty]  {\includegraphics[width=0.4\linewidth]{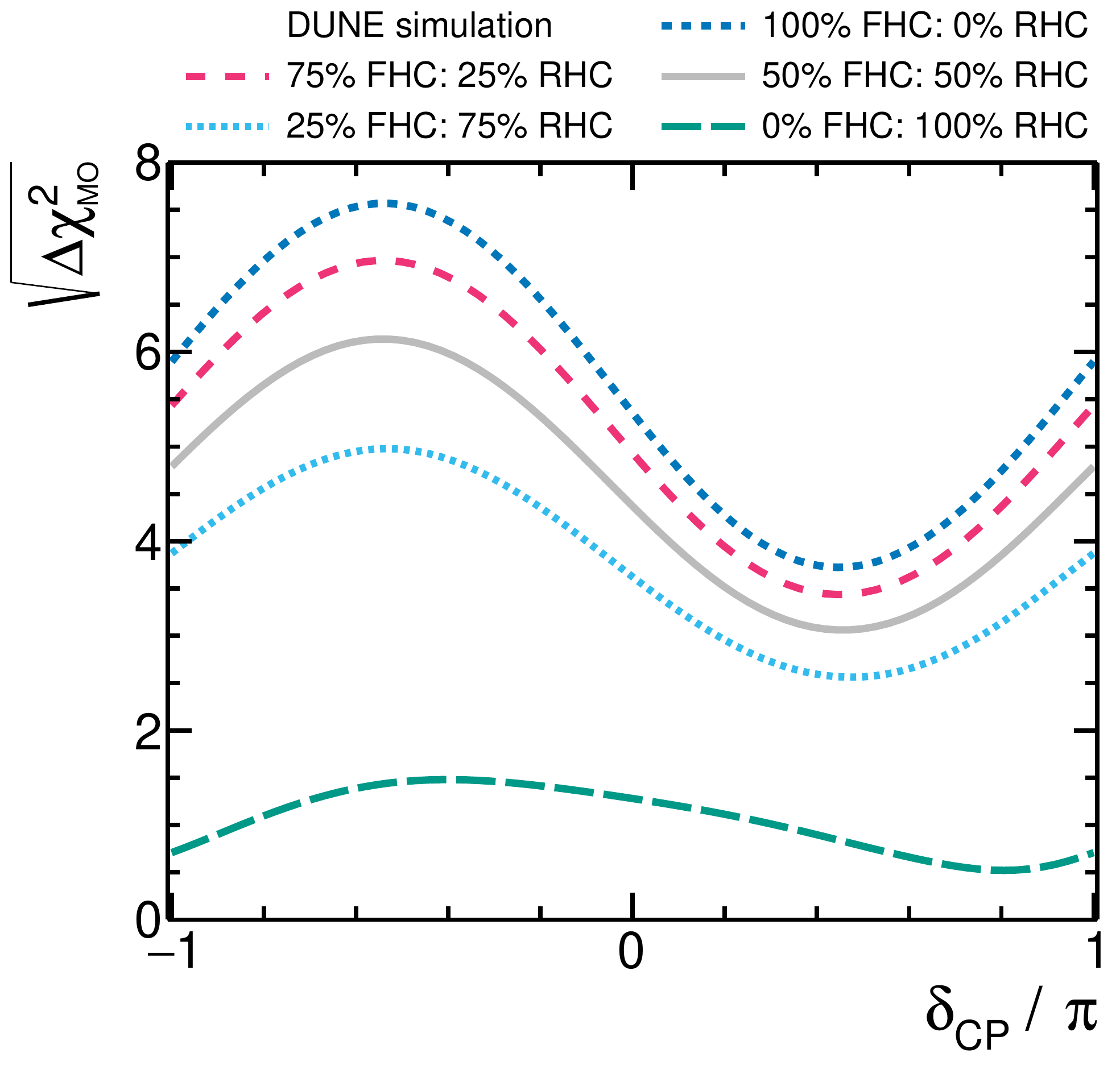}}
  \subfloat[IO, with $\theta_{13}$-penalty]  {\includegraphics[width=0.4\linewidth]{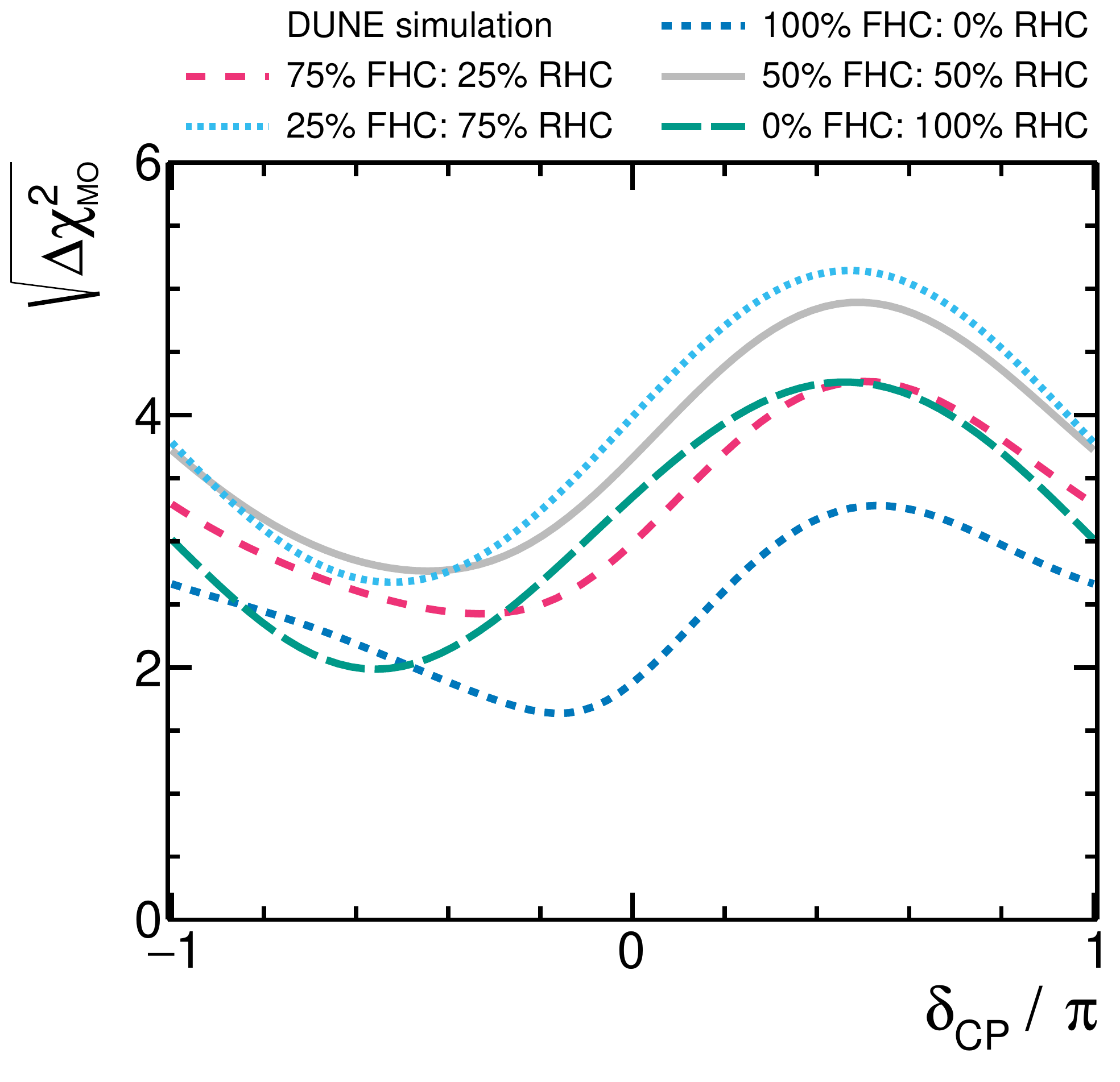}}\\
  \subfloat[NO, no $\theta_{13}$-penalty]    {\includegraphics[width=0.4\linewidth]{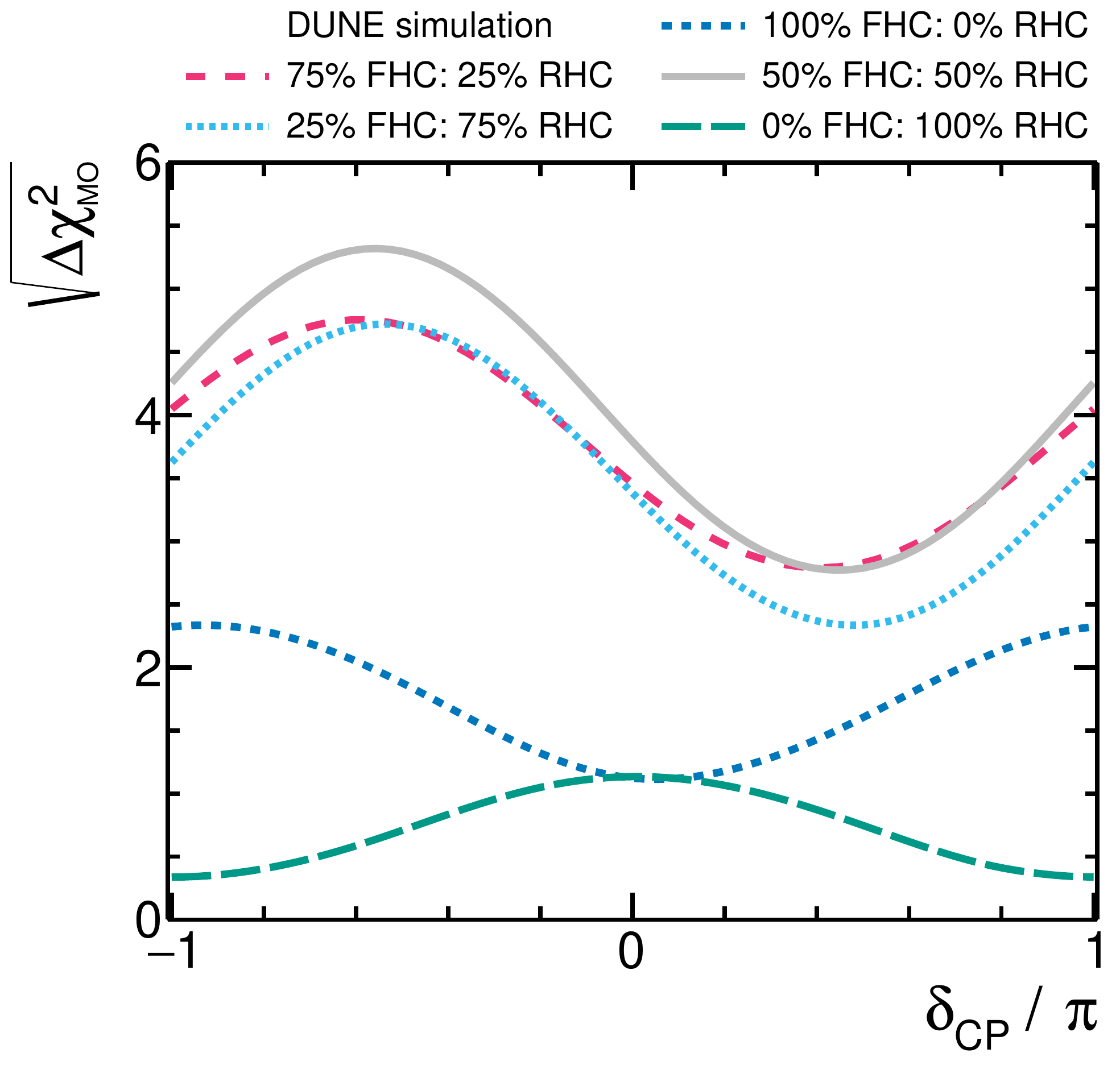}}
  \subfloat[IO, no $\theta_{13}$-penalty]    {\includegraphics[width=0.4\linewidth]{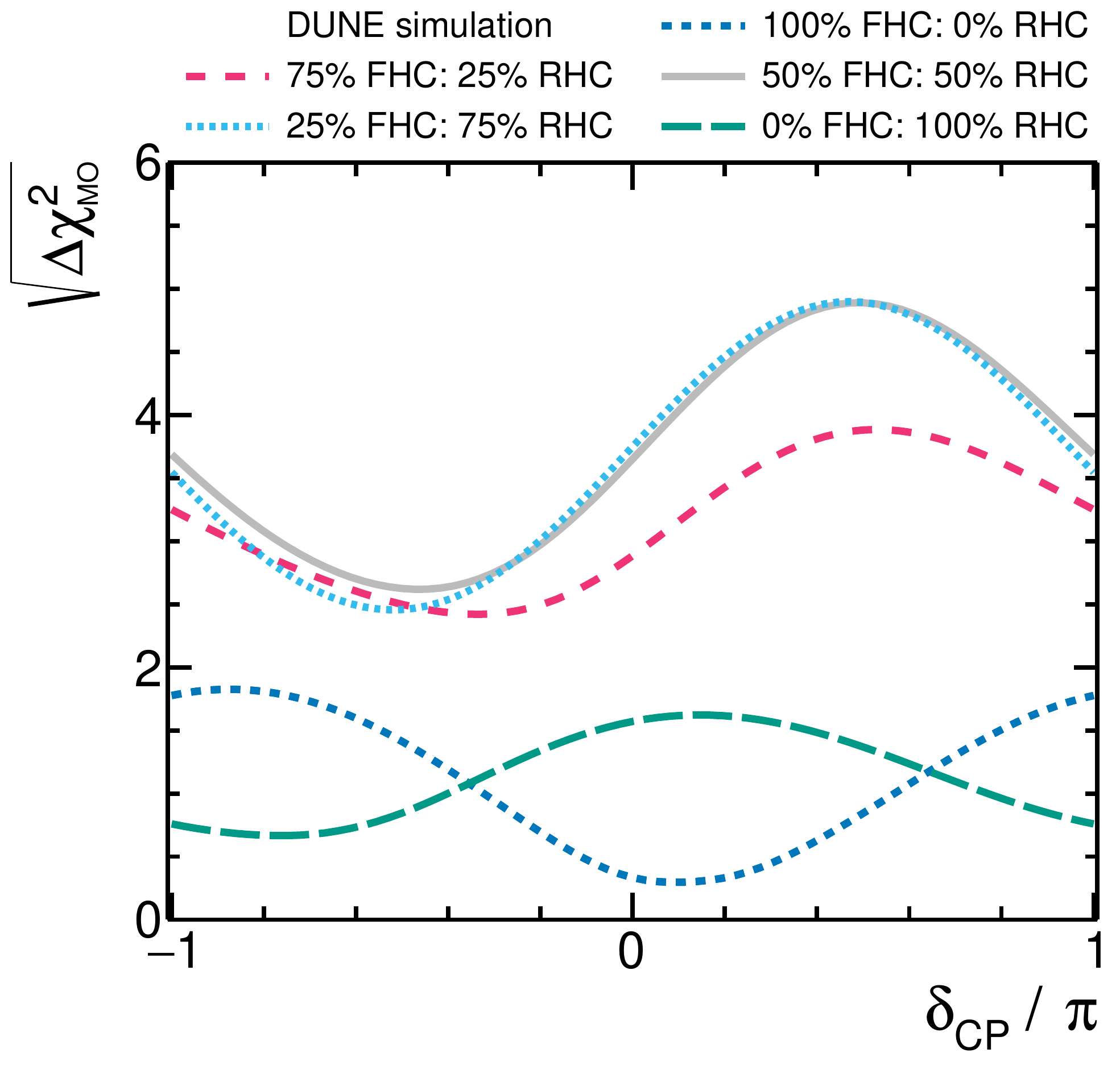}}
  \caption{The Asimov mass ordering sensitivity as a function of the true value of \deltacp, for a total exposure of 24 kt-MW-yr with different fractions of FHC and RHC running, with and without a $\theta_{13}$ penalty applied in the fit. Results are shown for both true normal and inverted ordering, with the true oscillation parameter values set to the NuFIT 4.0 best fit point in each ordering (see Table~\ref{tab:oscpar_nufit}).}
  \label{fig:run_opt_mh}
\end{figure*}
Figure~\ref{fig:run_opt_mh} shows DUNE's Asimov sensitivity to the mass ordering for a total 24 kt-MW-yr far detector exposure, with different fractions of FHC and RHC running, and the same four true oscillation parameter sets. A 24 kt-MW-yr exposure is used in Figure~\ref{fig:run_opt_mh} as it is around the exposure at which DUNE's median mass ordering significance exceeds 5$\sigma$ for some vales of \deltacp~\cite{Abi:2020qib}. For each point tested, all oscillation and nuisance parameters are allowed to vary, and two fits are carried out, one using each ordering. The difference in the best-fit $\chi^{2}$ values is calculated:
\begin{linenomath*}
  \begin{equation}
    \dchisqMO = \chi^{2}_{\mathrm{IO}} - \chi^{2}_{\mathrm{NO}},
    \label{eq:mh_chi2}
  \end{equation}
\end{linenomath*}
\noindent and the square root of the difference is used as the figure of merit on the y-axis in Figure~\ref{fig:run_opt_mh}. There are some caveats associated with this figure of merit, which are discussed in Section~\ref{sec:mh_sens}. 

It is clear from Figure~\ref{fig:run_opt_mh} that the mass ordering sensitivity has a strong dependence on the fraction of running in each beam mode. As in the CPV case, the effect is very different with and without the reactor $\theta_{13}$ constraint included. If the true ordering is normal and the reactor $\theta_{13}$ penalty is applied, the sensitivity increases significantly with increasing FHC running, with a full 1$\sigma$ increase in the sensitivity between equal beam running and 100\% FHC for most values of \deltacp. Conversely, if the ordering is inverted, 100\% FHC running would degrade the sensitivity by $\geq$1$\sigma$ for all values of \deltacp at the NuFIT 4.0 best fit point. Overall, the sensitivity to the inverted ordering is improved by a more equal split between the beam modes. It is clear that 100\% RHC running gives poor sensitivity for all values tested. 

Without the reactor $\theta_{13}$ constraint, the greatest sensitivity is obtained with close to an equal split of FHC and RHC running, and the sensitivity is significantly reduced with 100\% FHC running. This is because of a degeneracy between the effect of $\theta_{13}$ and the mass ordering on the rate of \nue appearance in FHC mode. If the mass ordering is normal, the \nue rate in FHC will be enhanced; without the reactor constraint, this excess can be accommodated by increasing the value of $\theta_{13}$.

\begin{figure*}[htbp]
  \centering
  \subfloat[CPV, with $\theta_{13}$-penalty] {\includegraphics[width=0.4\linewidth]{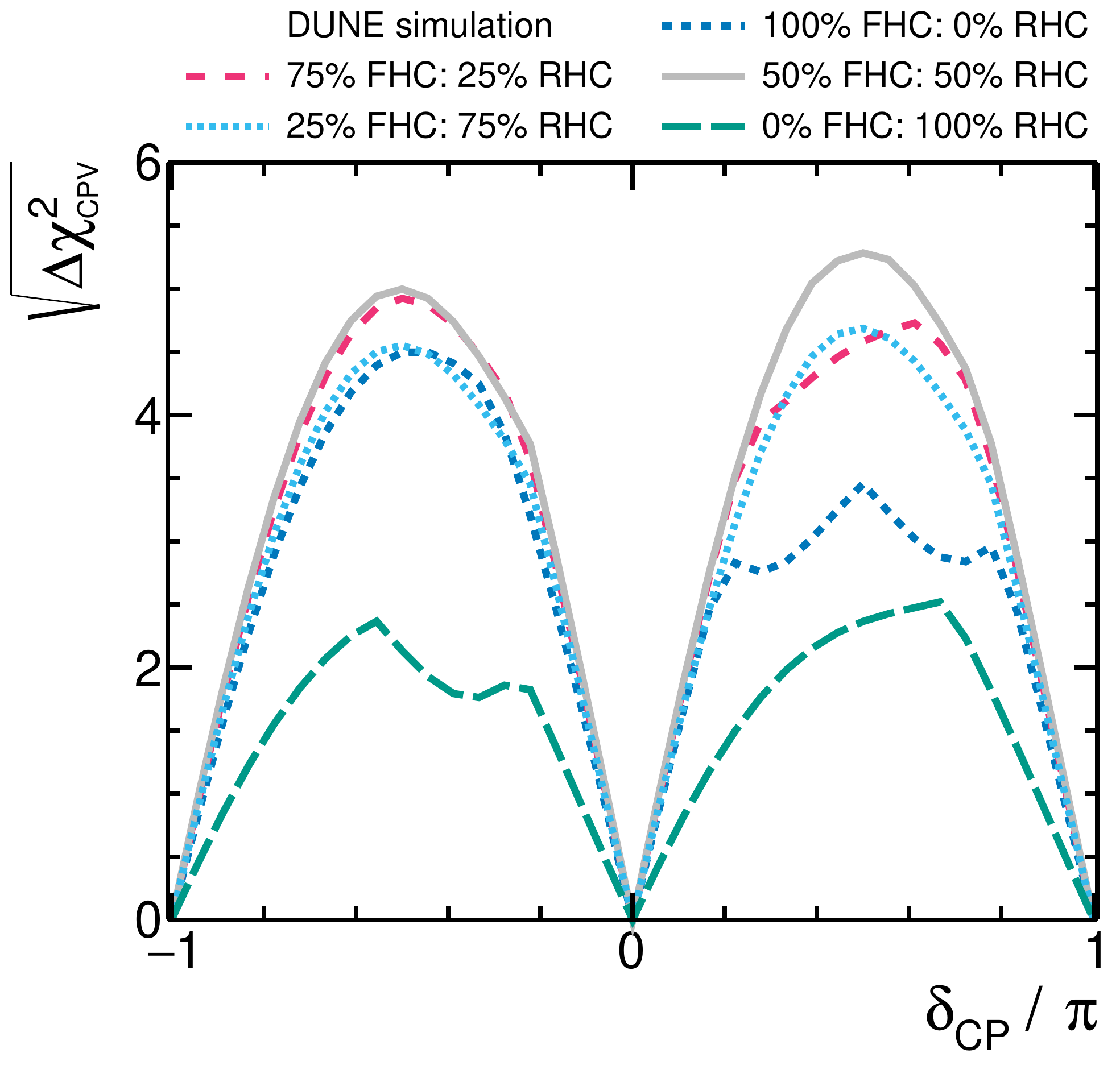}}
  \subfloat[CPV, no $\theta_{13}$-penalty]   {\includegraphics[width=0.4\linewidth]{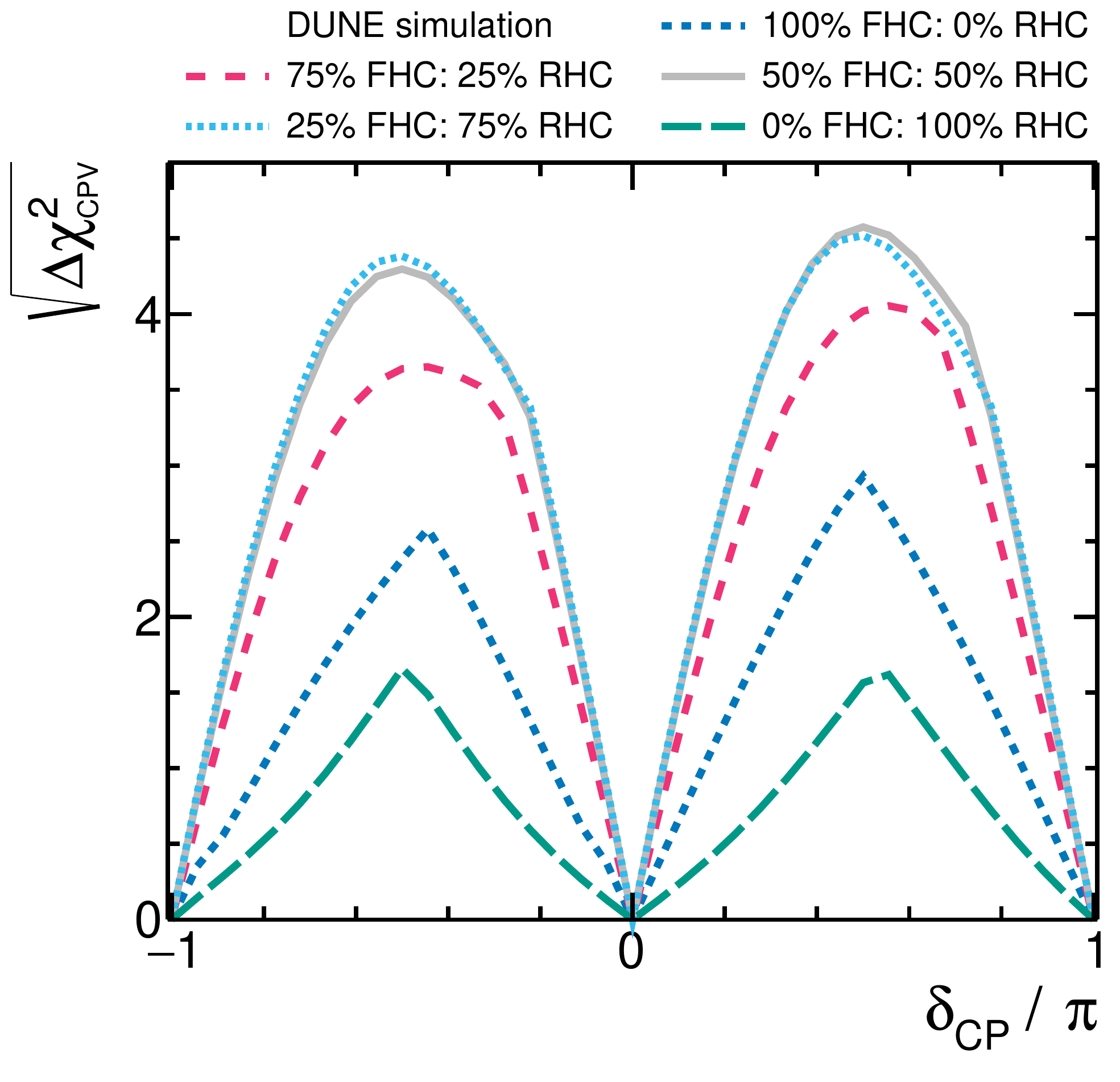}}\\
  \subfloat[MO, with $\theta_{13}$-penalty]  {\includegraphics[width=0.4\linewidth]{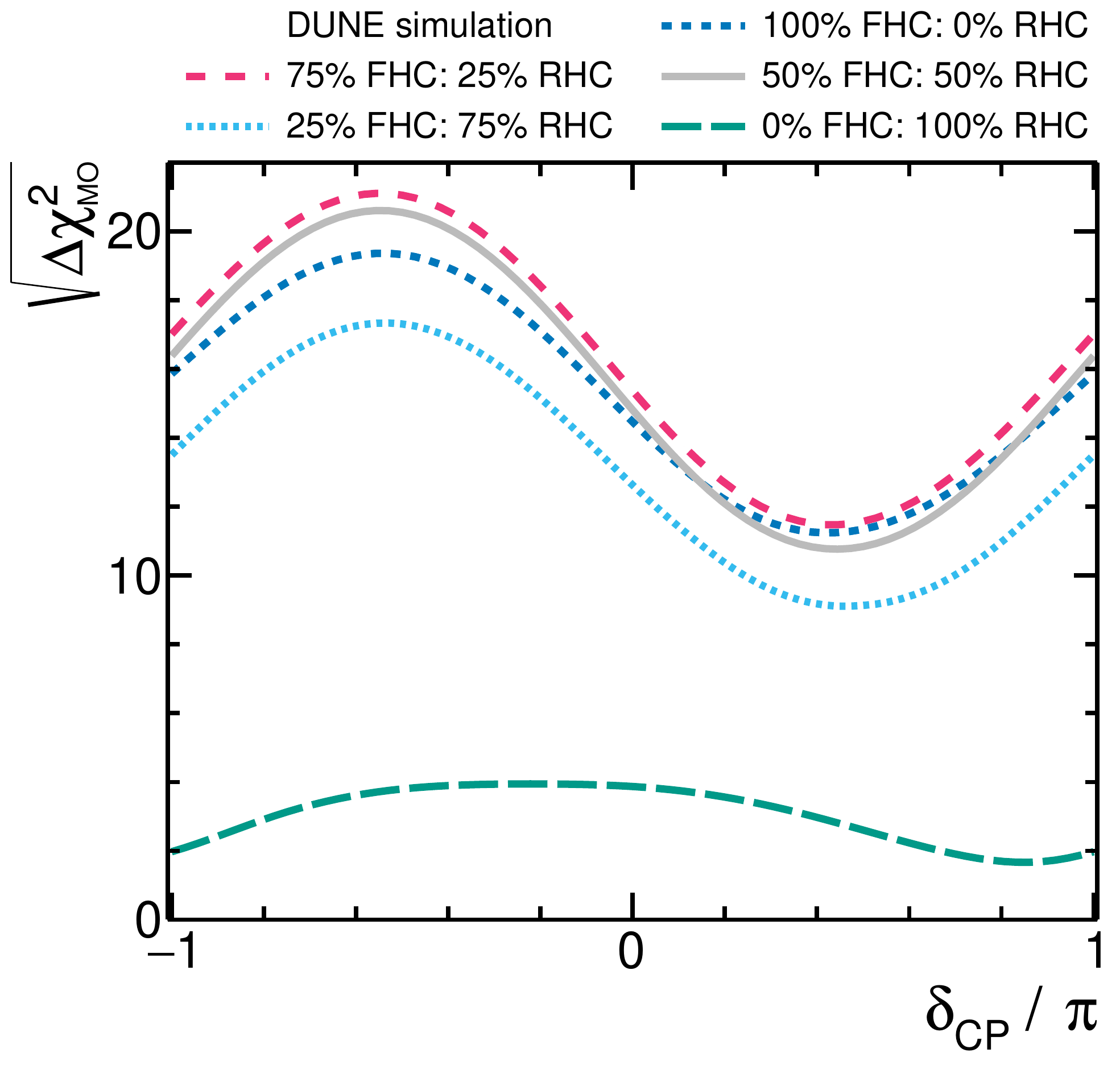}}
  \subfloat[MO, no $\theta_{13}$-penalty]    {\includegraphics[width=0.4\linewidth]{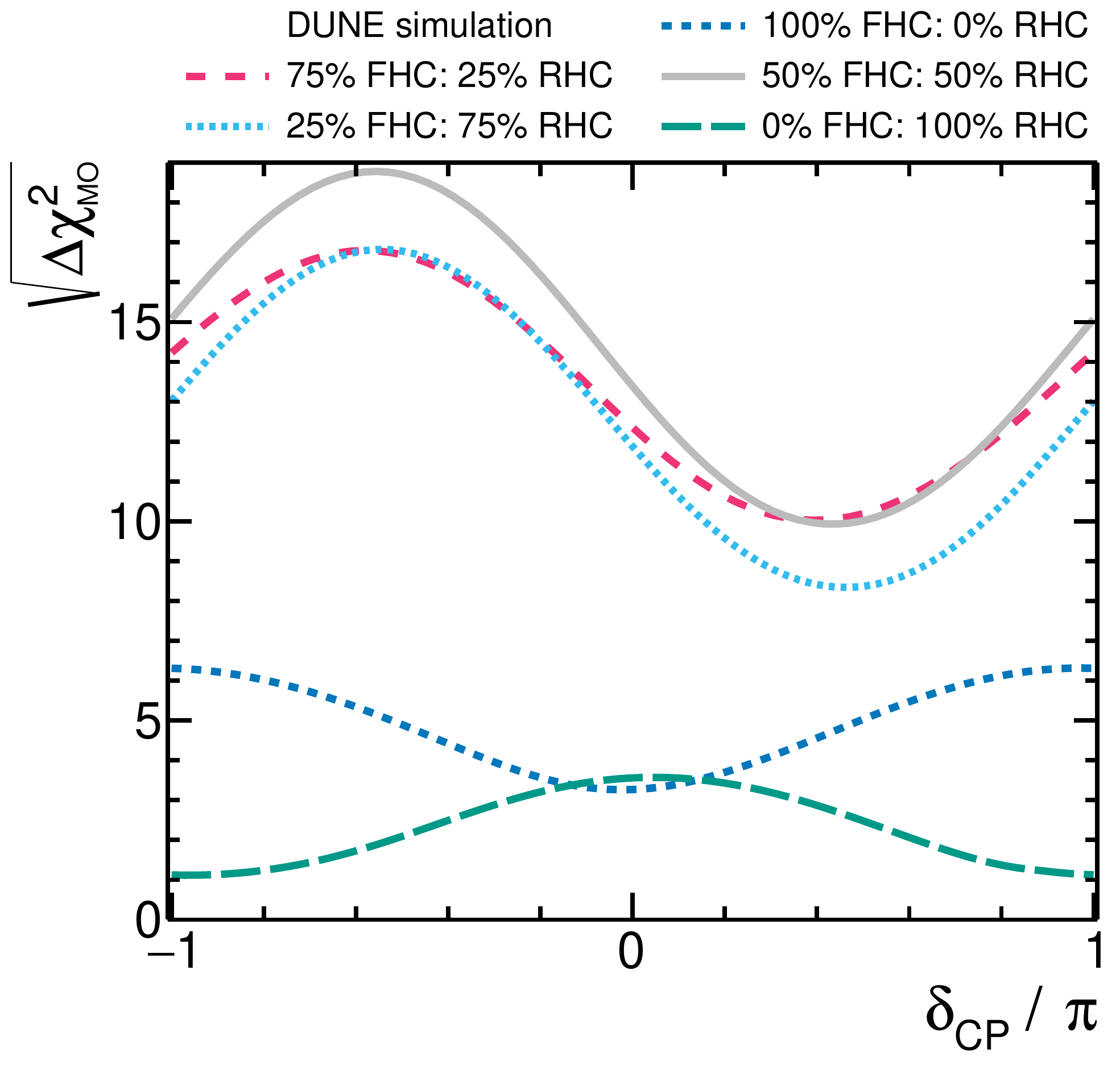}}
  \caption{The Asimov CPV and mass ordering sensitivities as a function of the true value of \deltacp, for a total exposure of 336 kt-MW-yr with different fractions of FHC and RHC running, with and without a $\theta_{13}$ penalty applied in the fit. Results are shown for both true normal ordering only, with the true oscillation parameter values set to the NuFIT 4.0 NO best fit point (see Table~\ref{tab:oscpar_nufit}).}
  \label{fig:run_opt_336ktmwyr}
\end{figure*}

For comparison, Figure~\ref{fig:run_opt_336ktmwyr} shows the Asimov CPV and mass ordering sensitivities, with and without the reactor $\theta_{13}$ constraint included, for true normal ordering only, for a large exposure of 336 kt-MW-yr, with different fractions of FHC and RHC running. At large exposures, running with strongly enhanced FHC no longer improves the sensitivity over equal beam mode running, with or without the $\theta_{13}$ penalty applied, for either CPV or mass ordering determination. This can be understood because the enhancement to the statistics that enhanced FHC brings is no longer as important to the sensitivity, and DUNE is able to place a constraint on the value of $\theta_{13}$ with its own data.

Overall, the sensitivity to CPV and the mass ordering is dependent on the division of running time between FHC and RHC, but a choice that increases the sensitivity in some region of parameter space can severely decrease the sensitivity in other regions. If there is strong reason to favor, for example, normal over inverted ordering when DUNE starts to take data, Figure~\ref{fig:run_opt_mh} shows that this could be more rapidly verified by running with more FHC data than RHC data, as the reactor $\theta_{13}$ constraint will be used in the main low exposure analysis. However, if this choice is wrong, this might cause DUNE to take longer to reach the same significance. Clearly this is an important consideration which should be revisited shortly before DUNE begins to collect data. Similarly, the CPV sensitivity shown in Figure~\ref{fig:run_opt_cpv} might be optimized if there is a strong reason to favor gaining sensitivity for $\deltacp > 0$ or $\deltacp < 0$, at a cost of reducing the sensitivity to CPV if \deltacp has the other sign. But, it is clear from Figures~\ref{fig:run_opt_cpv} and~\ref{fig:run_opt_mh} that equal running in FHC and RHC gives a close to optimal sensitivity across all of the parameter space, and as such is a reasonable {\it a priori} choice of run plan for studies of the DUNE sensitivity. Additionally, it is clear from Figure~\ref{fig:run_opt_336ktmwyr} that the improvement in the sensitivity with unequal beam running is a feature at low exposures, but not at high exposures, particularly because at high exposures when DUNE is able to constrain all the oscillation parameters with precision~\cite{Abi:2020qib}, there is a stronger motivation to run a DUNE-only analysis, without relying on the reactor $\theta_{13}$ measurement.

%% file: sections/cp_sens.tex
\FloatBarrier
\section{CP violation sensitivity}
\label{sec:cp_sens}

In this section, CPV sensitivity results are presented. For simplicity, only true NO will be shown unless explicitly stated. In all cases, a joint ND+FD fit is performed, and a $\theta_{13}$ penalty is always applied to incorporate the reactor measurement, as described in Section~\ref{sec:analysis_framework}. An equal split between FHC and RHC running is assumed based on the results obtained in Section~\ref{sec:run_plan_opt}. Asimov sensitivities, as shown in Section~\ref{sec:run_plan_opt}, are instructive but do not give information on how the expected sensitivity may vary with statistical or systematic uncertainties, or for variations in the other oscillation parameters of interest.

\begin{figure}[htbp]
  \centering
  \includegraphics[width=1\linewidth, trim={0cm 0cm 0cm 2.3cm}, clip]{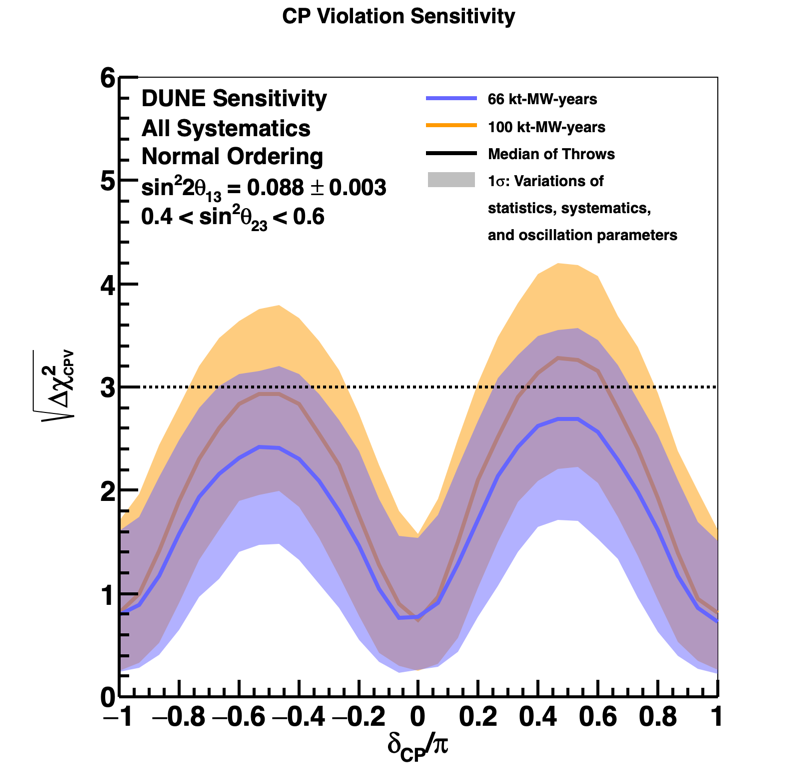}\\
  \includegraphics[width=1\linewidth, trim={0cm 0cm 0cm 2.3cm}, clip]{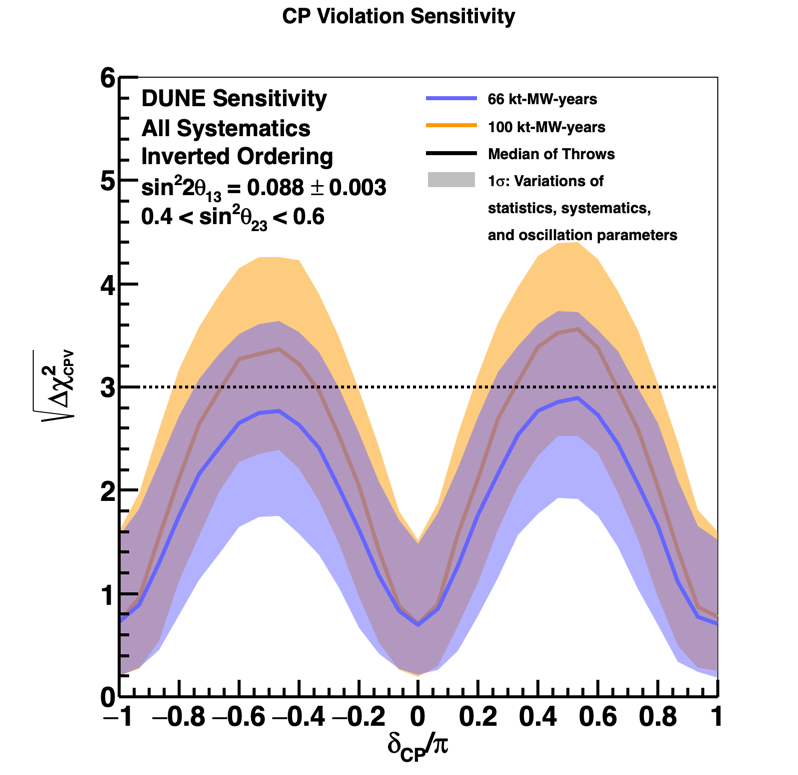}
  \caption{Significance of the DUNE determination of CP-violation ($\deltacp \neq \{0,\pm\pi\}$) as a function of the true value of \deltacp, for 66 kt-MW-yr (blue) and 100 kt-MW-yr (orange) exposures, for normal (top) and inverted (bottom) orderings. The width of the transparent bands cover 68\% of fits in which random throws are used to simulate systematic, oscillation parameter and statistical variations, with independent fits performed for each throw constrained by prior uncertainties. The solid lines show the median significance.}
  \label{fig:cpv_bands}
\end{figure}
Figure~\ref{fig:cpv_bands} shows the significance with which CPV ($\deltacp \neq \{0,\pm\pi\}$) can be observed for both NO and IO, for exposures of 66 and 100 kt-MW-yr.  The sensitivity metric used is the square root of the difference between the best fit $\chi^{2}$ values obtained for a CP-conserving fit and one where \deltacp is allowed to vary, as shown in Equation~\ref{eq:cpv_chi2}, which is calculated for each throw of the systematic, other oscillation parameters and statistics. 

The sensitivity shown in Figure~\ref{fig:cpv_bands} has a characteristic double peak structure because the significance of a CPV measurement decreases around CP-conserving values. The systematic and statistical variations mean that all throws have $\dchisqCPV \geq 0$, and therefore neither the median significance nor the band showing the central 68\% of throws reach exactly 0 at CP-conserving values. This is entirely expected, it simply means that random variations in the data will cause us to obtain a 1$\sigma$ measurement of CPV $\approx$32\% of the time for CP-conserving values. Median significances are slightly higher for IO than for NO, and by exposures of 100 kt-MW-yr, the median significance exceeds 3$\sigma$ for the maximal CP-violating values of $\pm\pi/2$. This presentation of the CPV sensitivity was followed in Ref.~\cite{Abi:2020qib}, and is very informative at high exposures. Around CP-conserving values ($\deltacp = \{0,\pm\pi\}$), the distribution of the sensitivity metric $\sqrt{\dchisqCPV}$ is non-Gaussian for all exposures. Additionally, at lower exposures, as shown in Figure~\ref{fig:cpv_bands}, the distribution of $\sqrt{\dchisqCPV}$ around maximally CP-violating values of $\deltacp = \pm\pi/2$ is increasingly non-Gaussian, making the spread in sensitivity harder to interpret with this presentation.

The CPV significance in Figure~\ref{fig:cpv_bands} (and previously in Ref.~\cite{Abi:2020qib}) is calculated using constant \dchisq critical values, where $\dchisqCPV \leq 1, 4, 9$ corresponds to a significance of 1, 2 and 3$\sigma$ for one degree of freedom. This assumption holds when Wilks' theorem can be applied~\cite{wilks}, but can lead to incorrect coverage where it cannot. It is known to break down for low-statistics samples, around physical boundaries, in the case of cyclic parameters, and where there are significant degeneracies. It is likely that a constant \dchisq treatment will break down for \deltacp, where all of these issues apply, as has indeed been shown by the T2K Collaboration~\cite{Abe:2021gky}.

The Feldman-Cousins method~\cite{Feldman:1997qc} is a brute force numerical method to calculate confidence intervals with correct coverage. A large number of toy experiments are produced, where the parameter(s) of interest (here \deltacp) is set to a desired true value, all other systematic and oscillation parameters are thrown, as described in Section~\ref{sec:analysis_framework}, and a statistical throw is made, for the two ND samples and four FD sample used in the analysis. Then two fits are performed, one where the parameter(s) of interest are fixed to the true value, and another where the test statistic is minimized with respect to the parameter(s) of interest. In both fits, all other parameters are allowed to vary. For each throw, the profile likelihood ratio \dchisqFC is calculated using the minimum $\chi^{2}$ values for those two fits, as in Equation~\ref{eq:dchisq_fc}.
\begin{linenomath*}
  \begin{equation}
    \dchisqFC = \chi^{2}(\theta_{\mathrm{true}}) - \min_{\theta}\chi^{2}(\theta)
    \label{eq:dchisq_fc}
  \end{equation}
\end{linenomath*}
\begin{figure}[htbp]
  \centering
  \includegraphics[width=0.8\columnwidth]{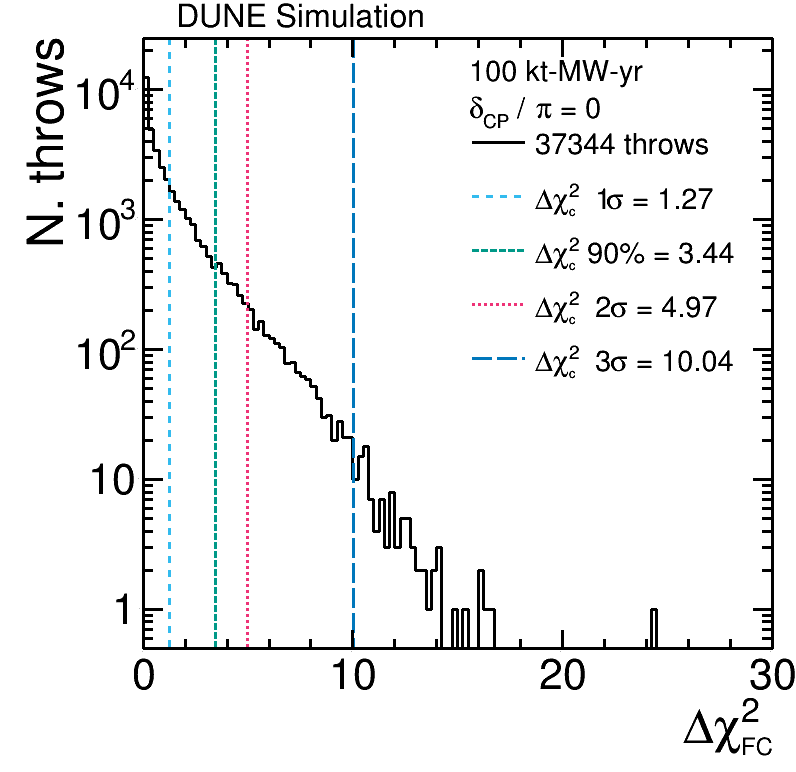}
  \caption{Distribution of \dchisqFC values, calculated using Equation~\ref{eq:dchisq_fc}, for a large number of throws with true $\deltacp = 0$, and a 100 kt-MW-yr exposure. The \dchisqcrit values (vertical lines) obtained using the Feldman-Cousins method show the \dchisqFC value below which 68.27\% (1$\sigma$), 90\%, 95.45\% (2$\sigma$) and 99.73\% (3$\sigma$) of throws reside, with the calculated values given in the legend. The number of throws used is also given.}
  \label{fig:fc_throws}
\end{figure}
The distribution of these throws is used to calculate the \dchisqFC value that gives the desired coverage, with the appropriate fraction of toys above/below the calculated value. These are labelled {\it critical values}, and are denoted \dchisqcrit. A distribution of \dchisqFC values is shown in Figure~\ref{fig:fc_throws} for an example ND+FD analysis with a 100kt-MW-yr exposure at the far detector, equal FHC and RHC run fractions, and the reactor $\theta_{13}$ constraint applied. In Figure~\ref{fig:fc_throws}, the \dchisqcrit values corresponding to for 68.27\% (1$\sigma$), 90\%, 95.45\% (2$\sigma$) and 99.73\% (3$\sigma$) of the throws are indicated. The \dchisqcrit values were only calculated up to the 3$\sigma$ level due to the very large number of throws required for higher confidence levels.

An uncertainty on the value of \dchisqcrit obtained from the toy throw distribution (e.g., Figure~\ref{fig:fc_throws}), is obtained using a bootstrap rethrowing method~\cite{rice2006mathematical}. The empirical PDF obtained from the throws is treated as the true PDF, and $B$ independent samples of size $n$ are drawn from it, where $n$ is the total number of throws used to build the empirical PDF. Each throw can be drawn multiple times in this method, so the ensemble of throws is different in each sample. Then, the standard deviation $s_{\hat{\vartheta}}$, on the \dchisqcrit values of interest, $\vartheta$, are calculated for each of the $B$ samples using:
\begin{linenomath*}
\begin{equation}
  s_{\hat{\vartheta}} = \sqrt{\frac{1}{B-1} \sum^{B}_{i=0} (\vartheta_{i}^{*} - \bar{\vartheta}^{*})^{2}},
  \label{eq:fc_uncertainty}
\end{equation}
\end{linenomath*}
where $\vartheta_{i}^{*}$ denotes the calculated \dchisqcrit value of interest for each of the samples, and $\bar{\vartheta}^{*}$ is their average value.

\begin{figure}[htbp]
  \centering
  \includegraphics[width=0.8\columnwidth]{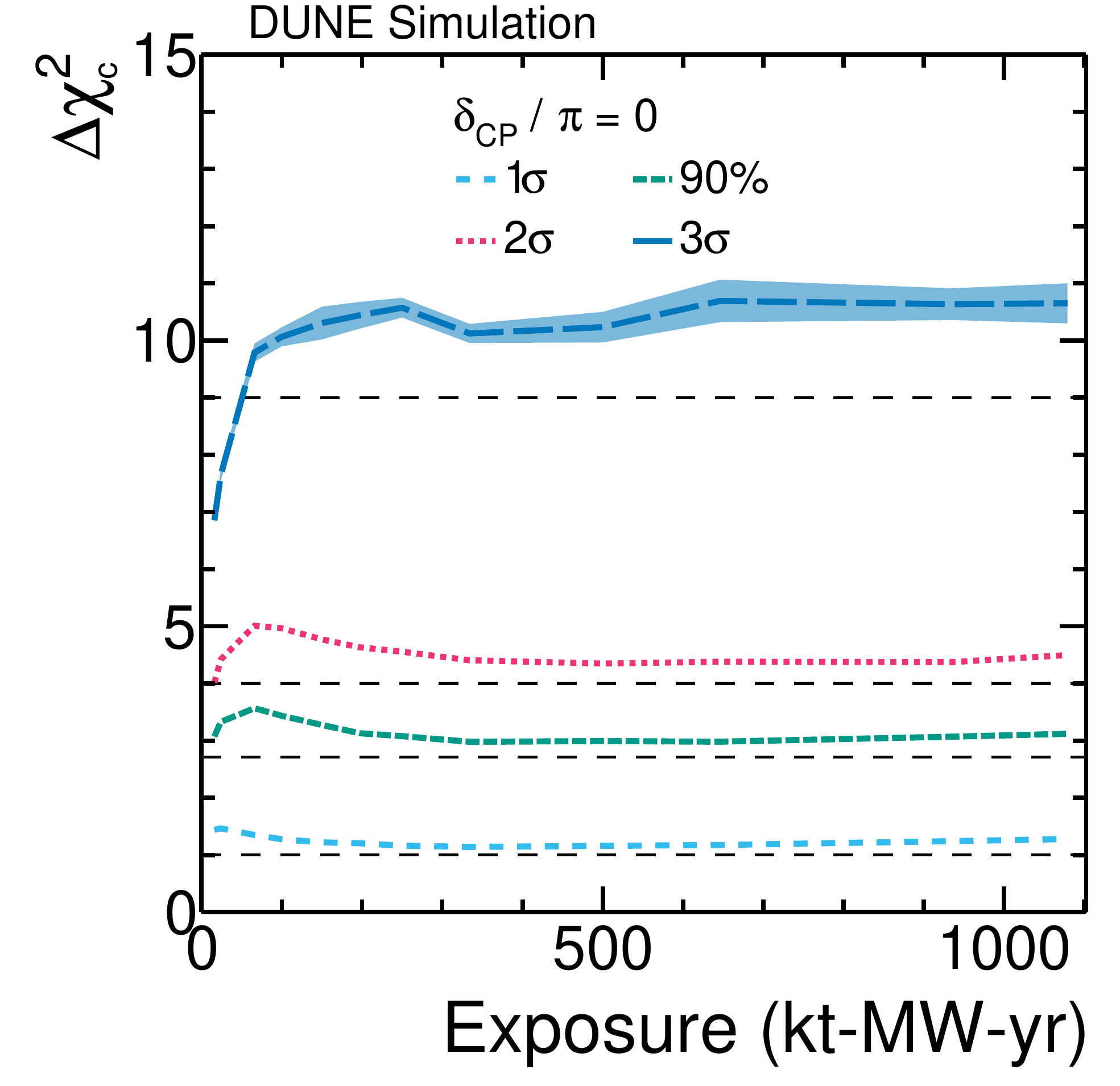}
  \caption{The \dchisqcrit values corresponding to 68.27\% (1$\sigma$), 90\%, 95.45\% (2$\sigma$) and 99.73\% (3$\sigma$) of throws, shown for true $\deltacp = 0$, as a function of exposure. A linear $x$-axis scale is used to highlight the stability of \dchisqcrit values for large exposures. The uncertainty on the \dchisqcrit values is obtained using Equation~\ref{eq:fc_uncertainty}, and is indicated as the shaded line. To guide the eye, horizontal dashed lines are included which indicate the 1$\sigma$, 90\%, 2$\sigma$ and 3$\sigma$ \dchisq values assumed using the constant-\dchisq method, with one degree of freedom. The distribution of throws used produced to calculate the \dchisqcrit values shown are given in Figure~\ref{fig:fc_throws_exp}.}
  \label{fig:fc_vs_exp}
\end{figure}
Figure~\ref{fig:fc_vs_exp} shows the evolution of the \dchisqcrit values as a function of exposure for $\deltacp = 0$, the relevant value for CPV sensitivity, for an ND+FD analysis with equal FHC and RHC running and the reactor $\theta_{13}$ constraint applied.
For all significance levels tested, the \dchisqcrit rise quickly as a function of exposure, and stabilize at values slightly higher than those suggested by the constant \dchisq method by exposures of $\approx$100 kt-MW-yr. The initial rise in the \dchisqcrit values is due to the low statistics at those exposures. Overall, this implies that the CPV significance is slightly weaker than what would be inferred from $\sigma = \sqrt{\dchisqCPV}$, as used for example in Figure~\ref{fig:cpv_bands}. Crucially, there is no constant increase in the \dchisqcrit values over time as has been reported by the T2K Collaboration~\cite{Abe:2021gky}. Details on the number of toy throws used at each point of Figure~\ref{fig:fc_vs_exp} are given in the Appendix, and the toy throw distributions from which the \dchisqcrit values and their uncertainties were calculated are shown in Figure~\ref{fig:fc_throws_exp}.

\begin{figure}[htbp]
  \centering
  \subfloat[100 kt-MW-yr]  {\label{fig:fc_vs_dcp_100}\includegraphics[width=0.8\columnwidth]{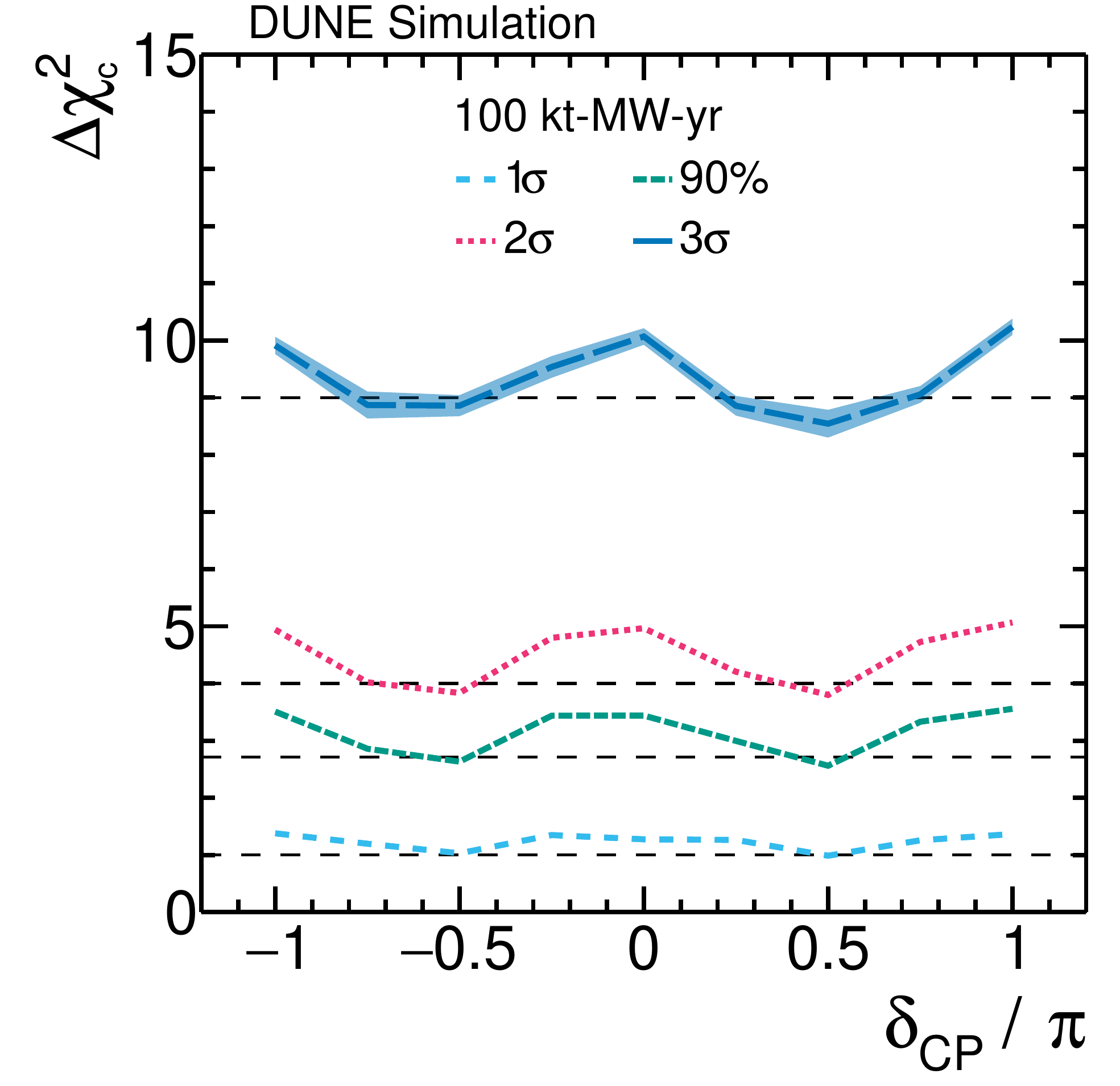}}\\
  \subfloat[336 kt-MW-yr]  {\label{fig:fc_vs_dcp_336}\includegraphics[width=0.8\columnwidth]{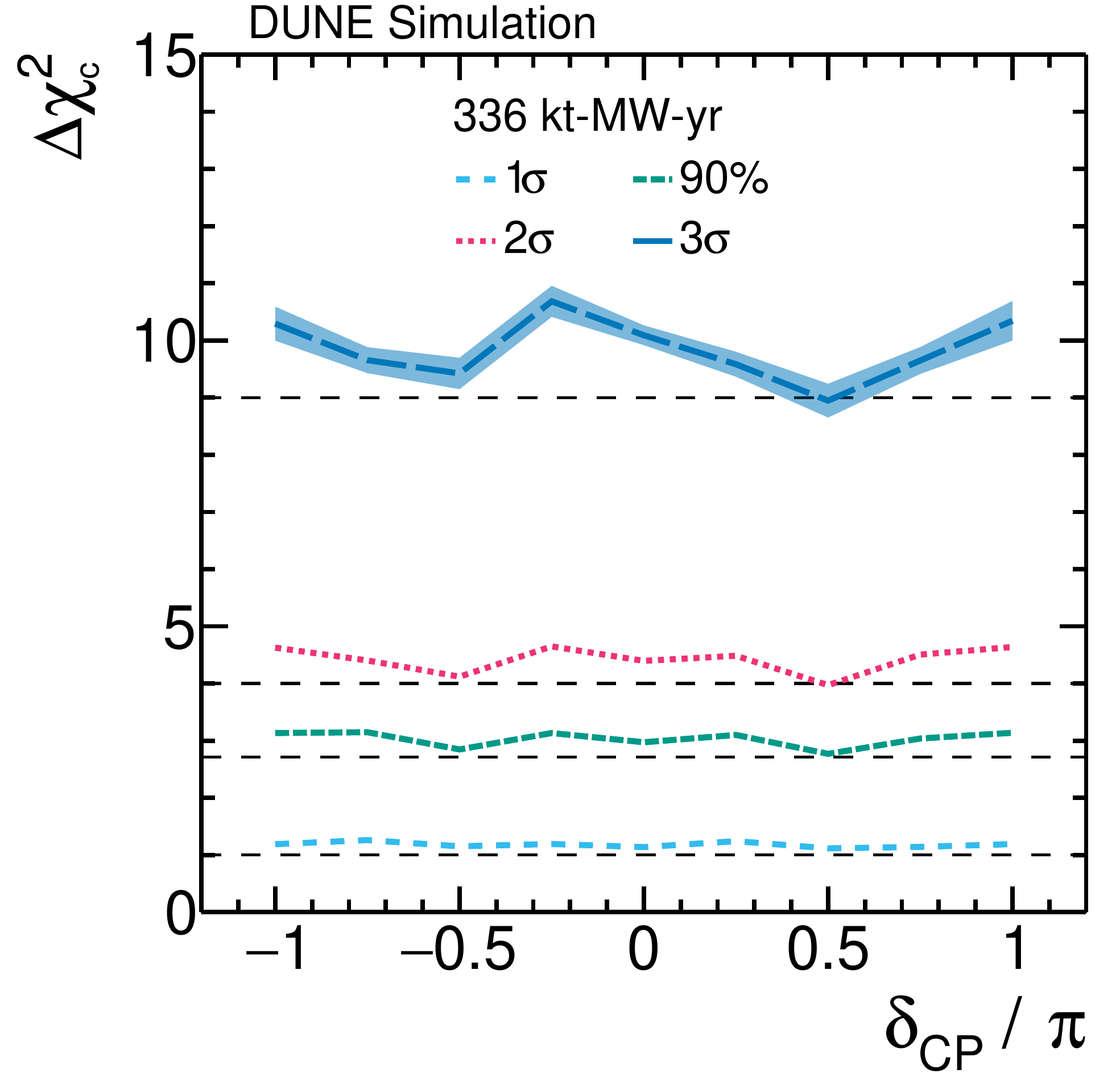}}
  \caption{The \dchisqcrit values corresponding to 68.27\% (1$\sigma$), 90\%, 95.45\% (2$\sigma$) and 99.73\% (3$\sigma$) of throws, shown as a function of true \deltacp, for exposures of 100 kt-MW-yr and 336 kt-MW-yr. The uncertainty on the \dchisqcrit values is obtained using Equation~\ref{eq:fc_uncertainty}, and is indicated as the shaded line. To guide the eye, horizontal dashed lines are included which indicate the 1$\sigma$, 90\%, 2$\sigma$ and 3$\sigma$ \dchisqCPV values assumed using the constant-\dchisq method, with one degree of freedom. The distribution of throws used produced to calculate the \dchisqcrit values shown are given in Figure~\ref{fig:fc_throws_100kt-MW-yr} (Figure~\ref{fig:fc_throws_336kt-MW-yr}) for 100 kt-MW-yr (336 kt-MW-yr).}
  \label{fig:fc_vs_dcp}
\end{figure}
As \deltacp is a cyclical parameter, with physical boundaries at $\pm\pi$, it is interesting to see how the \dchisqcrit values evolve as a function of it. Figure~\ref{fig:fc_vs_dcp} shows the \dchisqcrit as a function of true \deltacp, for an ND+FD analysis with equal FHC and RHC running including the reactor $\theta_{13}$ constraint, for both 100 kt-MW-yr and 336 kt-MW-yr exposures. There is a noticeable, although not large, depression in the \dchisqcrit values at $\deltacp = \pm\pi/2$ for all significance levels considered. This effect is larger at the lower, 100 kt-MW-yr, exposure, and is larger at higher significance levels. It is also clear from Figure~\ref{fig:fc_vs_dcp} that the \dchisqcrit behaviour is very similar at $\deltacp = \pm\pi/2$ as at $\deltacp = 0$. Although the \dchisqcrit values are relevant for CPV sensitivity, this evolution of the \dchisqcrit values with \deltacp will be important for estimating DUNE's \deltacp resolution. Details on the number of toy throws used at each point of Figure~\ref{fig:fc_vs_dcp} are given in the Appendix, and the toy throw distributions used to calculate the \dchisqcrit values and uncertainties are shown for the 100 kt-MW-yr (336 kt-MW-yr) test points in Figure~\ref{fig:fc_throws_100kt-MW-yr} (Figure~\ref{fig:fc_throws_336kt-MW-yr}).

\begin{figure*}[htbp]
  \centering
  \captionsetup[subfloat]{captionskip=-4pt}
  \subfloat[24 kt-MW-yr]  {\includegraphics[width=0.33\linewidth]{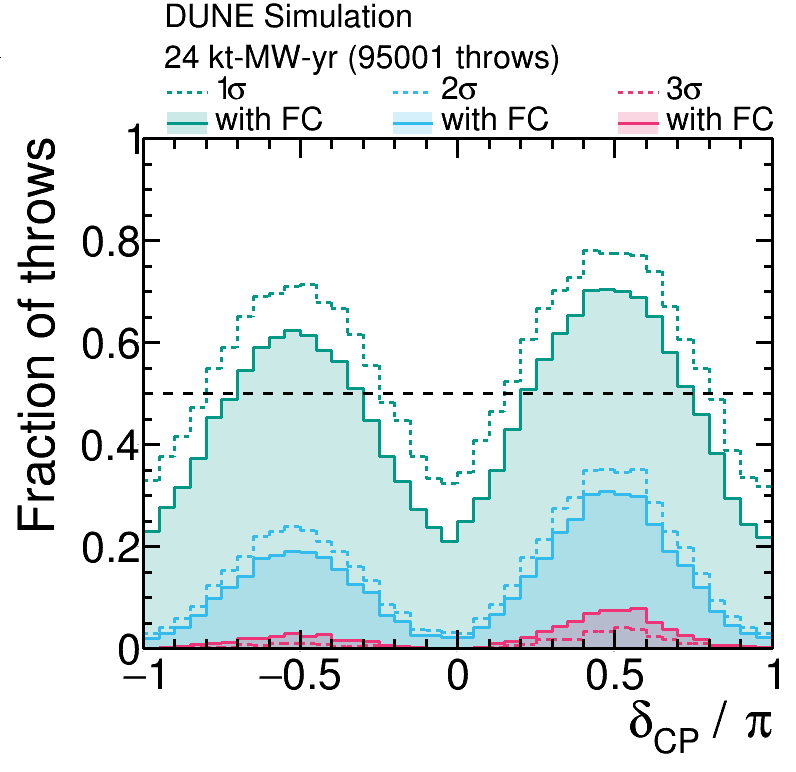}}
  \subfloat[66 kt-MW-yr]  {\includegraphics[width=0.33\linewidth]{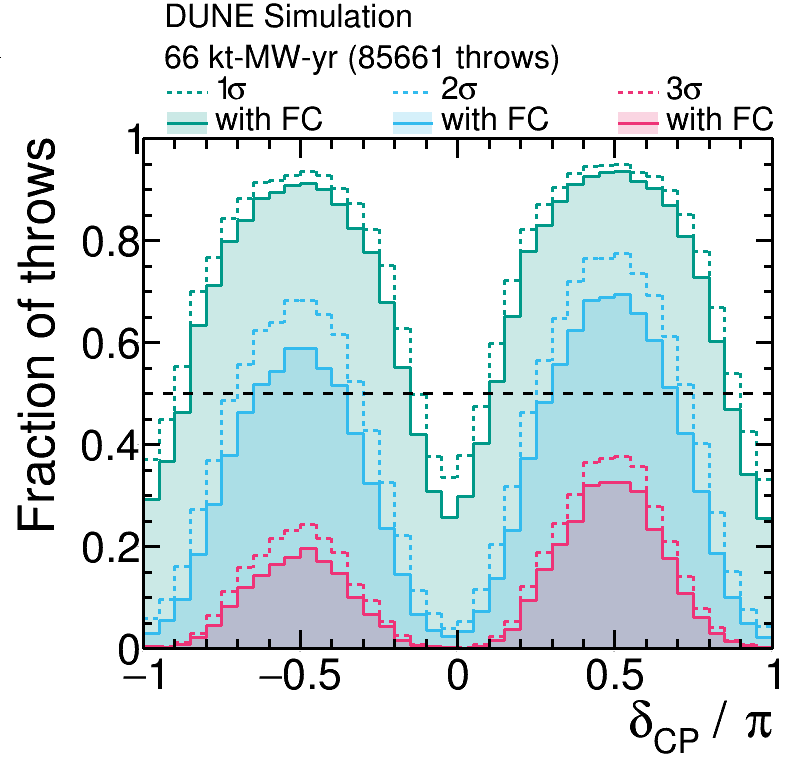}}
  \subfloat[100 kt-MW-yr] {\includegraphics[width=0.33\linewidth]{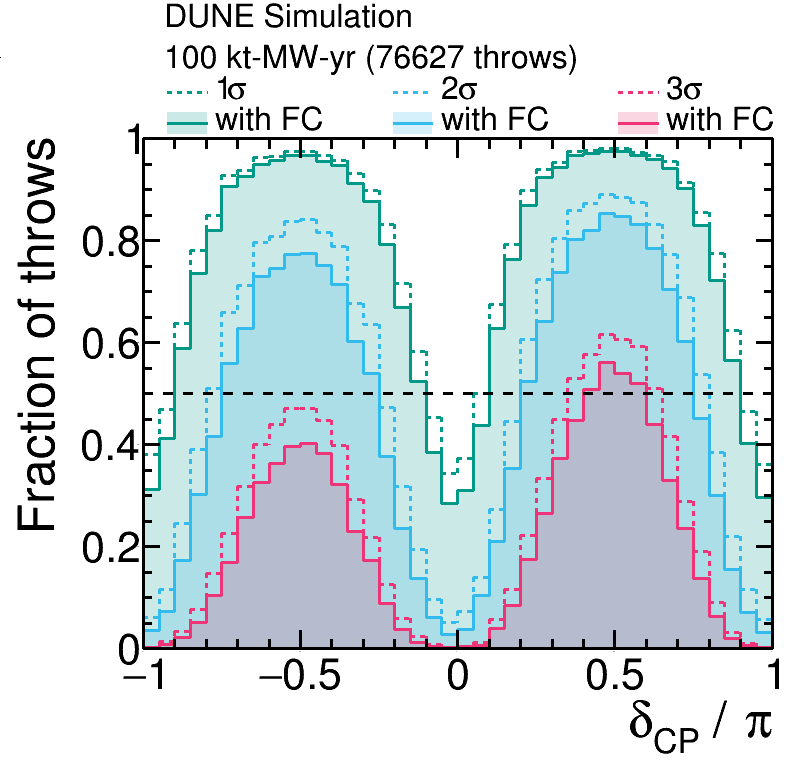}}\\
  \subfloat[150 kt-MW-yr] {\includegraphics[width=0.33\linewidth]{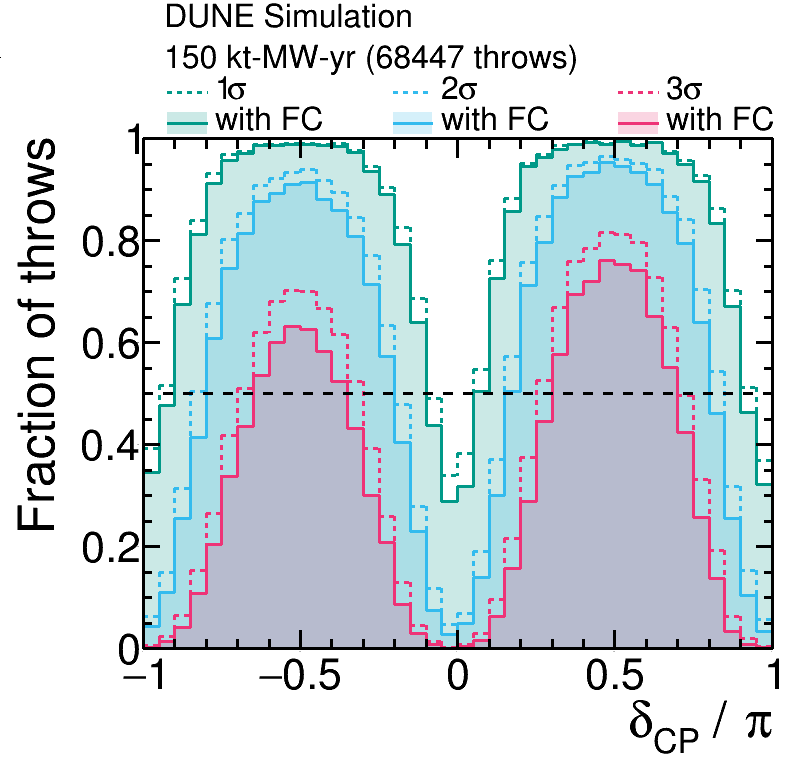}}
  \subfloat[197 kt-MW-yr] {\includegraphics[width=0.33\linewidth]{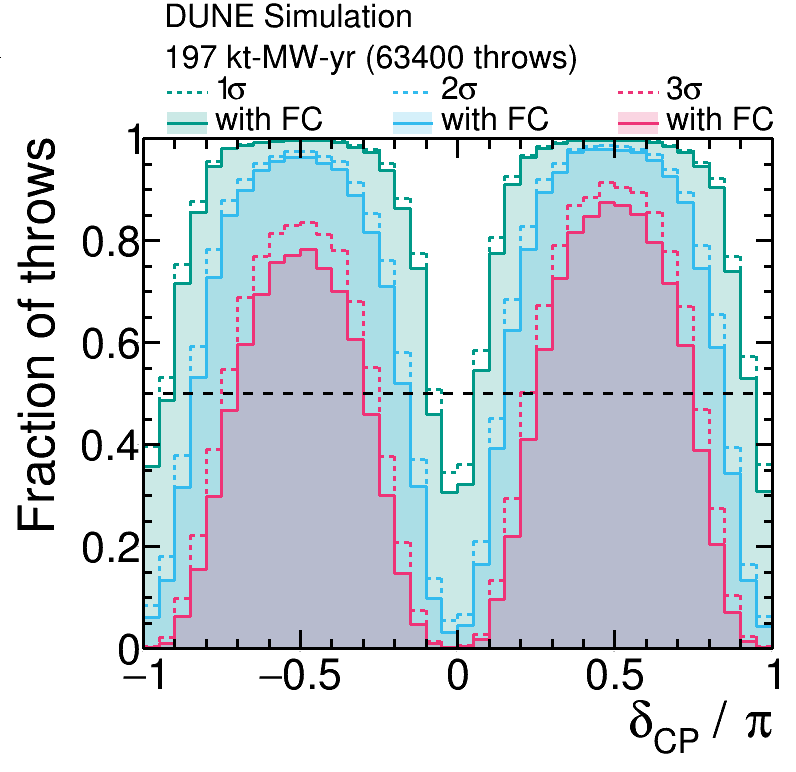}}
  \subfloat[336 kt-MW-yr] {\includegraphics[width=0.33\linewidth]{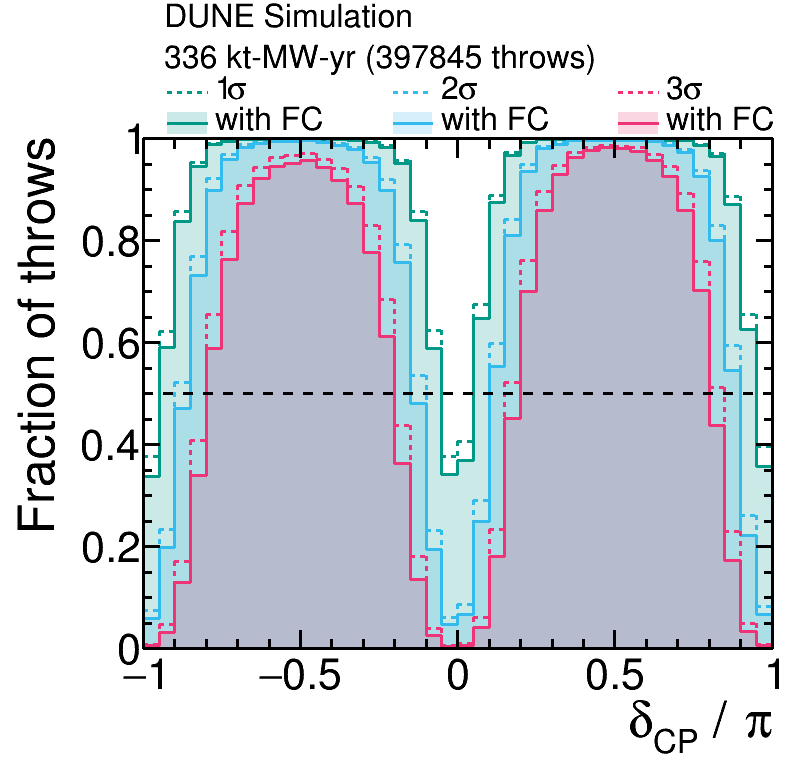}}
  \caption{Fraction of throws for which significance of DUNE's CP-violation test ($\deltacp \neq \{0,\pm\pi\}$) exceeds 1--3$\sigma$, calculated using both the FC (shaded histograms) and constant-\dchisq (dashed lines) methods, as a function of the true value of \deltacp. Shown for NO, for a number of different exposures. The number of throws used to make each figure is also shown.}
  \label{fig:cpv_over_time_fc}
\end{figure*}
DUNE's CPV sensitivity is calculated using Equation~\ref{eq:cpv_chi2} from an ensemble of throws of all systematic, other oscillation parameters and statistics. Figure~\ref{fig:cpv_over_time_fc} shows the fraction of throws for which DUNE would observe a CPV significance above a discrete threshold, as a function of the true value of \deltacp, for 1--3$\sigma$ significances and for a variety of exposures. The shaded histograms show the complete treatment including FC, while the dashed histograms show the constant-\dchisq treatment, to show the deviation from Wilks' theorem, and to facilitate comparison at higher significances where the FC treatment becomes computationally prohibitive.

The point at which the median significance (50\% of throws) passes different significance thresholds can be easily read from the figures, and can be compared with those shown in Figure~\ref{fig:cpv_bands}. The same double peak structure seen in Figure~\ref{fig:cpv_bands} can be observed. The median significance for measuring CPV exceeds 3$\sigma$ after $\approx$100 kt-MW-yr, but a significant fraction of throws exceed 3$\sigma$ at shorter exposures. For a 336 kt-MW-yr exposure, the fraction of throws for which the significance is less than 3$\sigma$ at maximal values of \deltacp is very small. In general, the effect of the Feldman-Cousins correction is to reduce the fraction of toy throws that cross each significance threshold (with respect to the constant-\dchisq result), by a maximum of $\approx$10\%, but the exact fraction changes as a function of true \deltacp value and exposure. An exception to this general trend is the 3$\sigma$ behaviour at 24 kt-MW-yr, the lowest exposure shown, where the significance increases. This is due to fall in the 3$\sigma$ \dchisqcrit value towards the lowest exposures observed in Figure~\ref{fig:fc_vs_exp}. The number of throws carried out at each exposure is indicated on each plot. The number of throws decreases as a function of exposure because fixed computing resources were used for each configuration, and the time for the ensemble of fits carried out for each throw to complete increases slightly with exposure. The final 336 kt-MW-yr exposure has more throws because it was generated for the analysis presented in Ref.~\cite{Abi:2020qib}, where more than one projection was considered --- requiring more throws to sample the space.
\begin{figure*}[htbp]
  \centering
  \subfloat[$\deltacp = -\pi/2$] {\includegraphics[width=0.8\columnwidth]{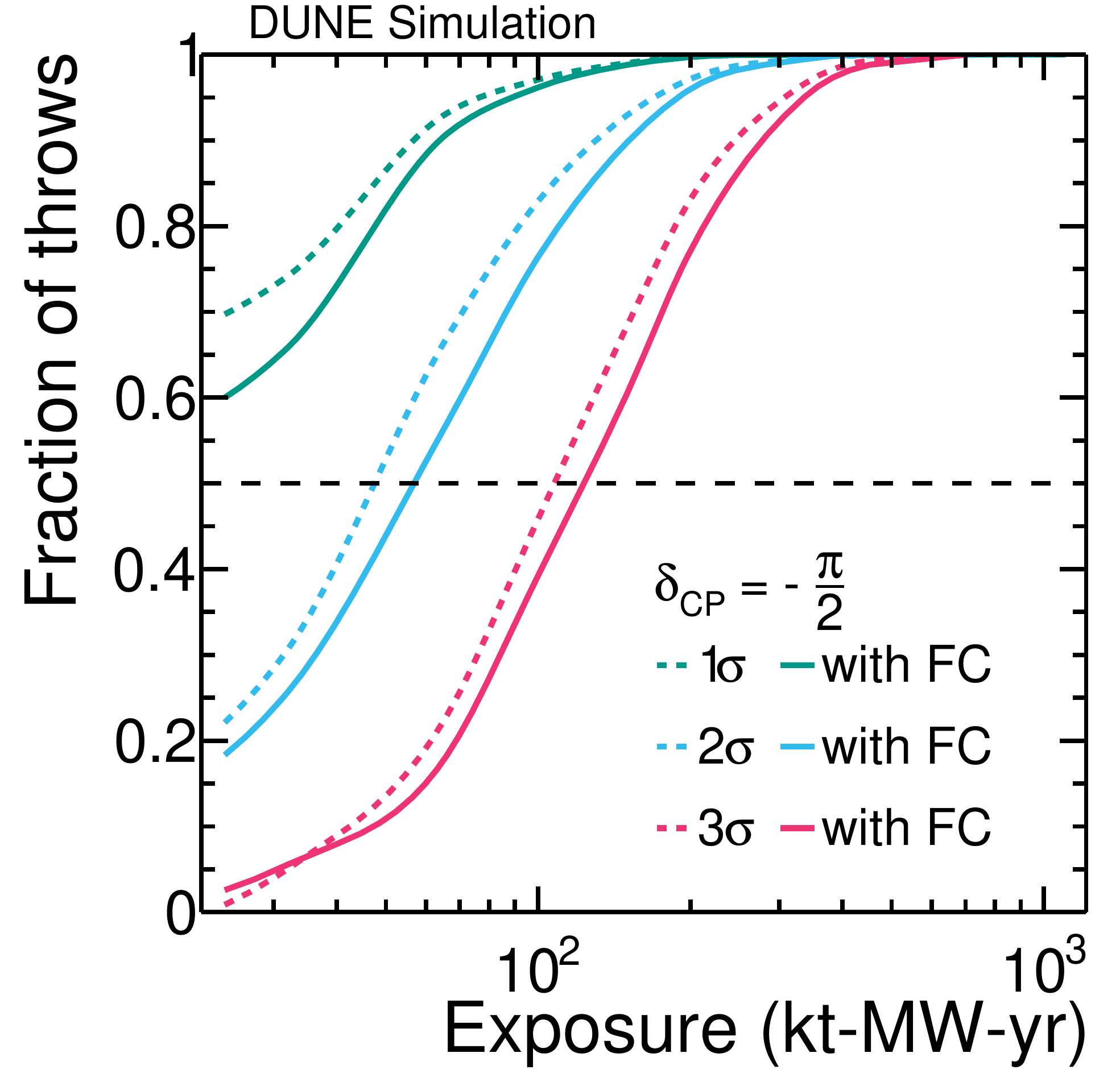}}
  \subfloat[50\% of \deltacp values] {\includegraphics[width=0.8\columnwidth]{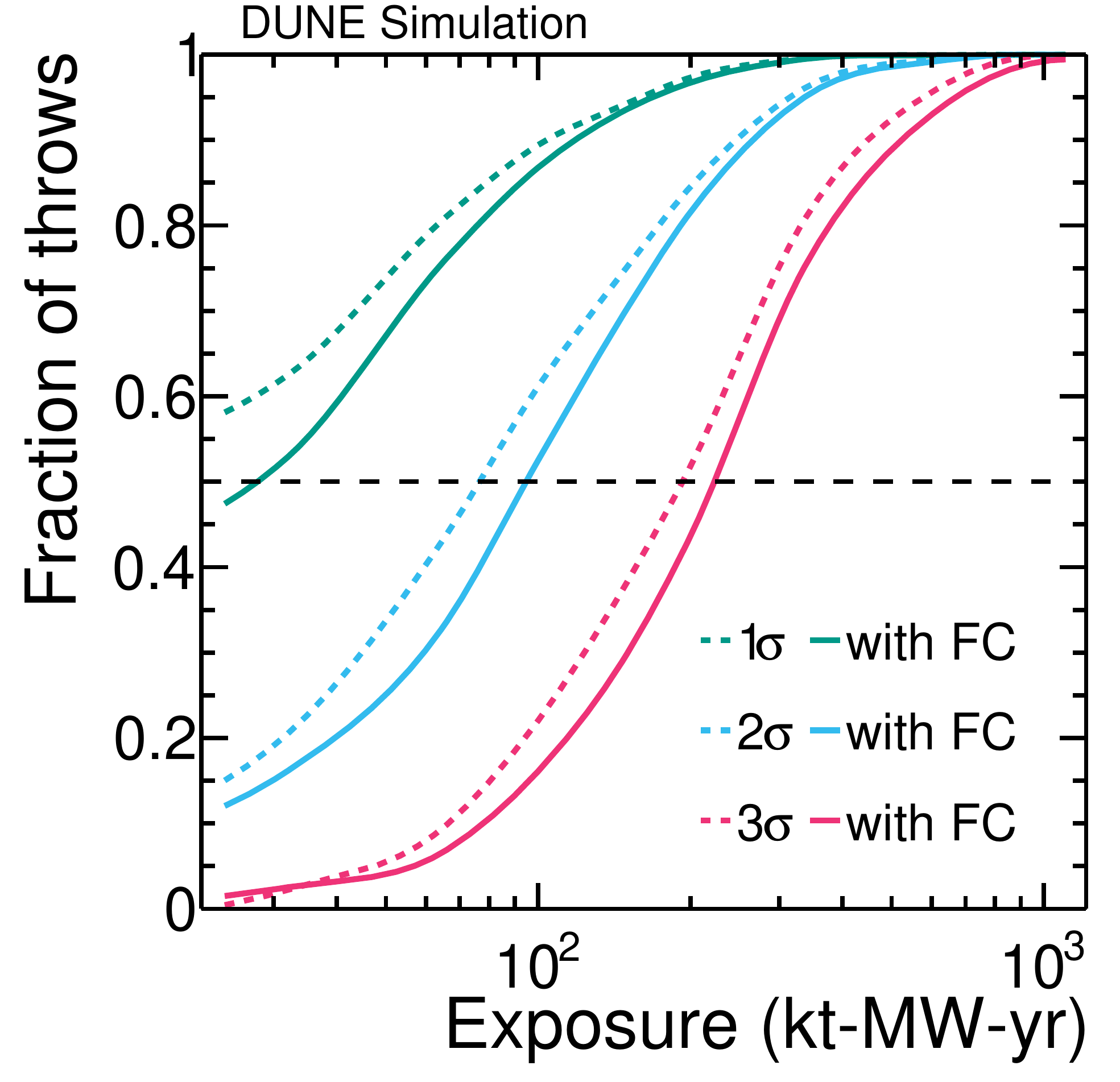}}
\caption{Fraction of throws for which the significance of DUNE's CP-violation test ($\deltacp \neq \{0,\pm\pi\}$) exceeds 1--3$\sigma$, for $\deltacp = -\pi/2$ and for 50\% of \deltacp values, calculated with the FC (solid lines) and constant-\dchisq (dashed lines) methods, as a function of exposure.}
  \label{fig:cpv_vs_exp_fc}
\end{figure*}

Figure~\ref{fig:cpv_vs_exp_fc} shows the fraction of throws which exceed different significance thresholds at the maximal \deltacp violation value of $\deltacp = -\pi/2$, and for 50\% of \deltacp values as a function of exposure, with and without FC corrections, for 1--3$\sigma$ significance values. Figure~\ref{fig:cpv_vs_exp_fc} was produced using the same throws used for Figure~\ref{fig:cpv_over_time_fc}, with additional points from higher exposures used in Ref.~\cite{Abi:2020qib}, but not shown in Figure~\ref{fig:cpv_over_time_fc} (646 kt-MW-yr and 1104 kt-MW-yr). After $\approx$200 kt-MW-yr, the median significance (including FC correction) for 50\% of the \deltacp range is greater than 3$\sigma$. It is clear from Figure~\ref{fig:cpv_vs_exp_fc} that the effect of the FC correction is not large, and $\approx$10\% longer exposures are required for the median expected significance to cross each threshold than without correction, at both $\deltacp = -\pi/2$ and for the 50\% range of \deltacp values.

\begin{figure*}[htbp]
  \centering
  \subfloat[24 kt-MW-yr] {\includegraphics[width=0.33\linewidth]{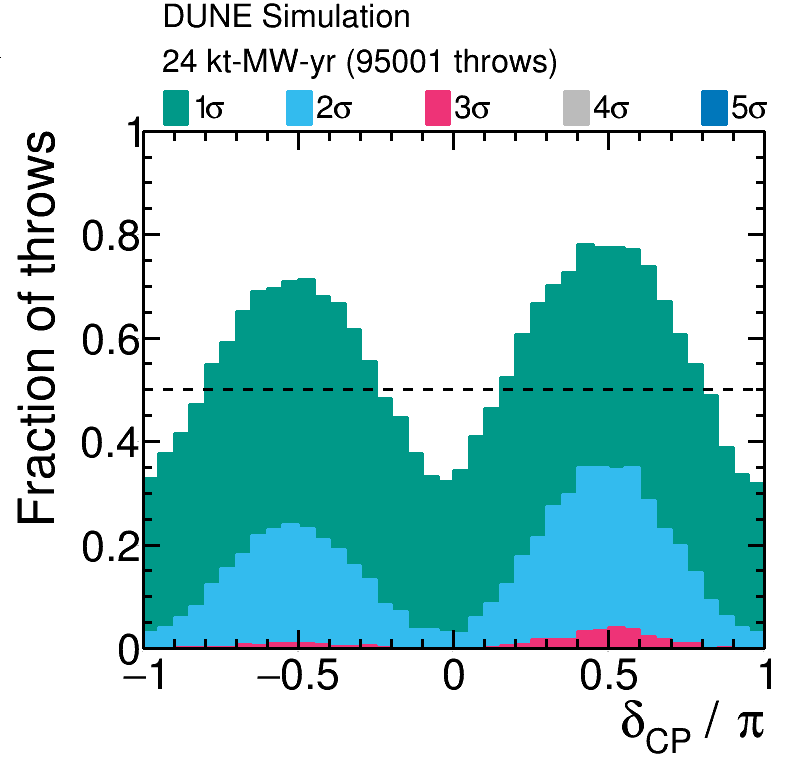}}
  \subfloat[66 kt-MW-yr] {\includegraphics[width=0.33\linewidth]{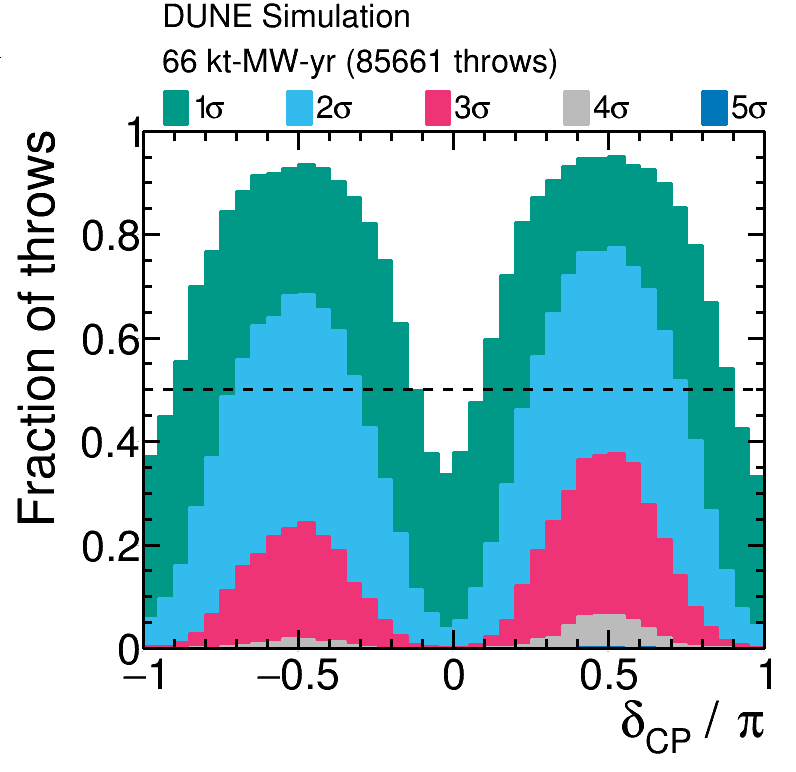}}
  \subfloat[100 kt-MW-yr]{\includegraphics[width=0.33\linewidth]{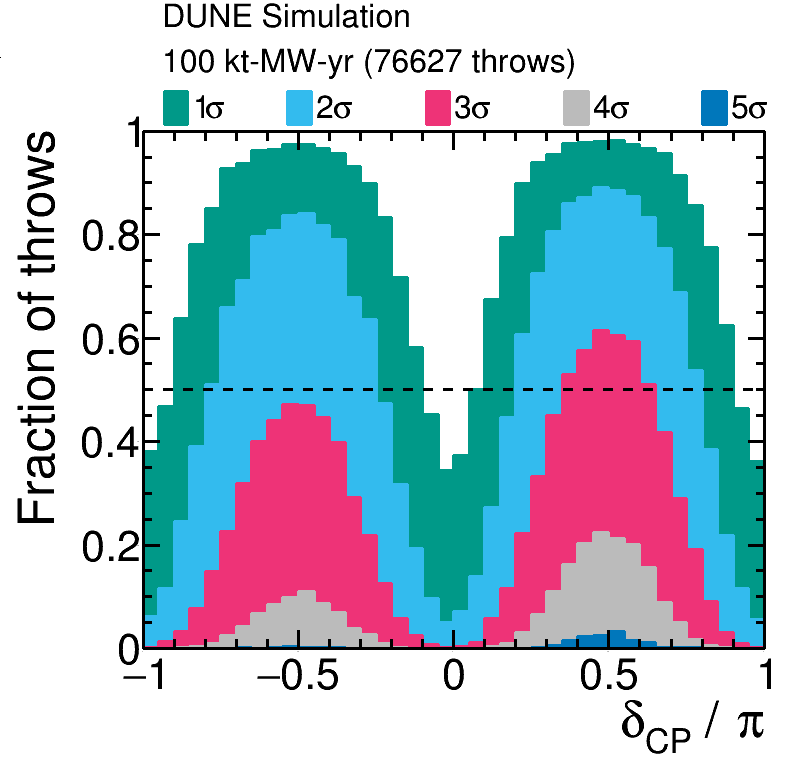}}\\
  \subfloat[150 kt-MW-yr]{\includegraphics[width=0.33\linewidth]{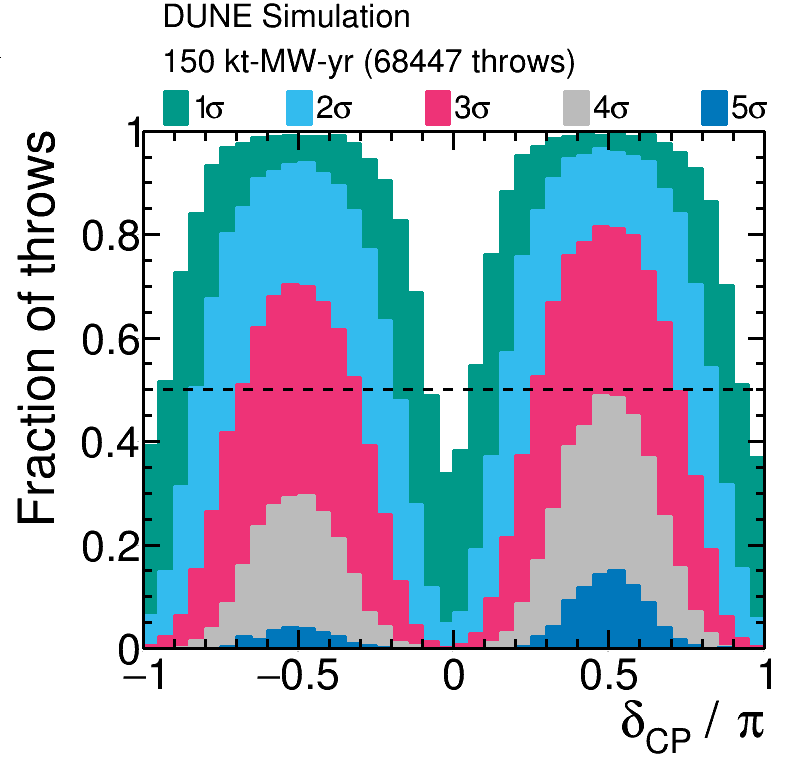}}
  \subfloat[197 kt-MW-yr]{\includegraphics[width=0.33\linewidth]{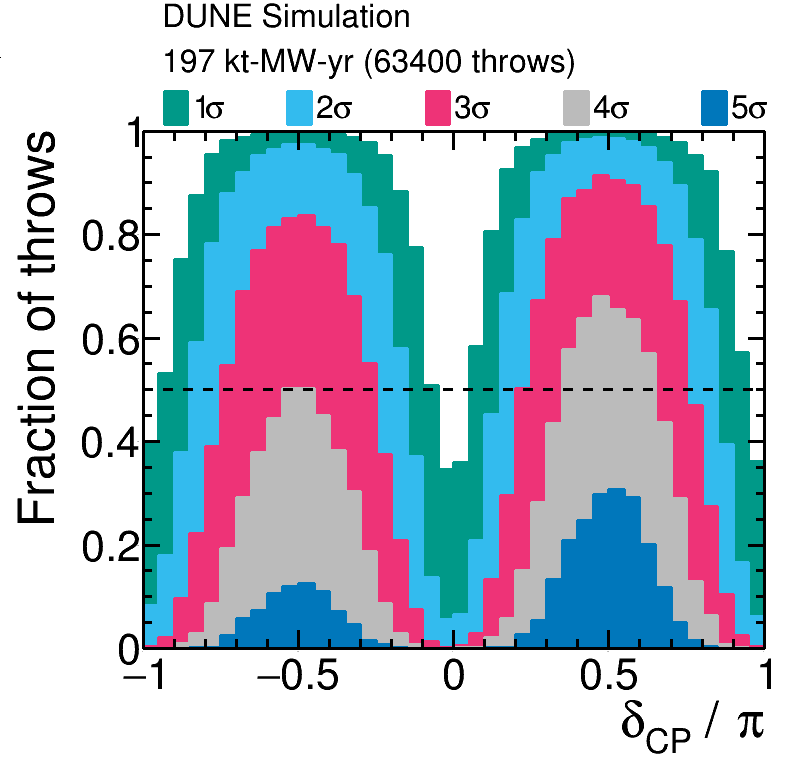}}
  \subfloat[336 kt-MW-yr]{\includegraphics[width=0.33\linewidth]{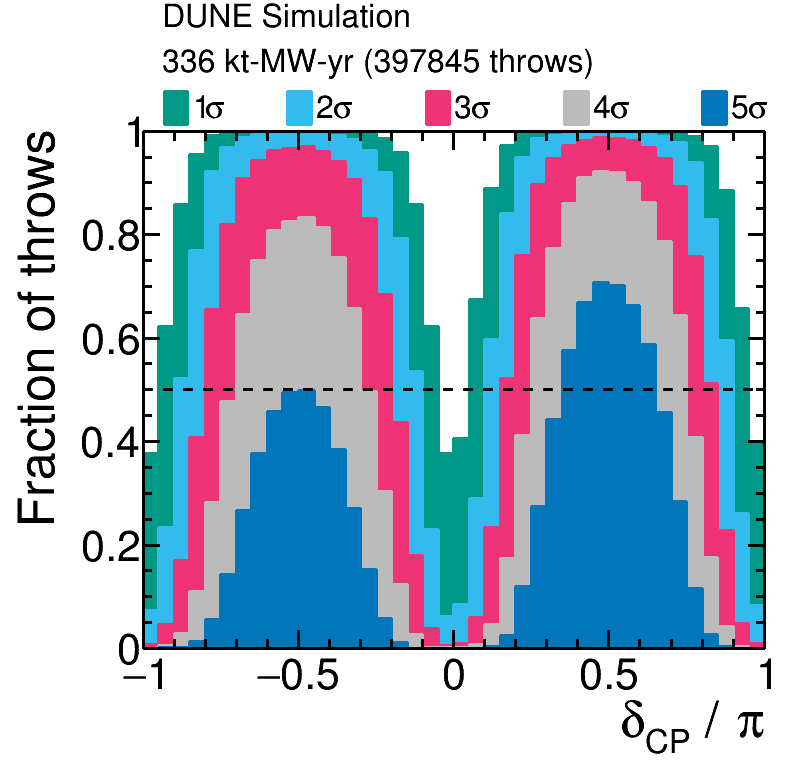}}
  \caption{Fraction of throws for which the significance of DUNE's CP-violation test ($\deltacp \neq \{0,\pm\pi\}$) exceeds 1--5$\sigma$, as a function of the true value of \deltacp. Shown for NO, for a number of different exposures. The number of throws used to make each figure is also shown.}
  \label{fig:cpv_over_time}
\end{figure*}
\begin{figure*}[htbp]
  \centering
  \subfloat[$\deltacp = -\pi/2$]     {\includegraphics[width=0.8\columnwidth]{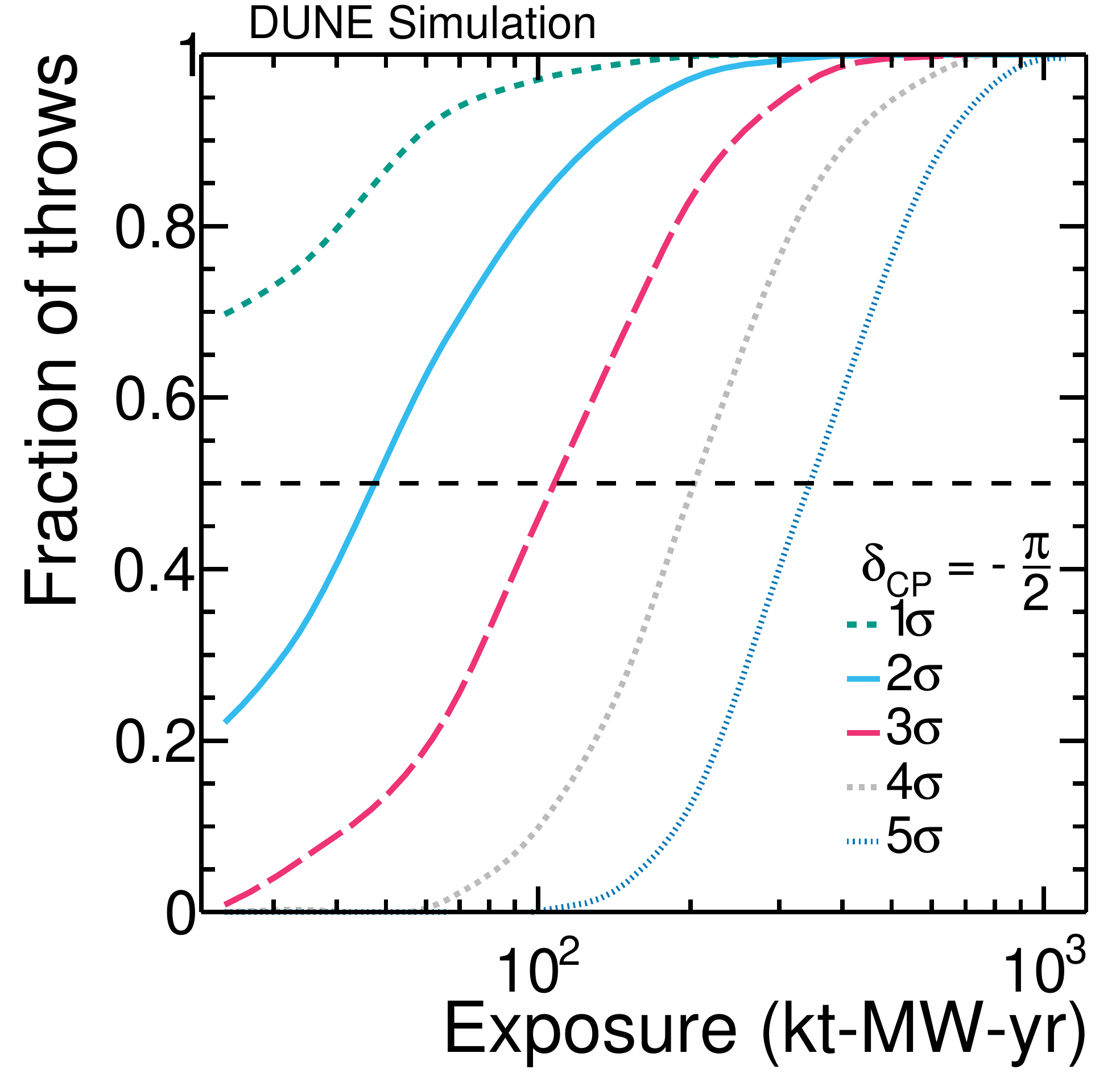}}
  \subfloat[50\% of \deltacp values] {\includegraphics[width=0.8\columnwidth]{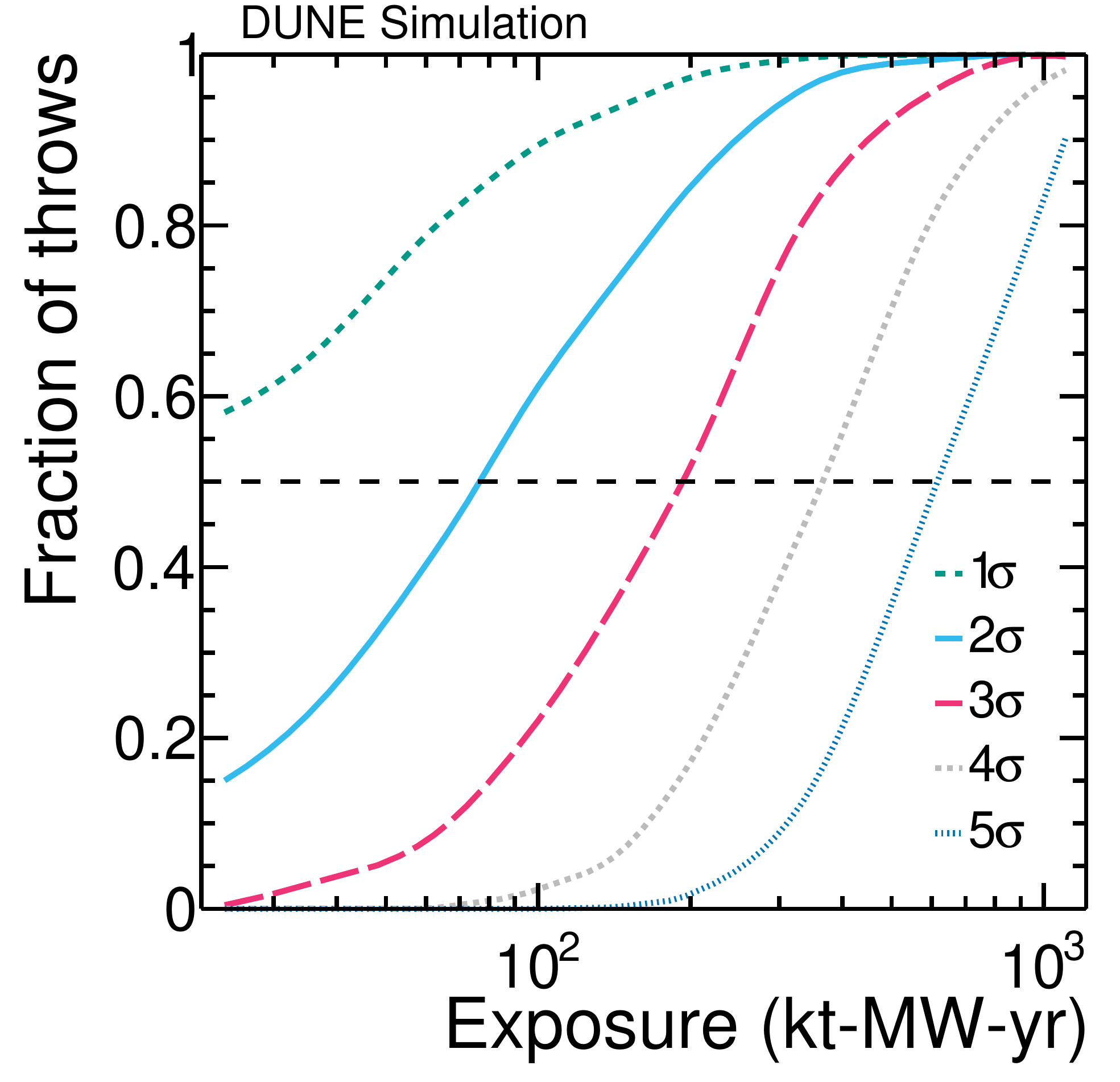}}
  \caption{Fraction of throws for which the significance of DUNE's CP-violation test ($\deltacp \neq \{0,\pm\pi\}$) exceeds 1--5$\sigma$, both assuming $\deltacp = -\pi/2$, and for 50\% of \deltacp values, shown as a function of exposure, for NO.}
  \label{fig:cpv_vs_exp}
\end{figure*}

Calculating \dchisqcrit values above 3$\sigma$ using the FC method is challenging due to the large number of throws to explore the tails of the \dchisqFC distribution and prohibitive computational cost. In Figure~\ref{fig:cpv_over_time}, the fraction of throws that exceed 1--5$\sigma$ significance, calculated only with the constant-\dchisq method are shown in order to explore DUNE's sensitivity at higher significance levels. All the caveats described above relating to the constant-\dchisq method still apply. Figure~\ref{fig:cpv_over_time} shows that although the median significance to CPV does not exceed 5$\sigma$ until $\approx$336 kt-MW-yr, there are significant fractions of throws at lower exposures which reach $5\sigma$ significance. Figure~\ref{fig:cpv_vs_exp} shows the fraction of throws which exceed different significance thresholds at the maximal CP-violating value of $\deltacp = -\pi/2$, and for 50\% of all \deltacp values, as a function of exposure. By $\approx$200 kt-MW-yr, where the median significance for 50\% of the \deltacp range is greater than 3$\sigma$, the sensitivity at $\deltacp = -\pi/2$ exceeds 4$\sigma$.

\FloatBarrier

%% file: sections/mh_sens.tex
\section{Neutrino mass ordering sensitivity}
\label{sec:mh_sens}

In this section, the toy-throwing approach described in Section~\ref{sec:analysis_framework} is used to explore the neutrino mass ordering sensitivity as a function of exposure in detail. In all cases, a joint ND+FD fit is performed, and the reactor $\theta_{13}$ constraint is always applied, as described in Section~\ref{sec:analysis_framework}. An equal split between FHC and RHC running is assumed based on the results obtained in Section~\ref{sec:run_plan_opt}.

\begin{figure}[htbp]
  \centering
  \includegraphics[width=1\linewidth, trim={0cm 0cm 0cm 2.3cm}, clip]{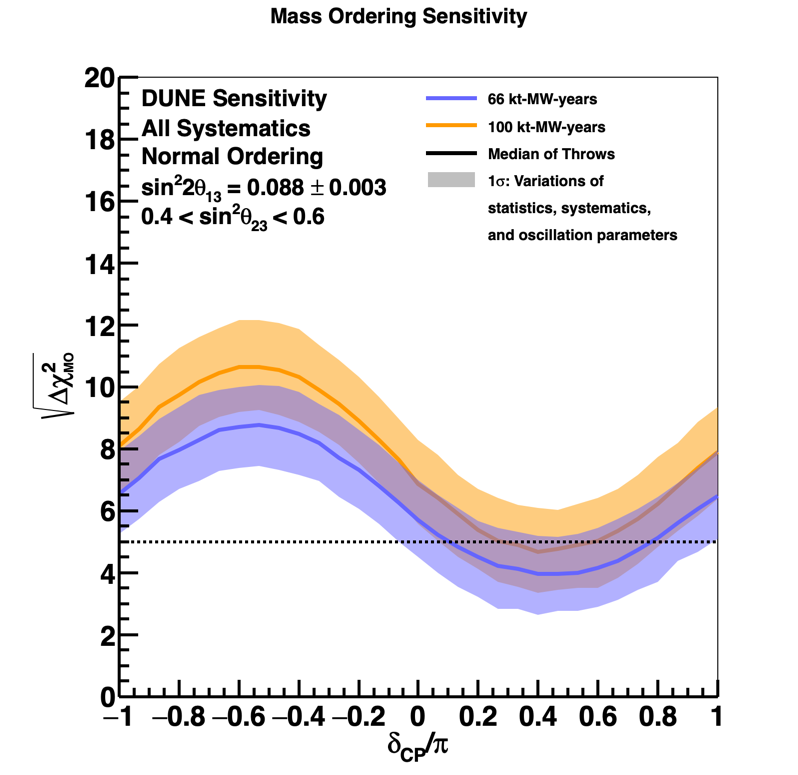}
  \includegraphics[width=1\linewidth, trim={0cm 0cm 0cm 2.3cm}, clip]{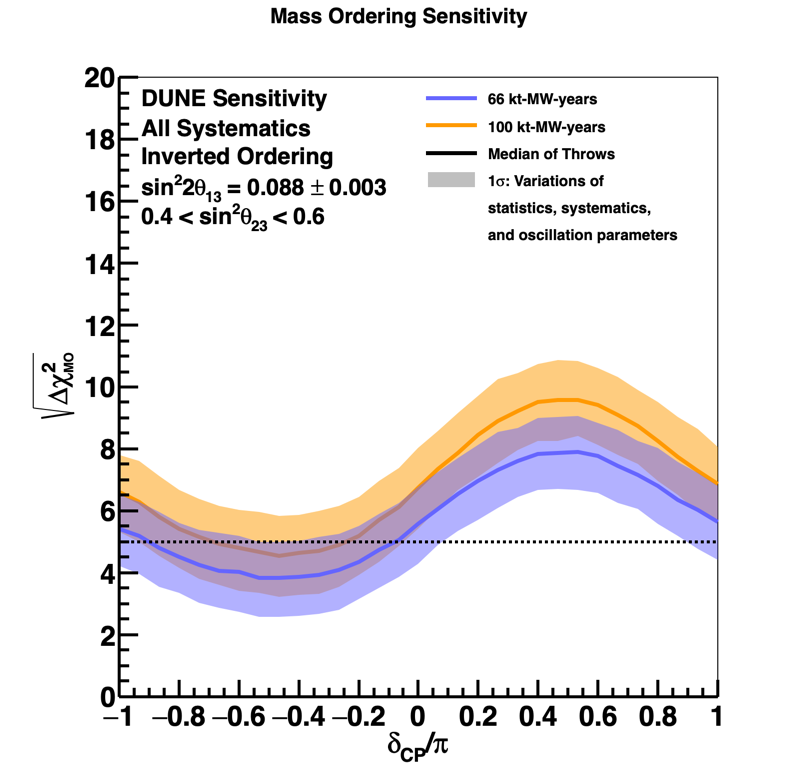}
  \caption{Significance of the DUNE determination of the neutrino mass ordering, as a function of the true value of \deltacp, for 66 kt-MW-yr (blue) and 100 kt-MW-yr (orange) exposures. The width of the transparent bands cover 68\% of fits in which random throws are used to simulate systematic, oscillation parameter and statistical variations, with independent fits performed for each throw constrained by prior uncertainties. The solid lines show the median significance.}
  \label{fig:mh_bands}
\end{figure}
Figure~\ref{fig:mh_bands} shows the significance with which the neutrino mass ordering can be determined for both true NO and IO, for exposures of 66 and 100 kt-MW-yr. The sensitivity metric used is the square root of the difference between the best fit $\chi^{2}$ value obtained using each ordering, as shown in Equation~\ref{eq:mh_chi2}, which is calculated for each throw of the systematics, other oscillation parameters and statistics. The characteristic shape of the MH sensitivity in Figure~\ref{fig:mh_bands} results from near degeneracy between matter and CPV effects that occurs near $\deltacp=\pi/2$ ($\deltacp=-\pi/2$) for true normal (inverted) ordering. Dedicated studies have shown that special attention must be paid to the statistical interpretation of neutrino mass ordering sensitivities~\cite{Ciuffoli:2013rza,Qian:2012zn,Blennow:2013oma} because the \dchisqMO metric does not follow the expected chi-square distribution for one degree of freedom, so the interpretation of the $\sqrt{\dchisqMO}$ as the sensitivity is complicated.

\begin{figure*}[htbp]
  \centering
  \subfloat[6 kt-MW-yr]   {\includegraphics[width=0.33\linewidth]{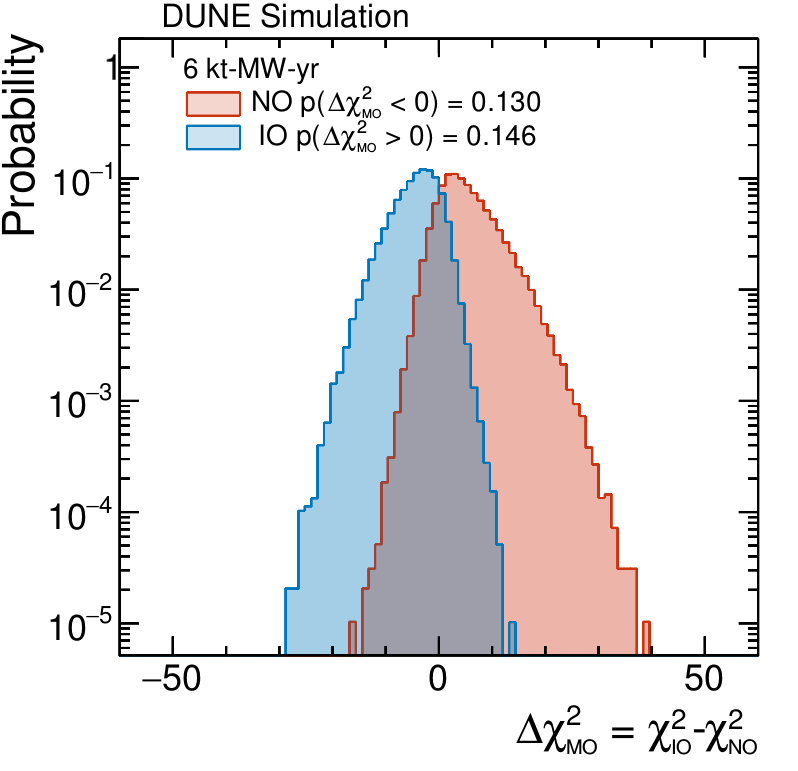}}
  \subfloat[12 kt-MW-yr]  {\includegraphics[width=0.33\linewidth]{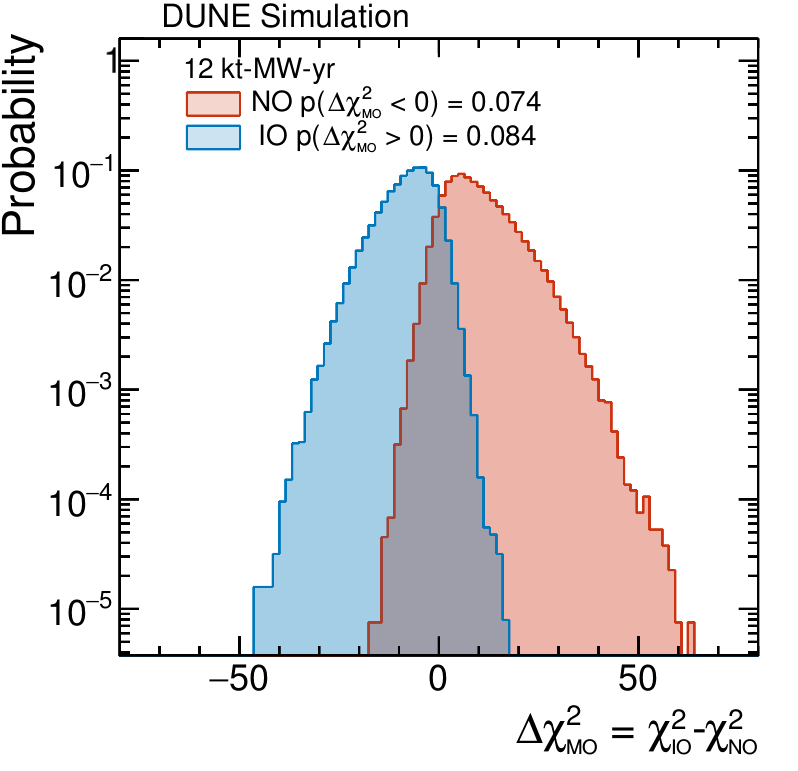}}
  \subfloat[24 kt-MW-yr]  {\includegraphics[width=0.33\linewidth]{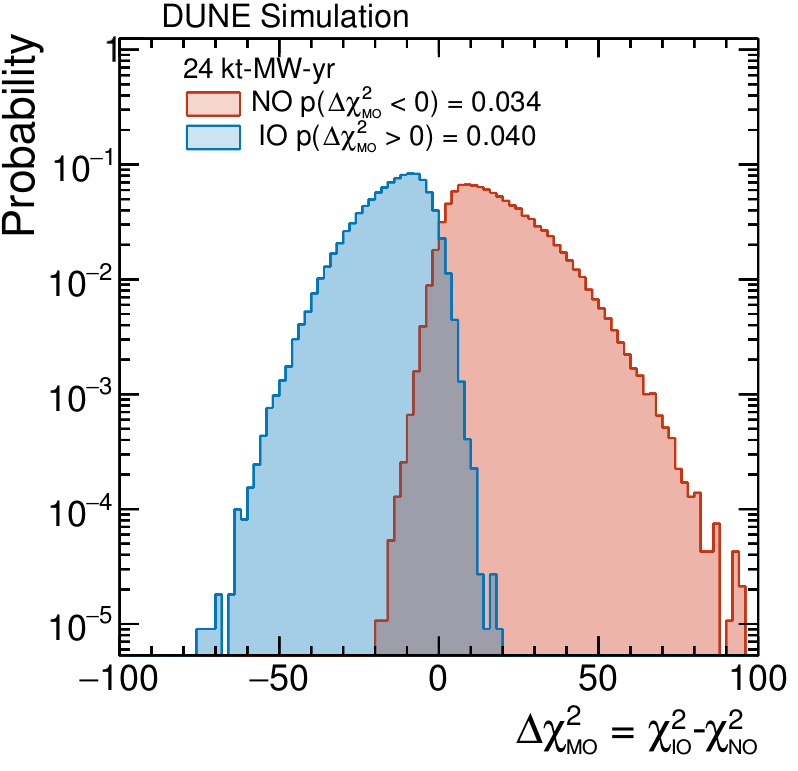}}\\
  \subfloat[66 kt-MW-yr]  {\includegraphics[width=0.33\linewidth]{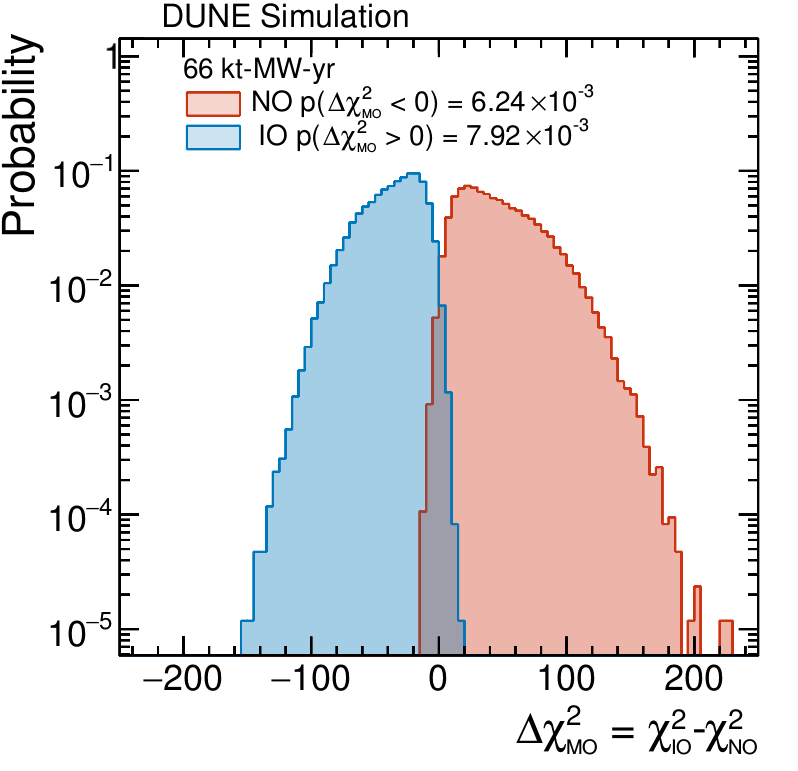}}
  \subfloat[100 kt-MW-yr] {\includegraphics[width=0.33\linewidth]{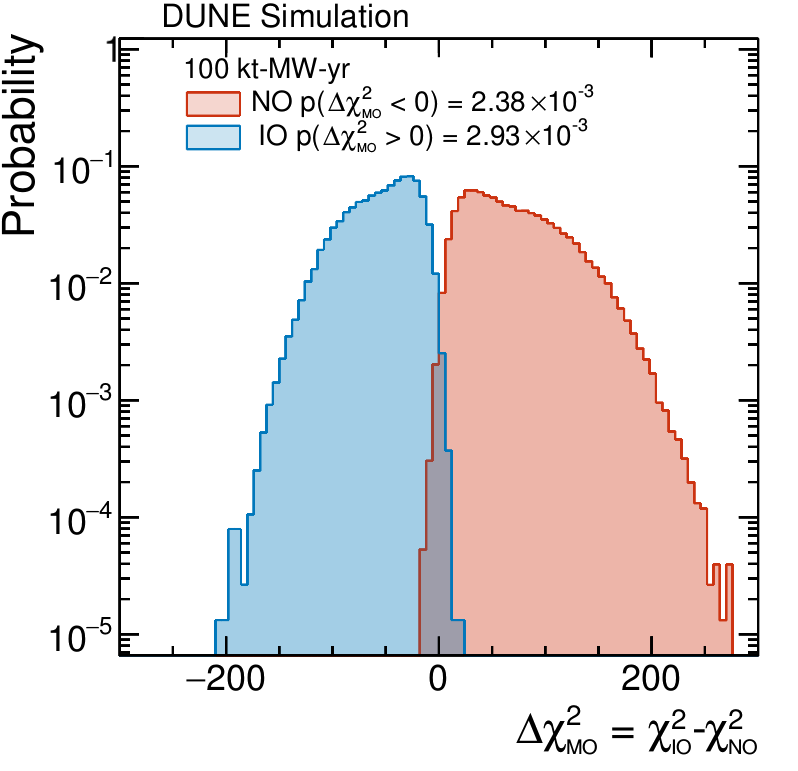}}
  \subfloat[336 kt-MW-yr] {\includegraphics[width=0.33\linewidth]{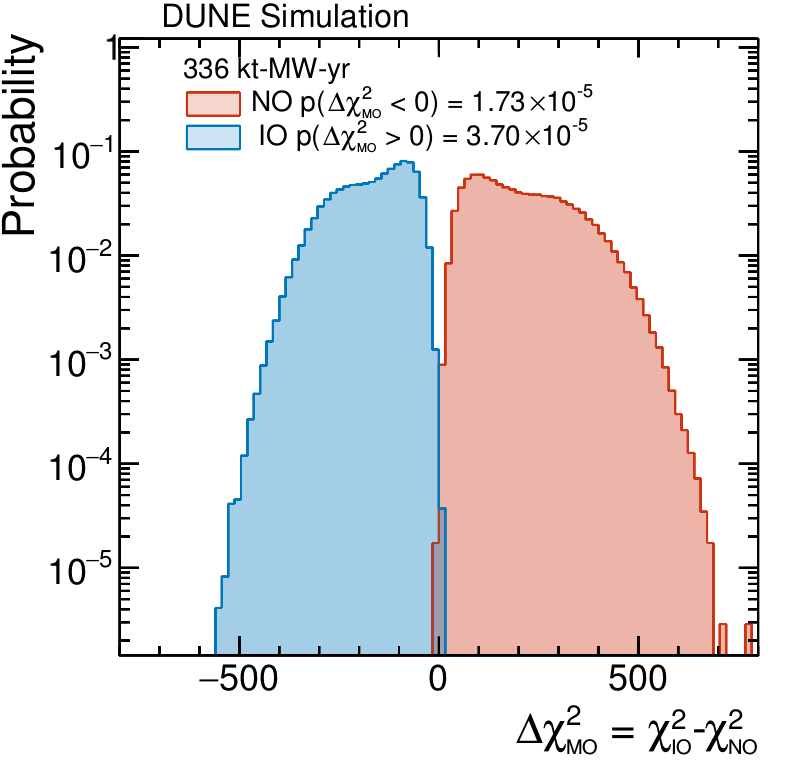}}
  \caption{The distribution of $\dchisqMO = \chi^{2}_{\mathrm{IO}} - \chi^{2}_{\mathrm{NO}}$ values shown for both true normal (red) and true inverted (blue) hierarchies built using random throws of the systematic parameters, the oscillation parameters and with statistical variations. In each case, the $\chi^{2}$ values are separately minimized with respect to all variable parameters before calculating the test statistic. The fraction of throws for which the value of \dchisqMO is greater than (less than) 0 is also given for inverted (normal) hierarchies. For each ordering and exposure, approximately 100,000 throws were used.}
  \label{fig:mh_comp_over_time}
\end{figure*}
Given the complications with the interpretation of significance for mass ordering determination, it is instructive to look at the distribution of the test-statistic (Equation~\ref{eq:mh_chi2}), which gives more information than the 68\% central band and median throw shown in Figure~\ref{fig:mh_bands}. Figure~\ref{fig:mh_comp_over_time} shows the distribution of \dchisqMO obtained for a large ensemble of throws, for both true normal and inverted orderings, for a number of different exposures. There is a uniform distribution of true \deltacp used in the throws at each exposure. The change in shape at higher exposures in Figure~\ref{fig:mh_comp_over_time} is due to the degeneracy between \deltacp and the effect of the mass ordering, and as might be expected from Figure~\ref{fig:mh_bands}, the separation between hierarchies is greater for some true values of \deltacp than others. This additional structure starts to become obvious from a $\approx$66 kt-MW-yr exposure, at which point the CPV sensitivity is not very strong (see Section~\ref{sec:cp_sens}). For all exposures, the shape of the throw distribution is highly non-Gaussian, which makes it difficult to apply simple corrections to the sensitivity of the sort described in Ref.~\cite{Blennow:2013oma}. As a result alternatives to $\sqrt{\dchisqMO}$ as a sensitivity metric are not explored, the full information is given in Figure~\ref{fig:mh_comp_over_time}.

\begin{figure}[htbp]
  \centering
  \includegraphics[width=0.8\linewidth]{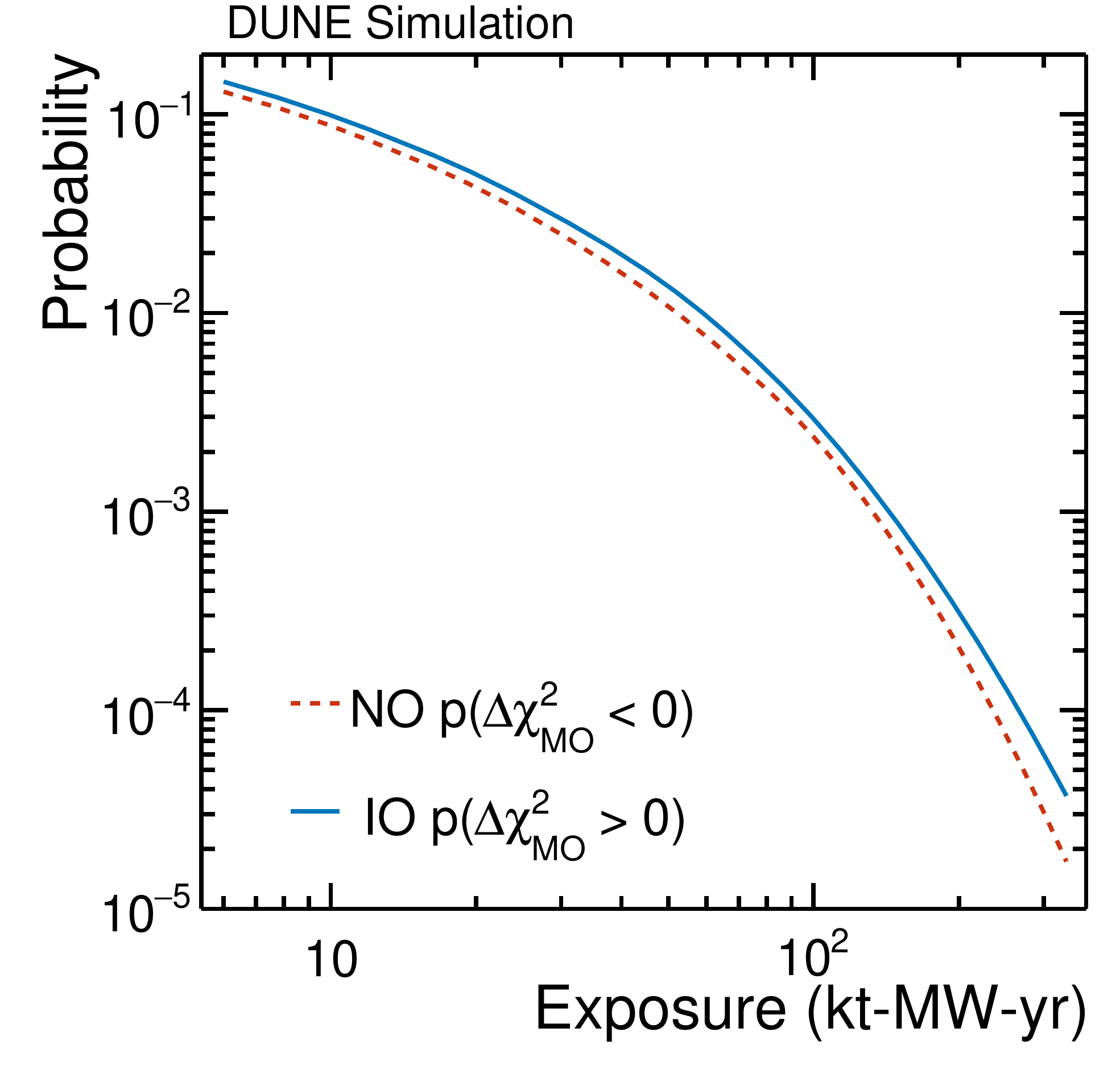}
  \caption{The probability for preferring the wrong neutrino mass ordering as a function of exposure, shown for both true NO and IO.}
  \label{fig:mh_wrong}
\end{figure}
Figure~\ref{fig:mh_comp_over_time} also indicates the probability for the test statistic \dchisqMO to be less (more) than zero from the toy throws for true normal (inverted) orderings at each exposure. This information is summarized in Figure~\ref{fig:mh_wrong}. This marks the proportion of toys which appear more like the incorrect ordering than the true ordering for the toy, and gives a sense of the ambiguity between the hierarchies, although it is not easily converted to a single number sensitivity. It is clear from Figures~\ref{fig:mh_comp_over_time} and~\ref{fig:mh_wrong} that DUNE is sensitive to the mass ordering even from very low ($\approx$12 kt-MW-yr) exposures, with a small probability for preferring the incorrect ordering. By exposures of 66 kt-MW-yr, the overlap between the orderings is very small with $\approx$1\% of toy throws which appear more like the incorrect ordering than the true ordering.

\begin{figure*}[htbp]
  \centering
  \subfloat[6 kt-MW-yr]   {\includegraphics[width=0.33\linewidth]{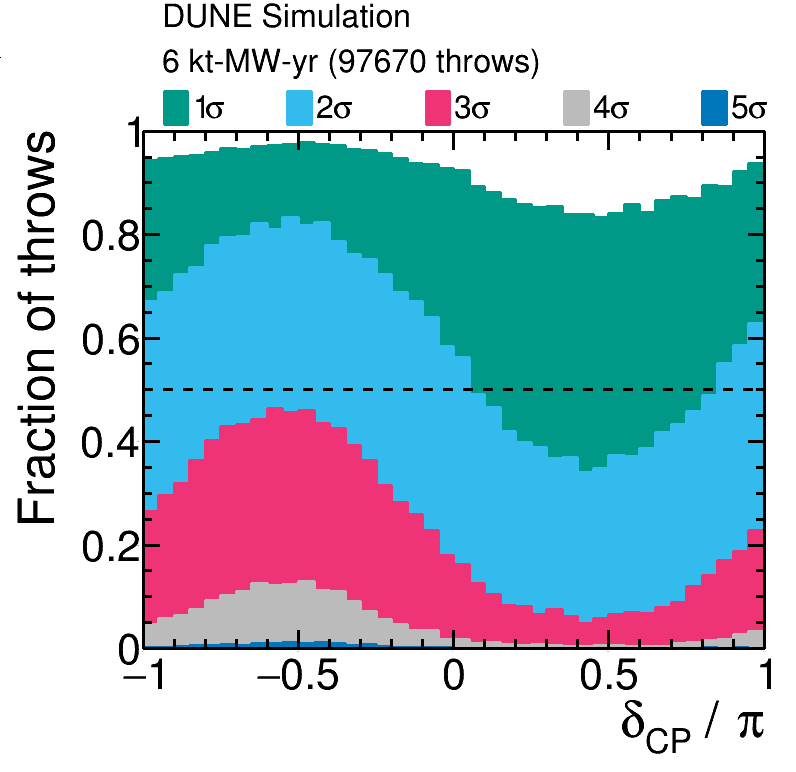}}
  \subfloat[12 kt-MW-yr]  {\includegraphics[width=0.33\linewidth]{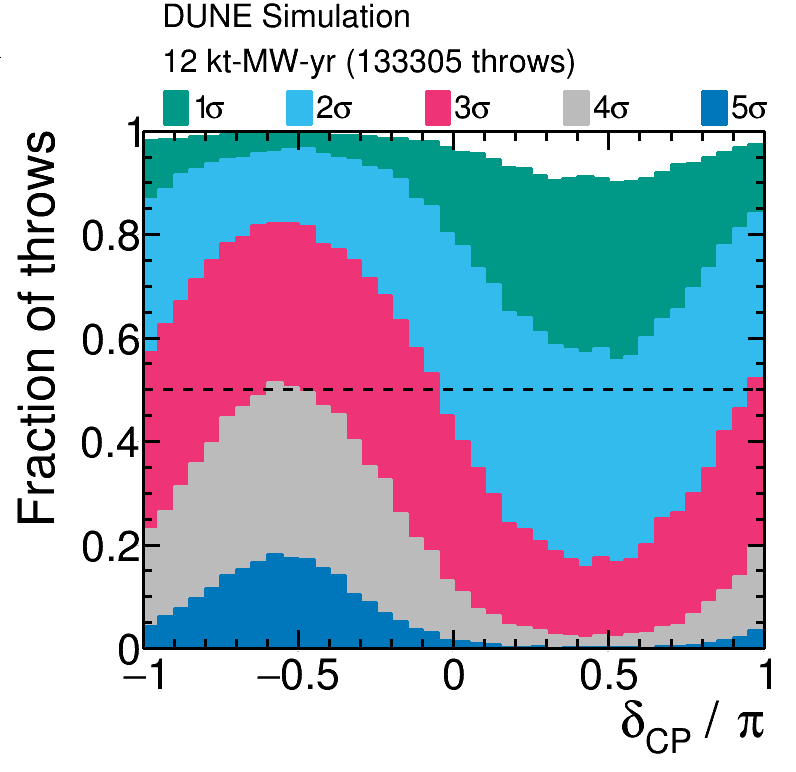}}
  \subfloat[24 kt-MW-yr]  {\includegraphics[width=0.33\linewidth]{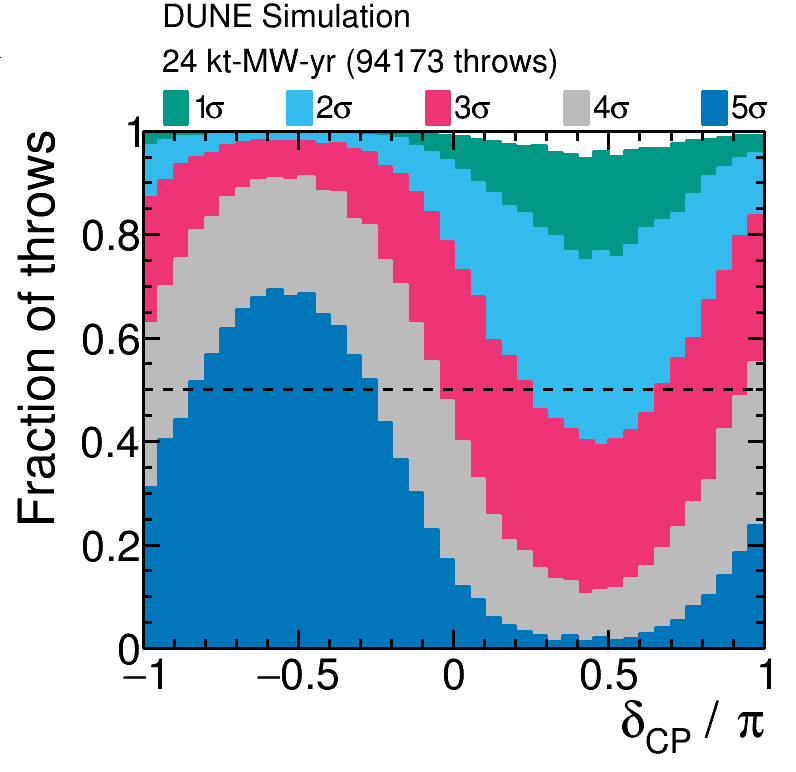}}\\
  \subfloat[66 kt-MW-yr]  {\includegraphics[width=0.33\linewidth]{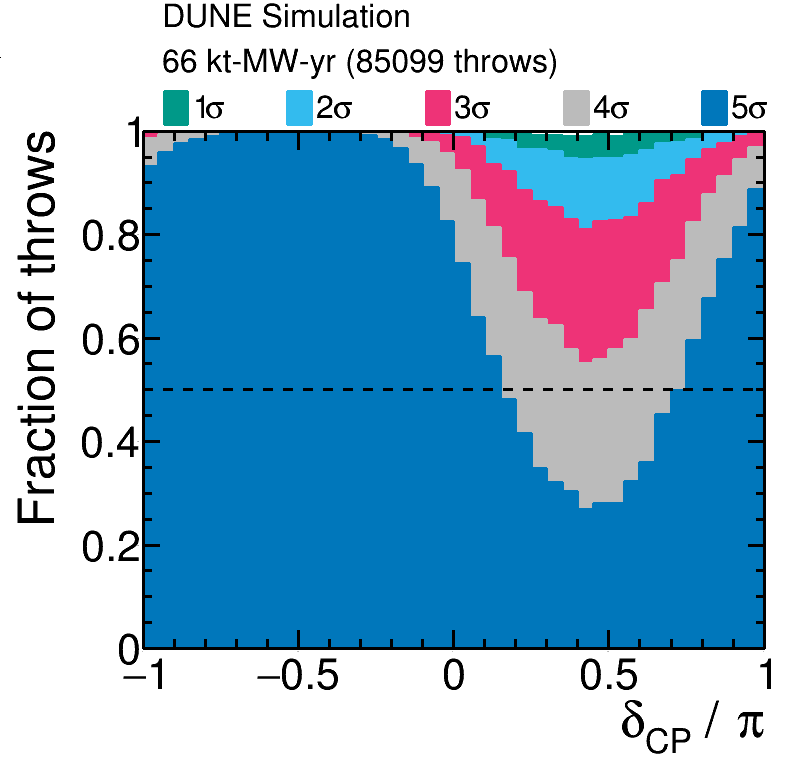}}
  \subfloat[100 kt-MW-yr] {\includegraphics[width=0.33\linewidth]{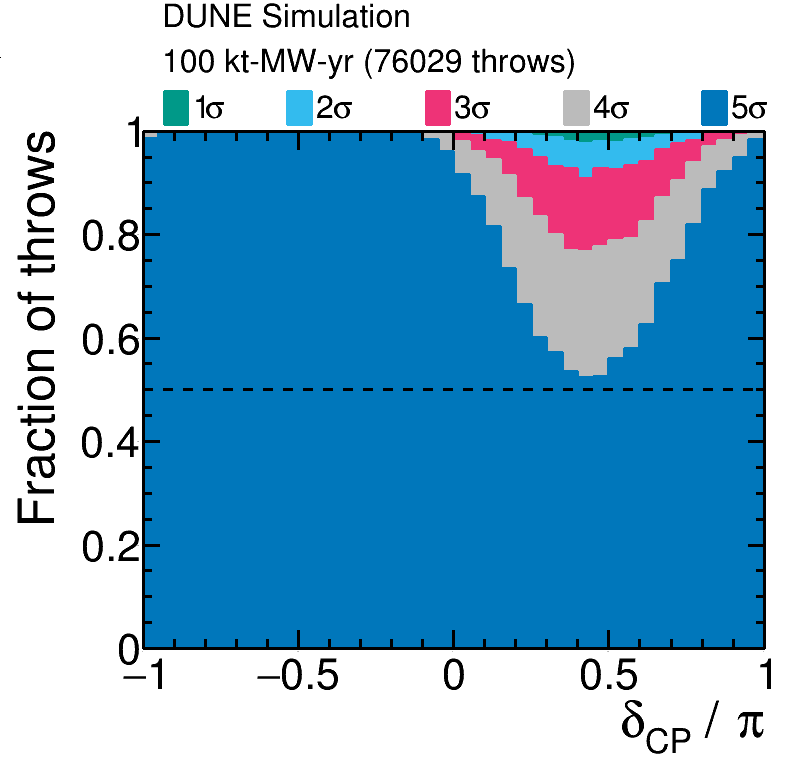}}
  \subfloat[336 kt-MW-yr] {\includegraphics[width=0.33\linewidth]{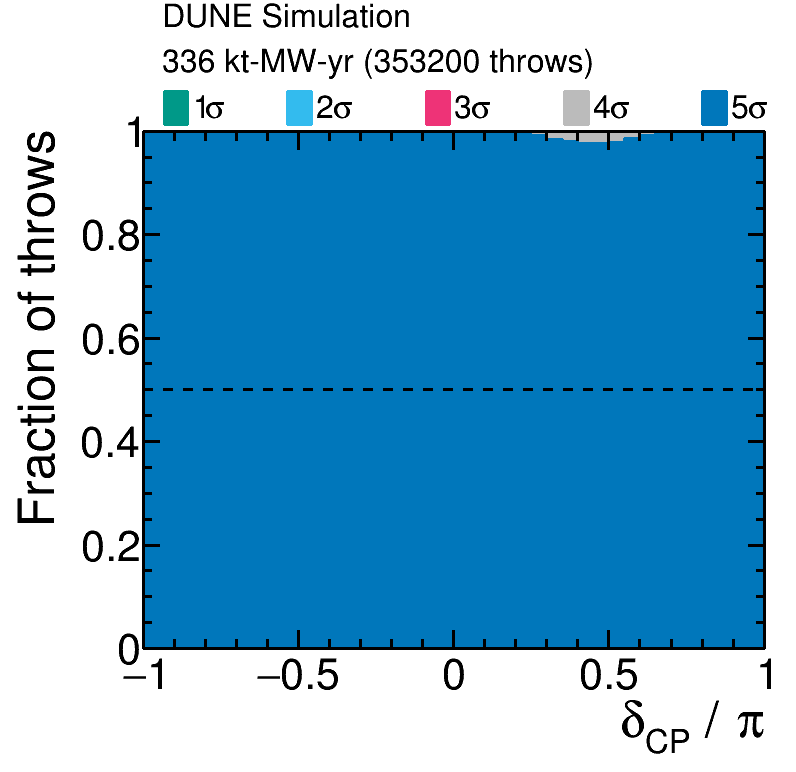}}
  \caption{Fraction of throws for which the DUNE sensitivity to the mass ordering exceeds 1--5$\sigma$ significance, as a function of the true value of \deltacp. Shown for NO, for a number of different exposures. The number of throws used to make each figure is also shown.}
  \label{fig:mh_nh_over_time}
\end{figure*}

Figure~\ref{fig:mh_nh_over_time} shows an alternative way to present the result of the throws as a function of \deltacp, which is complementary to Figure~\ref{fig:mh_bands}. The fraction of throws for which the simple figure of merit (the square-root of Equation~\ref{eq:mh_chi2}) exceeds different confidence levels are shown, for 1--5$\sigma$ significances, and a variety of exposures, all for true NO. The same throws are used as in Figures~\ref{fig:mh_comp_over_time}. Despite the caveats regarding the intepretation of $\sqrt{\dchisqMO}$ as units of $\sigma$, the general trend is clear, and provides more information about the expected DUNE sensitivity at low exposures. As with Figures~\ref{fig:cpv_over_time_fc} and~\ref{fig:cpv_over_time}, the point at which the median significance (50\% of throws) passes different significance thresholds can be easily read from the figures, and can be compared with those shown in Figure~\ref{fig:mh_bands}. The same general shape as a function of \deltacp as was observed in Figure~\ref{fig:mh_bands} can be seen. The general trend would be very similar in IO, reflected in the line $\deltacp = 0$, although a slightly longer exposure is required to reach the same sensitivity. The median significance for $\deltacp = -\pi/2$ exceeds 5$\sigma$ for 24 kt-MW-yr, at which point the fraction of throws for which the significance is 3$\sigma$ or smaller is only $\approx$2\%. By 66 kt-MW-yr, 100\% of the throws exceed 5$\sigma$ at $\deltacp = -\pi/2$. By 100 kt-MW-yr exposures, the median significance approaches 5 $\sigma$ for all true values of \deltacp. At long exposures of 336 kt-MW-yr, almost 100\% of the throws exceed 5$\sigma$ for all values of \deltacp.

%% file: sections/conclusion.tex
\section{Conclusion}
\label{sec:conclude}

In this work a detailed exploration of DUNE's sensitivity to CPV and the mass ordering at low exposures has been presented. The analysis uses the same framework, flux, cross section and detector models and selections as were used in Ref.~\cite{Abi:2020qib}, which showed the ultimate DUNE sensitivity to CPV, the neutrino mass ordering and other oscillation parameters, with large statistics samples after long exposures.

The effect of operating with different run plans, involving different ratios of FHC and RHC beam modes, on the mass ordering and CPV Asimov sensitivities was explored. It was found that for low exposures, the sensitivity to both CPV and the mass ordering can be increased for certain regions of parameter space, but at a cost to the sensitivity in other regions. This sensitivity increase is in part produced by leveraging the strong $\theta_{13}$ constraint available from reactor experiments. If there is a strong reason to favor the exploration of a given region of parameter space when DUNE begins to take data, this issue should be re-visited. However, with no strong motivation to focus on a given ordering or region of \deltacp parameter space, equal FHC and RHC beam running provides a close to optimal sensitivity across all of the parameter space, so was used for the subsequent detailed sensitivity studies. The increase in sensitivity for unequal beam running is also a feature of low exposure running, and degrades the sensitivity almost uniformly across the parameter space investigated for large exposures, with and without a $\theta_{13}$ constraint applied.

The studies presented here demonstrate that a full treatment of DUNE's sensitivity at low exposures supports the conclusions made in Refs.~\cite{Abi:2020qib} and~\cite{Abi:2020evt} using Asimov studies. In particular, the median CPV sensitivity is $\approx$3$\sigma$ for $\deltacp = \pm\pi/2$ after approximately a 100 kt-MW-yr FD exposure. Variations in the expected sensitivity around the median value were also explored. Additionally, it was shown that the CPV sensitivity is not significantly degraded when Feldman-Cousins corrections are included, leading to $\approx$10\% longer exposures to reach a given significance level. Crucially, it was found that after an initial low-exposure rise, the Feldman-Cousins \dchisqcrit do not change as a function of exposure, unlike the rise with exposure which has been observed by the T2K experiment~\cite{Abe:2021gky}.

It has also been shown that strong statements on the mass ordering can be expected with very short exposures of $\approx$12 kt-MW-yr, which supports the results shown in Refs.~\cite{Abi:2020qib} and~\cite{Abi:2020evt} with a more complete treatment of the systematic uncertainty.

Although the analysis used here makes no assumptions about the FD staging scenario, and results are given as a function of exposure only, the results are dependent on having a performant ND complex from the start of the experiment. In particular, the low-exposures necessary to make world-leading statements about the mass ordering can only be given with confidence with ND samples included in the fit. Additional samples of events from detectors other than ND-LAr in the DUNE ND complex are not explicitly included in this analysis, but there is an assumption that it will be possible to control the uncertainties to the level used in the analysis, and it should be understood that that implicitly relies on having a highly capable ND.

%% file: sections/acknowledgements.tex
\begin{acknowledgements}
This document was prepared by the DUNE collaboration using the
resources of the Fermi National Accelerator Laboratory 
(Fermilab), a U.S. Department of Energy, Office of Science, 
HEP User Facility. Fermilab is managed by Fermi Research Alliance, 
LLC (FRA), acting under Contract No. DE-AC02-07CH11359.
%
%
This work was supported by
CNPq,
FAPERJ,
FAPEG and 
FAPESP,                         Brazil;
CFI, 
IPP and 
NSERC,                          Canada;
CERN;
M\v{S}MT,                       Czech Republic;
ERDF, 
H2020-EU and 
MSCA,                           European Union;
CNRS/IN2P3 and
CEA,                            France;
INFN,                           Italy;
FCT,                            Portugal;
NRF,                            South Korea;
CAM, 
Fundaci\'{o}n ``La Caixa'',
Junta de Andaluc\'ia-FEDER, and 
MICINN,                         Spain;
SERI and 
SNSF,                           Switzerland;
T\"UB\.ITAK,                    Turkey;
The Royal Society and 
UKRI/STFC,                      United Kingdom;
DOE and 
NSF,                            United States of America.
This research used resources of the 
National Energy Research Scientific Computing Center (NERSC), 
a U.S. Department of Energy Office of Science User Facility 
operated under Contract No. DE-AC02-05CH11231.
\end{acknowledgements}

%% file: sections/fc_appendix.tex
\section{Feldman-Cousins throw distributions}\label{sec:fc_appendix}
\begin{figure*}[h]
  \centering
  \subfloat[24 kt-MW-yr] {\includegraphics[width=0.33\linewidth]{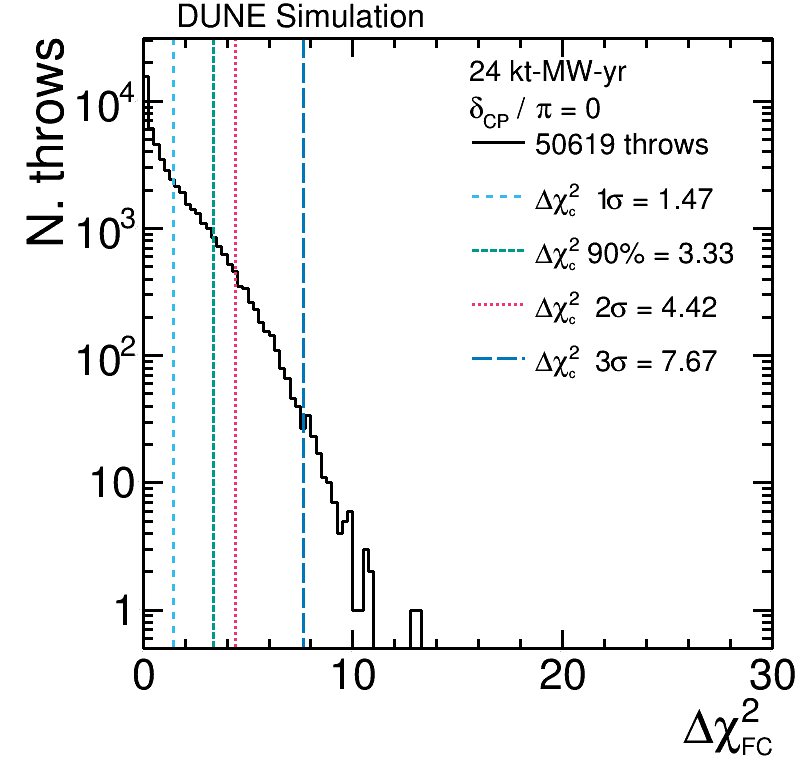}}
  \subfloat[66 kt-MW-yr] {\includegraphics[width=0.33\linewidth]{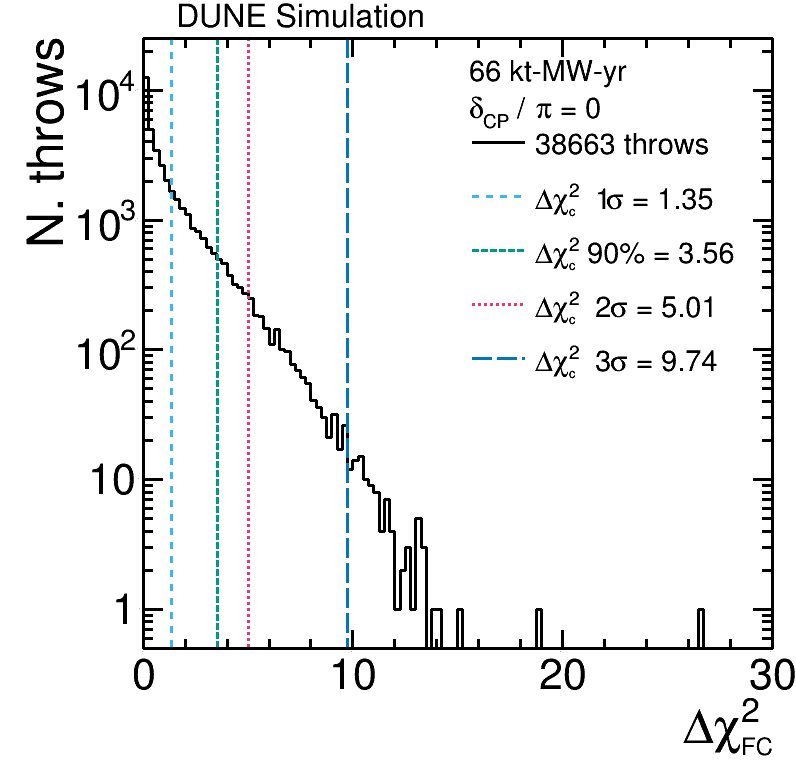}}
  \subfloat[100 kt-MW-yr]{\includegraphics[width=0.33\linewidth]{nh_FC_ndfd_100ktMWyr_dcp0.png}}\\
  \subfloat[150 kt-MW-yr]{\includegraphics[width=0.33\linewidth]{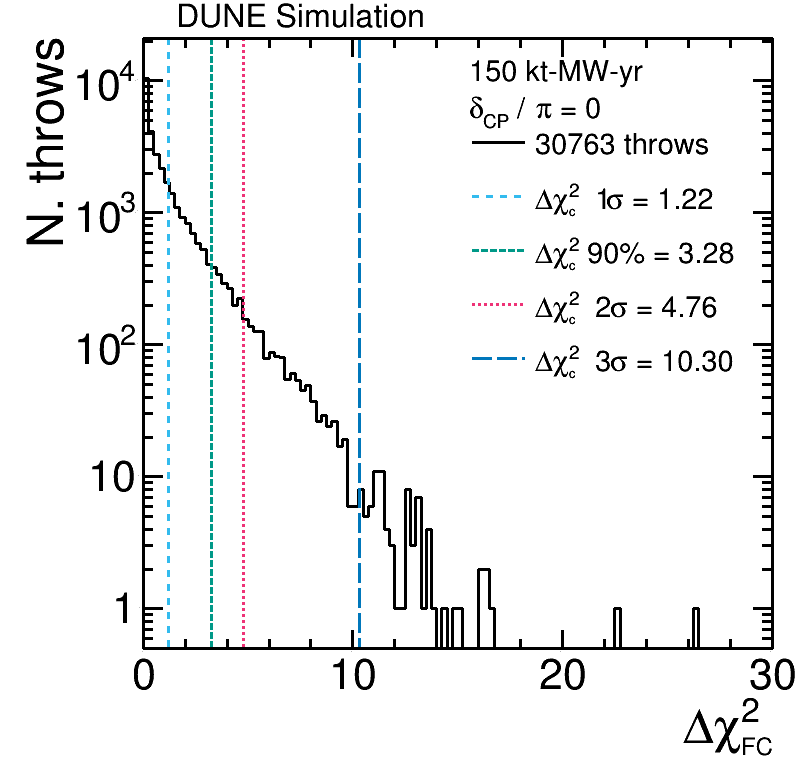}}
  \subfloat[197 kt-MW-yr]{\includegraphics[width=0.33\linewidth]{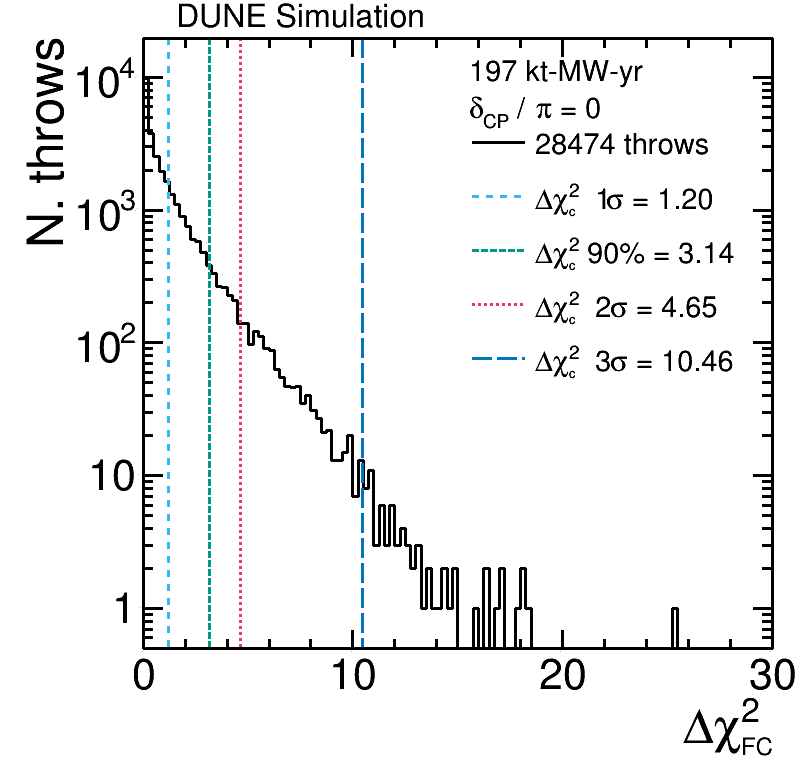}}
  \subfloat[336 kt-MW-yr]{\includegraphics[width=0.33\linewidth]{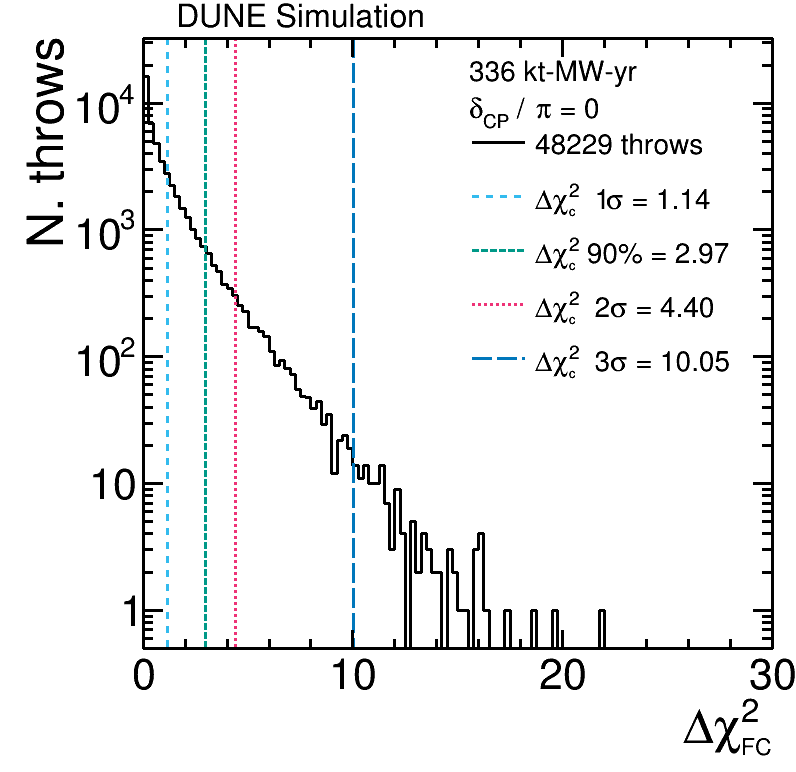}}\\
  \subfloat[500 kt-MW-yr]{\includegraphics[width=0.33\linewidth]{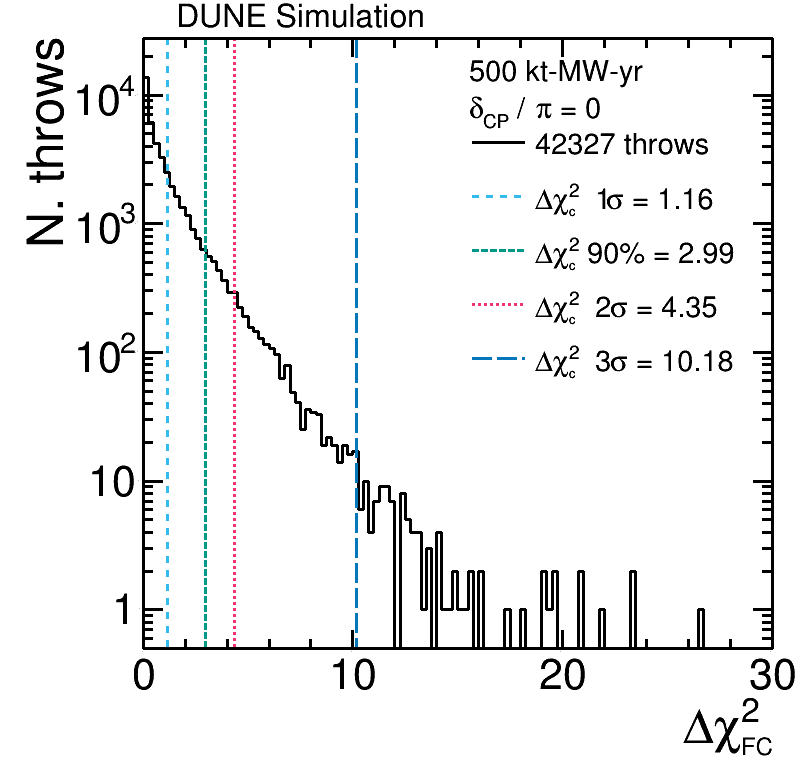}}
  \subfloat[646 kt-MW-yr]{\includegraphics[width=0.33\linewidth]{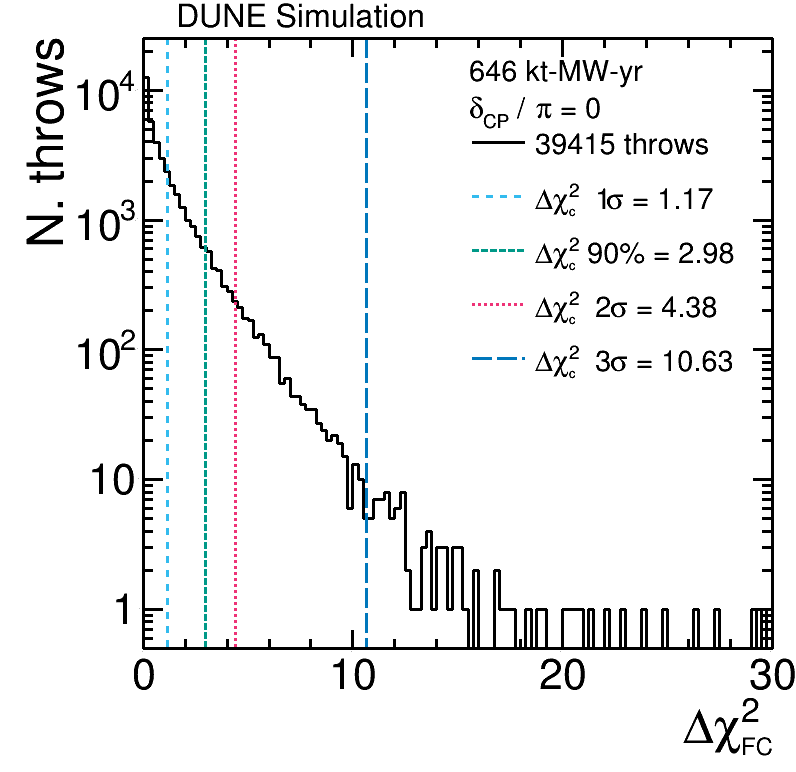}}
  \subfloat[936 kt-MW-yr]{\includegraphics[width=0.33\linewidth]{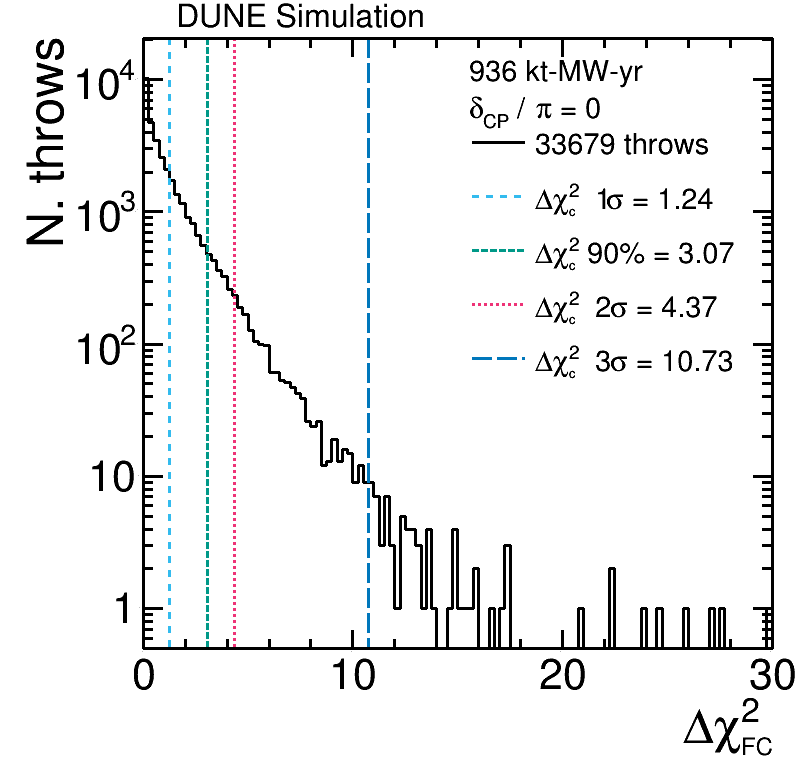}}
  \caption{Distribution of \dchisq values, calculated using Equation~\ref{eq:dchisq_fc}, for a large number of throws with true $\deltacp = 0$, for a variety of exposures. The \dchisqcrit values (vertical lines) obtained using the Feldman-Cousins method show the \dchisqFC value below which 68.27\% (1$\sigma$), 90\%, 95.45\% (2$\sigma$) and 99.73\% (3$\sigma$) of throws reside, with the calculated values given in the legend. The number of throws used is also given.}
  \label{fig:fc_throws_exp}
\end{figure*}
\begin{figure*}[h]
  \centering
  \subfloat[$\deltacp/\pi = -1$]    {\includegraphics[width=0.33\linewidth]{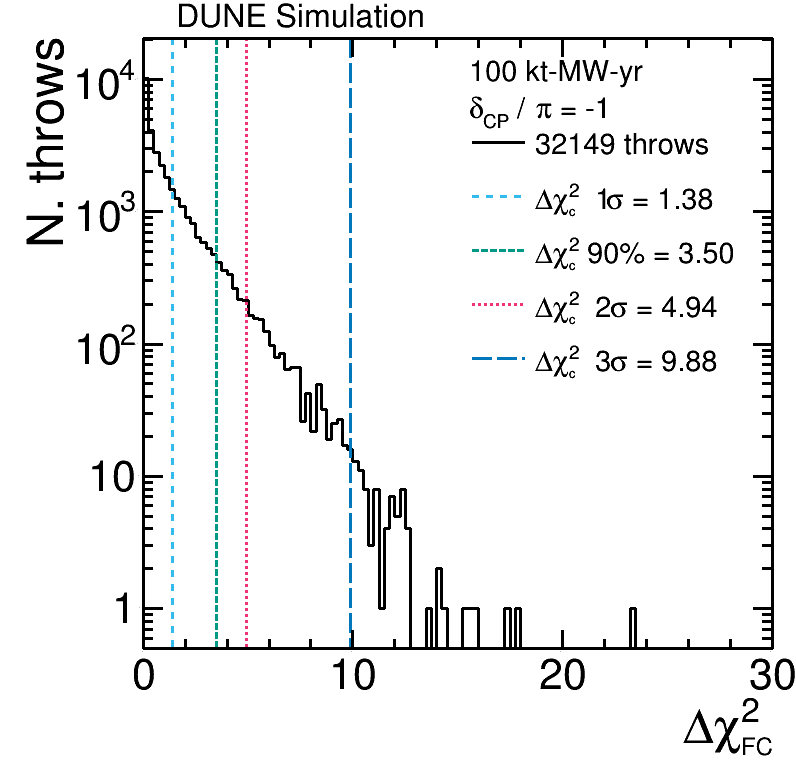}}
  \subfloat[$\deltacp/\pi = -0.75$] {\includegraphics[width=0.33\linewidth]{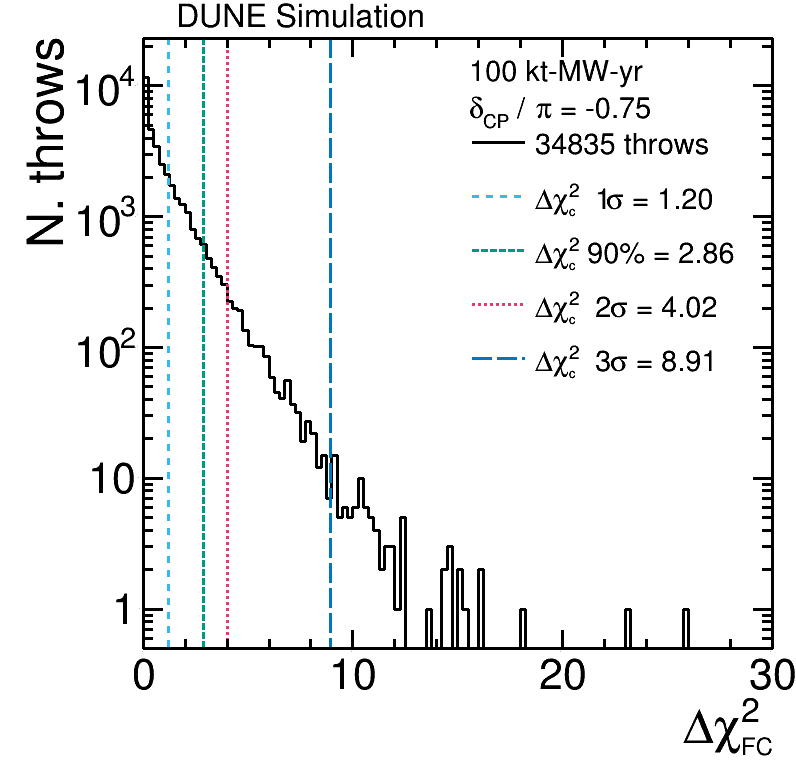}}
  \subfloat[$\deltacp/\pi = -0.5$]  {\includegraphics[width=0.33\linewidth]{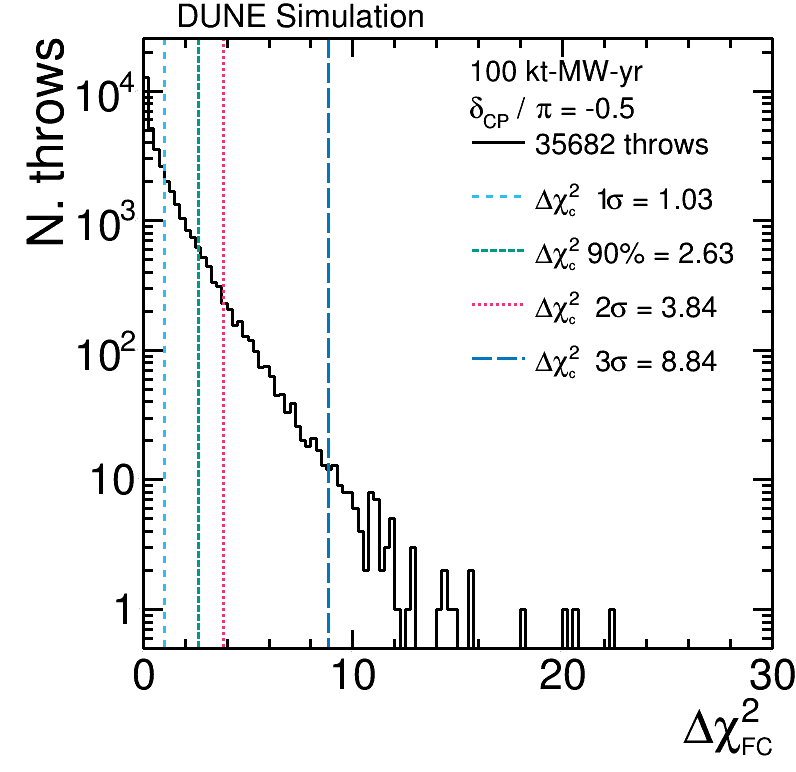}}\\
  \subfloat[$\deltacp/\pi = -0.25$] {\includegraphics[width=0.33\linewidth]{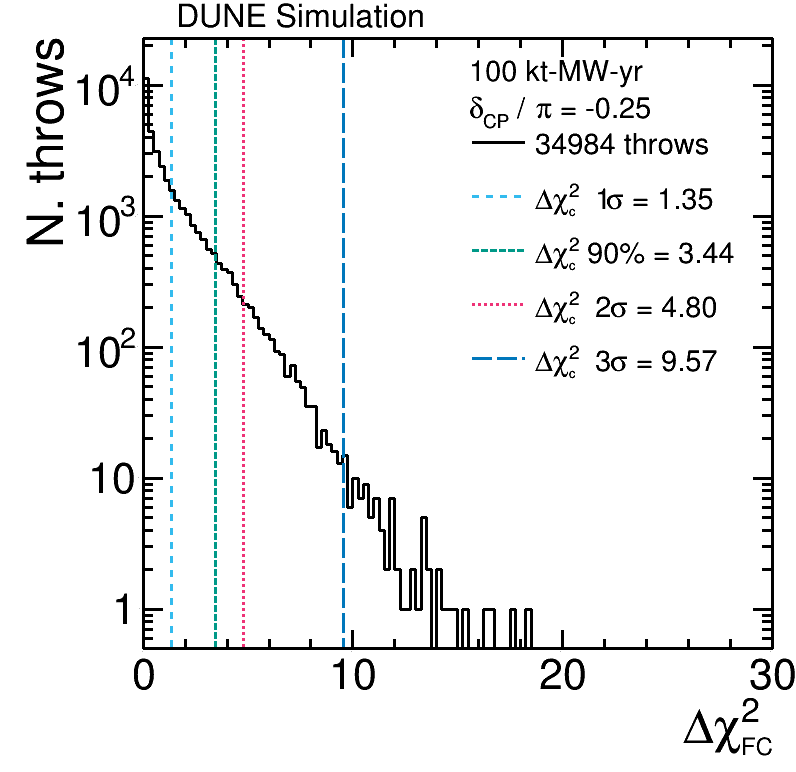}}
  \subfloat[$\deltacp/\pi = 0$]     {\includegraphics[width=0.33\linewidth]{{nh_FC_ndfd_100ktMWyr_dcp0}.png}}
  \subfloat[$\deltacp/\pi = 0.25$]  {\includegraphics[width=0.33\linewidth]{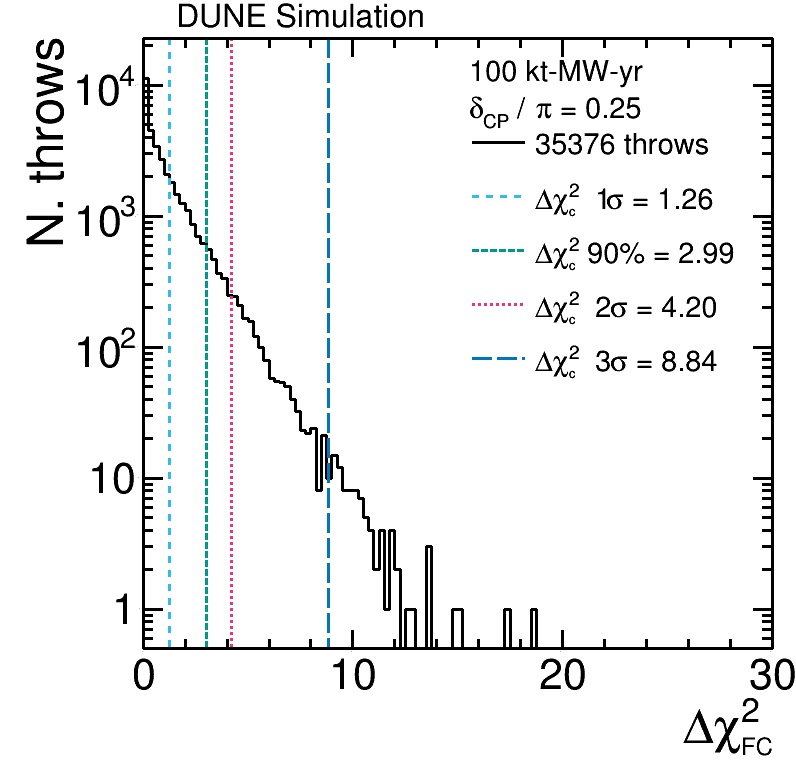}}\\
  \subfloat[$\deltacp/\pi = 0.5$]   {\includegraphics[width=0.33\linewidth]{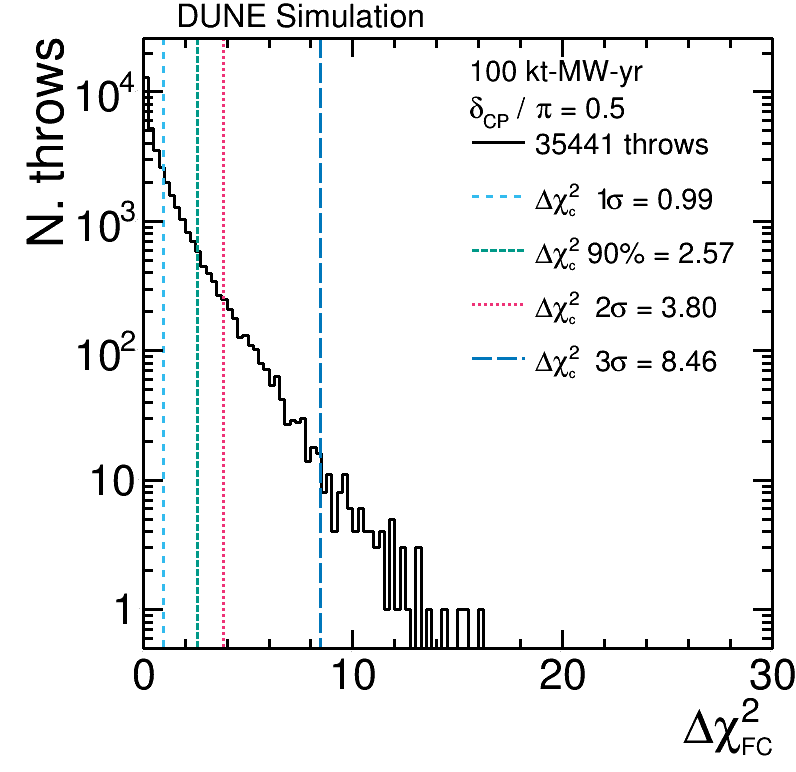}}
  \subfloat[$\deltacp/\pi = 0.75$]  {\includegraphics[width=0.33\linewidth]{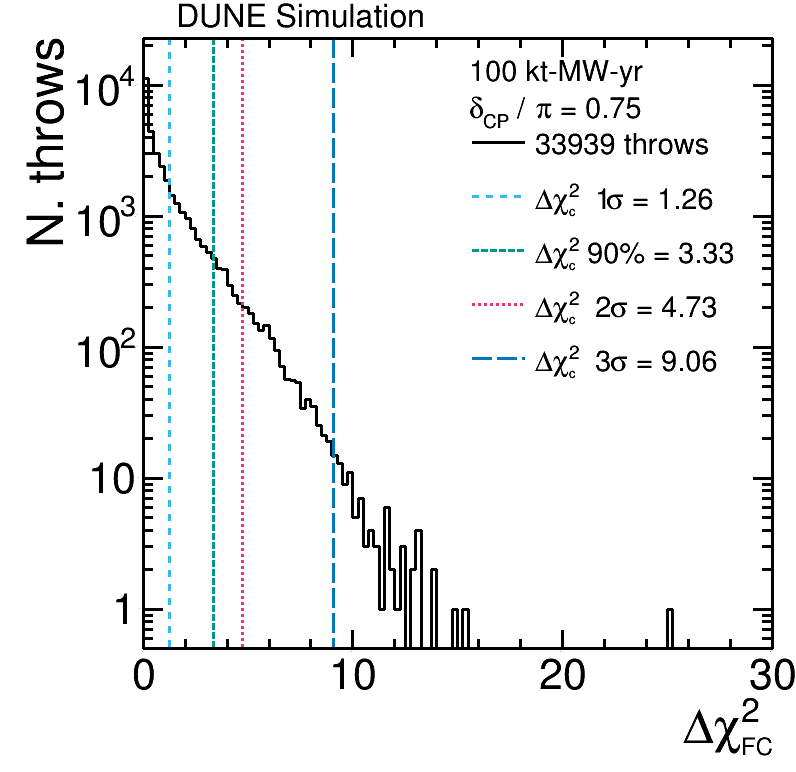}}
  \subfloat[$\deltacp/\pi = 1$]     {\includegraphics[width=0.33\linewidth]{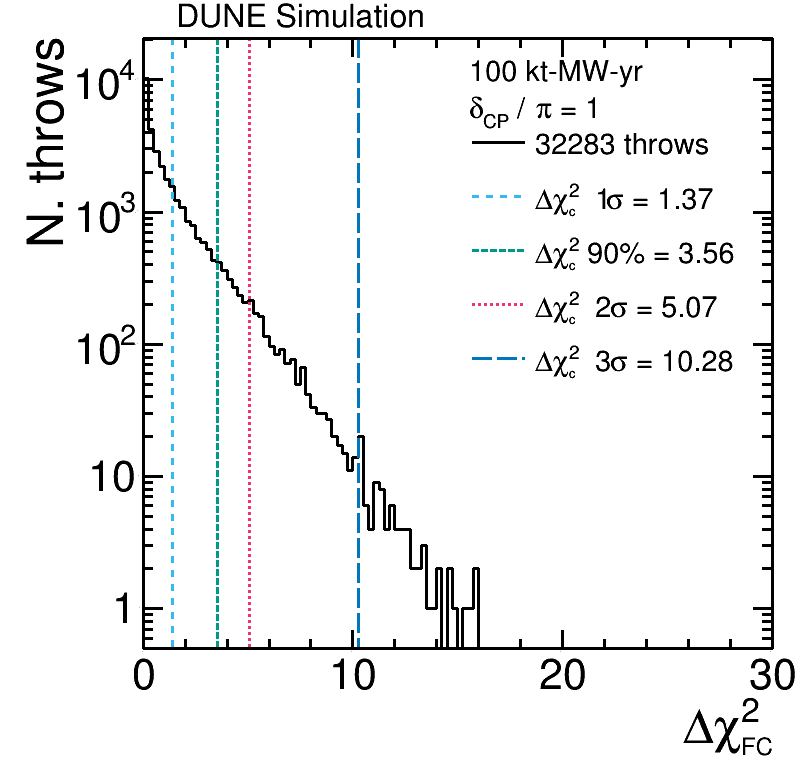}}
  \caption{Distribution of \dchisq values, calculated using Equation~\ref{eq:dchisq_fc}, for a large number of throws for 9 different values of true \deltacp, for a 100 kt-MW-yr exposure. The \dchisqcrit values (vertical lines) obtained using the Feldman-Cousins method show the \dchisqFC value below which 68.27\% (1$\sigma$), 90\%, 95.45\% (2$\sigma$) and 99.73\% (3$\sigma$) of throws reside, with the calculated values given in the legend. The number of throws used is also given.}
  \label{fig:fc_throws_100kt-MW-yr}
\end{figure*}
\begin{figure*}[h]
  \centering
  \subfloat[$\deltacp/\pi = -1$]    {\includegraphics[width=0.33\linewidth]{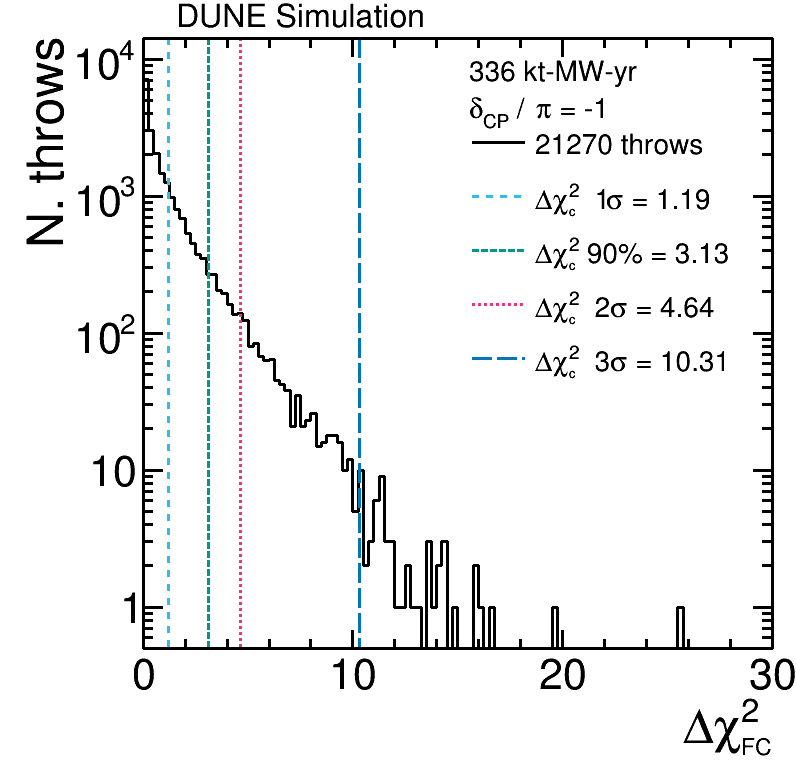}}
  \subfloat[$\deltacp/\pi = -0.75$] {\includegraphics[width=0.33\linewidth]{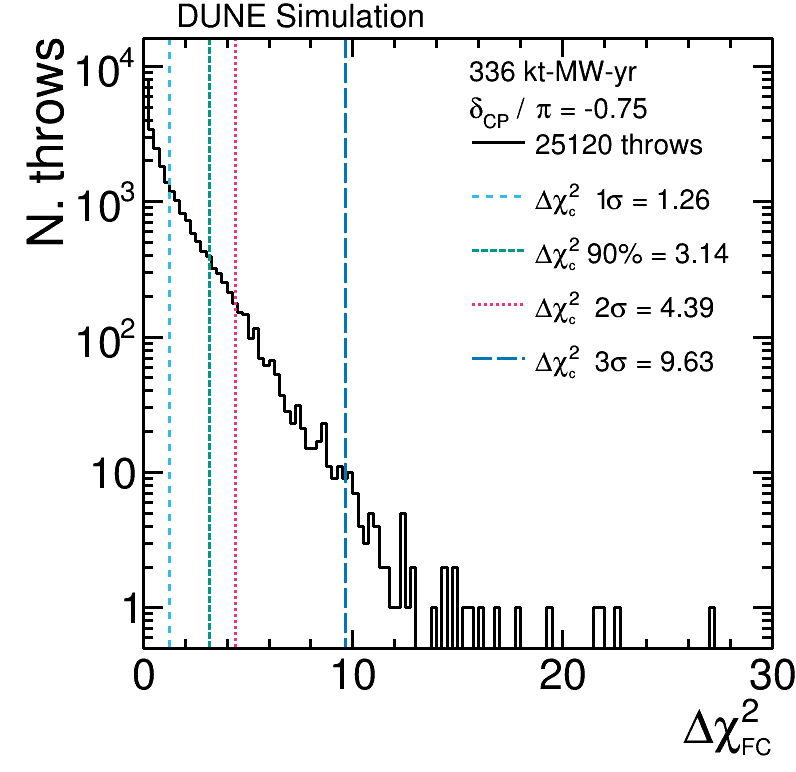}}
  \subfloat[$\deltacp/\pi = -0.5$]  {\includegraphics[width=0.33\linewidth]{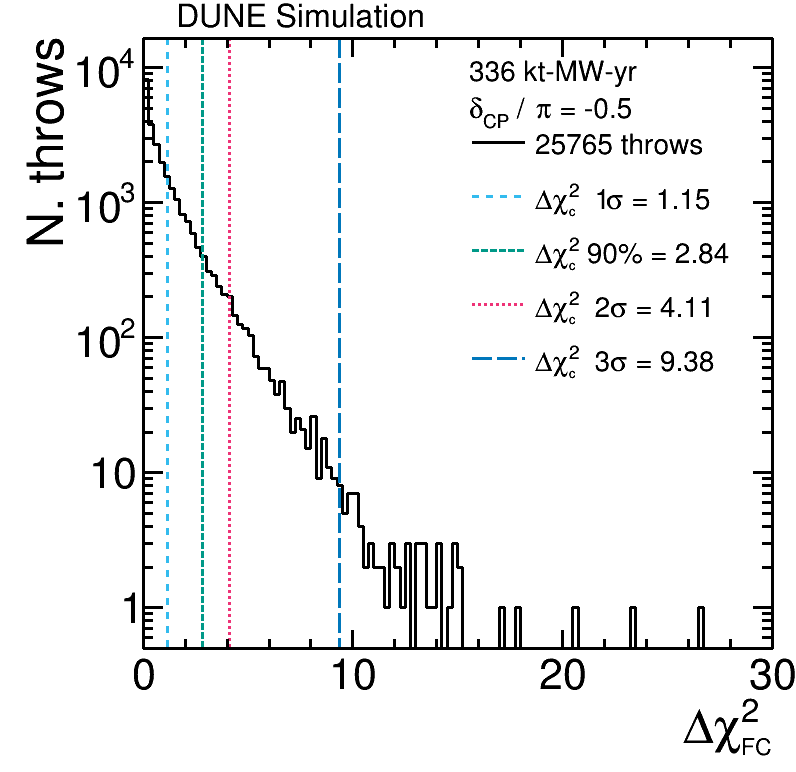}}\\
  \subfloat[$\deltacp/\pi = -0.25$] {\includegraphics[width=0.33\linewidth]{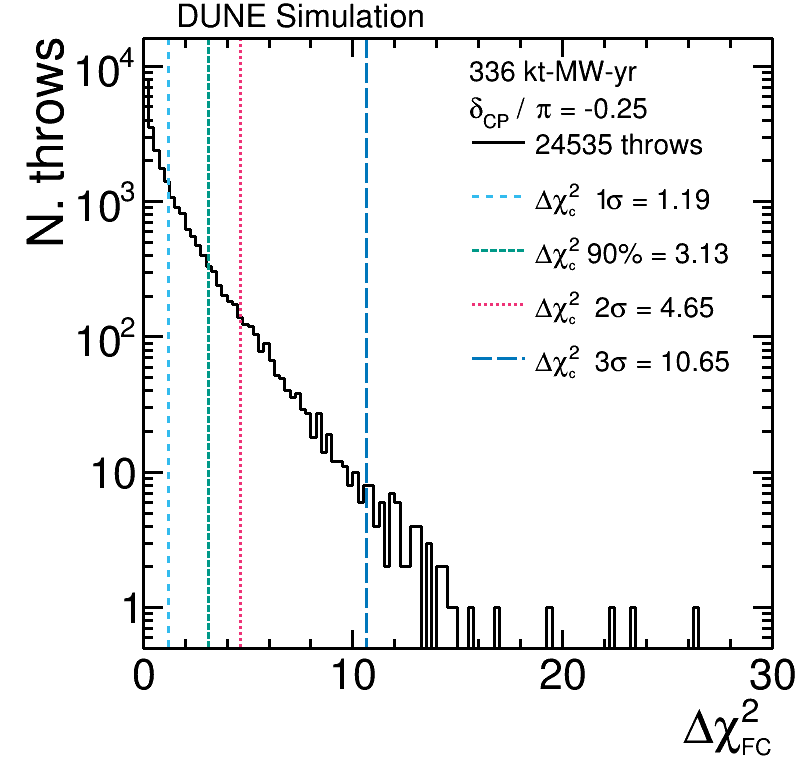}}
  \subfloat[$\deltacp/\pi = 0$]     {\includegraphics[width=0.33\linewidth]{{nh_FC_ndfd_336ktMWyr_dcp0}.png}}
  \subfloat[$\deltacp/\pi = 0.25$]  {\includegraphics[width=0.33\linewidth]{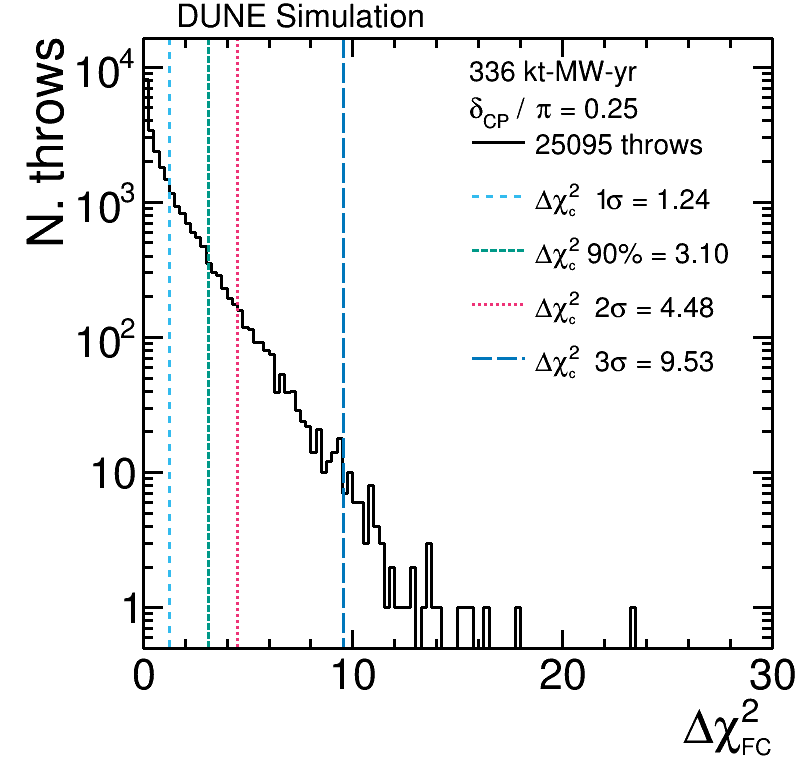}}\\
  \subfloat[$\deltacp/\pi = 0.5$]   {\includegraphics[width=0.33\linewidth]{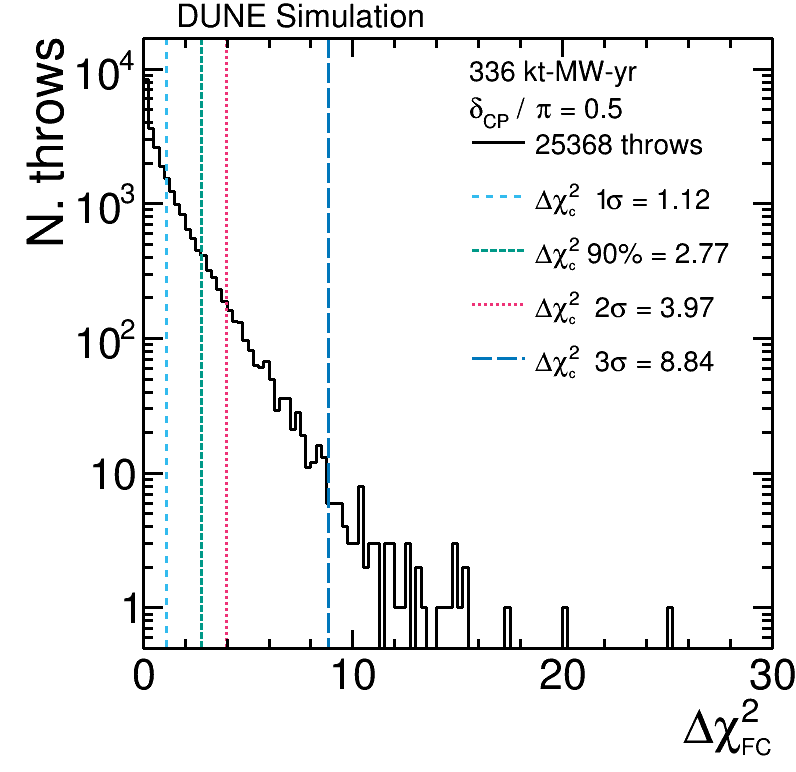}}
  \subfloat[$\deltacp/\pi = 0.75$]  {\includegraphics[width=0.33\linewidth]{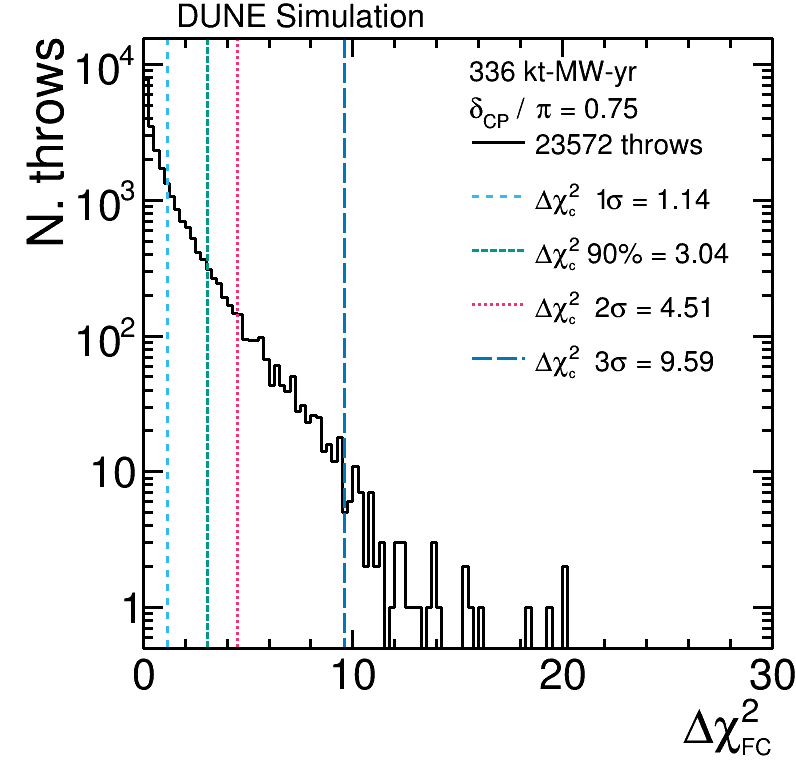}}
  \subfloat[$\deltacp/\pi = 1$]     {\includegraphics[width=0.33\linewidth]{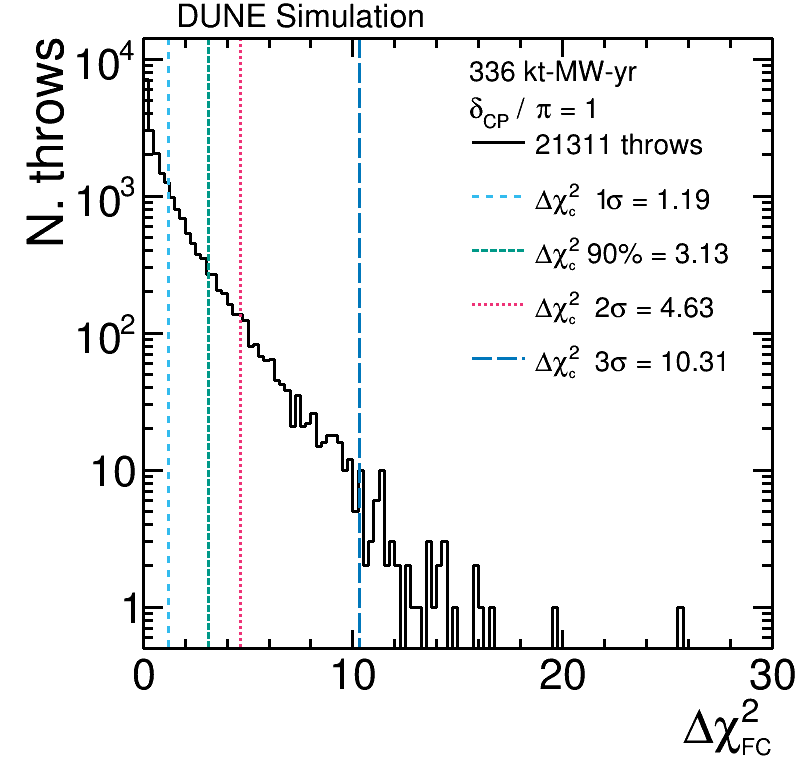}}
  \caption{Distribution of \dchisq values, calculated using Equation~\ref{eq:dchisq_fc}, for a large number of throws for 9 different values of true \deltacp, for a 336 kt-MW-yr exposure. The \dchisqcrit values (vertical lines) obtained using the Feldman-Cousins method show the \dchisqFC value below which 68.27\% (1$\sigma$), 90\%, 95.45\% (2$\sigma$) and 99.73\% (3$\sigma$) of throws reside, with the calculated values given in the legend. The number of throws used is also given.}
  \label{fig:fc_throws_336kt-MW-yr}
\end{figure*}
The distribution of throws used to calculate the \dchisqcrit values for Figure~\ref{fig:fc_vs_exp} for nine different exposures with $\deltacp = 0$ are shown in Figure~\ref{fig:fc_throws_exp}; for Figure~\ref{fig:fc_vs_dcp_100} for nine different values of \deltacp with an exposure of 100 kt-MW-yr in Figure~\ref{fig:fc_throws_100kt-MW-yr}; and for Figure~\ref{fig:fc_vs_dcp_336} for nine different values of \deltacp with an exposure of 336 kt-MW-yr in Figure~\ref{fig:fc_throws_336kt-MW-yr}. For each distribution shown in Figures~\ref{fig:fc_throws_exp},~\ref{fig:fc_throws_100kt-MW-yr} and~\ref{fig:fc_throws_336kt-MW-yr}, the calculated \dchisqcrit values corresponding to for 68.27\% (1$\sigma$), 90\%, 95.45\% (2$\sigma$) and 99.73\% (3$\sigma$) of the throws are given and indicated with a vertical line. The number of throws used is also given. The \dchisqcrit values were only calculated up to the 3$\sigma$ level due to the very large number of throws required for higher confidence levels.

%% file: main.bbl
\begin{thebibliography}{73}%
\makeatletter
\providecommand \@ifxundefined [1]{%
 \@ifx{#1\undefined}
}%
\providecommand \@ifnum [1]{%
 \ifnum #1\expandafter \@firstoftwo
 \else \expandafter \@secondoftwo
 \fi
}%
\providecommand \@ifx [1]{%
 \ifx #1\expandafter \@firstoftwo
 \else \expandafter \@secondoftwo
 \fi
}%
\providecommand \natexlab [1]{#1}%
\providecommand \enquote  [1]{``#1''}%
\providecommand \bibnamefont  [1]{#1}%
\providecommand \bibfnamefont [1]{#1}%
\providecommand \citenamefont [1]{#1}%
\providecommand \href@noop [0]{\@secondoftwo}%
\providecommand \href [0]{\begingroup \@sanitize@url \@href}%
\providecommand \@href[1]{\@@startlink{#1}\@@href}%
\providecommand \@@href[1]{\endgroup#1\@@endlink}%
\providecommand \@sanitize@url [0]{\catcode `\\12\catcode `\$12\catcode
  `\&12\catcode `\#12\catcode `\^12\catcode `\_12\catcode `\%12\relax}%
\providecommand \@@startlink[1]{}%
\providecommand \@@endlink[0]{}%
\providecommand \url  [0]{\begingroup\@sanitize@url \@url }%
\providecommand \@url [1]{\endgroup\@href {#1}{\urlprefix }}%
\providecommand \urlprefix  [0]{URL }%
\providecommand \Eprint [0]{\href }%
\providecommand \doibase [0]{http://dx.doi.org/}%
\providecommand \selectlanguage [0]{\@gobble}%
\providecommand \bibinfo  [0]{\@secondoftwo}%
\providecommand \bibfield  [0]{\@secondoftwo}%
\providecommand \translation [1]{[#1]}%
\providecommand \BibitemOpen [0]{}%
\providecommand \bibitemStop [0]{}%
\providecommand \bibitemNoStop [0]{.\EOS\space}%
\providecommand \EOS [0]{\spacefactor3000\relax}%
\providecommand \BibitemShut  [1]{\csname bibitem#1\endcsname}%
\let\auto@bib@innerbib\@empty
\bibitem [{\citenamefont {Abi}\ \emph {et~al.}(2020{\natexlab{a}})\citenamefont
  {Abi} \emph {et~al.}}]{Abi:2020wmh}%
  \BibitemOpen
  \bibfield  {author} {\bibinfo {author} {\bibfnamefont {B.}~\bibnamefont
  {Abi}} \emph {et~al.} (\bibinfo {collaboration} {DUNE}),\ }\href {\doibase
  10.1088/1748-0221/15/08/T08008} {\bibfield  {journal} {\bibinfo  {journal}
  {JINST}\ }\textbf {\bibinfo {volume} {15}},\ \bibinfo {pages} {T08008}
  (\bibinfo {year} {2020}{\natexlab{a}})},\ \Eprint
  {http://arxiv.org/abs/2002.02967} {arXiv:2002.02967 [physics.ins-det]}
  \BibitemShut {NoStop}%
\bibitem [{\citenamefont {Qian}\ and\ \citenamefont
  {Vogel}(2015)}]{Qian:2015waa}%
  \BibitemOpen
  \bibfield  {author} {\bibinfo {author} {\bibfnamefont {X.}~\bibnamefont
  {Qian}}\ and\ \bibinfo {author} {\bibfnamefont {P.}~\bibnamefont {Vogel}},\
  }\href {\doibase 10.1016/j.ppnp.2015.05.002} {\bibfield  {journal} {\bibinfo
  {journal} {Prog. Part. Nucl. Phys.}\ }\textbf {\bibinfo {volume} {83}},\
  \bibinfo {pages} {1} (\bibinfo {year} {2015})},\ \Eprint
  {http://arxiv.org/abs/1505.01891} {arXiv:1505.01891 [hep-ex]} \BibitemShut
  {NoStop}%
\bibitem [{\citenamefont {Fukugita}\ and\ \citenamefont
  {Yanagida}(1986)}]{Fukugita:1986hr}%
  \BibitemOpen
  \bibfield  {author} {\bibinfo {author} {\bibfnamefont {M.}~\bibnamefont
  {Fukugita}}\ and\ \bibinfo {author} {\bibfnamefont {T.}~\bibnamefont
  {Yanagida}},\ }\href {\doibase 10.1016/0370-2693(86)91126-3} {\bibfield
  {journal} {\bibinfo  {journal} {Phys. Lett.}\ }\textbf {\bibinfo {volume}
  {B174}},\ \bibinfo {pages} {45} (\bibinfo {year} {1986})}\BibitemShut
  {NoStop}%
\bibitem [{\citenamefont {Davidson}\ \emph {et~al.}(2008)\citenamefont
  {Davidson}, \citenamefont {Nardi},\ and\ \citenamefont
  {Nir}}]{Davidson:2008bu}%
  \BibitemOpen
  \bibfield  {author} {\bibinfo {author} {\bibfnamefont {S.}~\bibnamefont
  {Davidson}}, \bibinfo {author} {\bibfnamefont {E.}~\bibnamefont {Nardi}}, \
  and\ \bibinfo {author} {\bibfnamefont {Y.}~\bibnamefont {Nir}},\ }\href
  {\doibase 10.1016/j.physrep.2008.06.002} {\bibfield  {journal} {\bibinfo
  {journal} {Phys. Rept.}\ }\textbf {\bibinfo {volume} {466}},\ \bibinfo
  {pages} {105} (\bibinfo {year} {2008})},\ \Eprint
  {http://arxiv.org/abs/0802.2962} {arXiv:0802.2962 [hep-ph]} \BibitemShut
  {NoStop}%
\bibitem [{\citenamefont {Abi}\ \emph {et~al.}(2021{\natexlab{a}})\citenamefont
  {Abi} \emph {et~al.}}]{Abi:2020kei}%
  \BibitemOpen
  \bibfield  {author} {\bibinfo {author} {\bibfnamefont {B.}~\bibnamefont
  {Abi}} \emph {et~al.} (\bibinfo {collaboration} {DUNE}),\ }\href {\doibase
  10.1140/epjc/s10052-021-09007-w} {\bibfield  {journal} {\bibinfo  {journal}
  {Eur. Phys. J. C}\ }\textbf {\bibinfo {volume} {81}},\ \bibinfo {pages} {322}
  (\bibinfo {year} {2021}{\natexlab{a}})},\ \Eprint
  {http://arxiv.org/abs/2008.12769} {arXiv:2008.12769 [hep-ex]} \BibitemShut
  {NoStop}%
\bibitem [{\citenamefont {Abi}\ \emph {et~al.}(2021{\natexlab{b}})\citenamefont
  {Abi} \emph {et~al.}}]{Abi:2020lpk}%
  \BibitemOpen
  \bibfield  {author} {\bibinfo {author} {\bibfnamefont {B.}~\bibnamefont
  {Abi}} \emph {et~al.} (\bibinfo {collaboration} {DUNE}),\ }\href {\doibase
  10.1140/epjc/s10052-021-09166-w} {\bibfield  {journal} {\bibinfo  {journal}
  {Eur. Phys. J. C}\ }\textbf {\bibinfo {volume} {81}},\ \bibinfo {pages} {423}
  (\bibinfo {year} {2021}{\natexlab{b}})},\ \Eprint
  {http://arxiv.org/abs/2008.06647} {arXiv:2008.06647 [hep-ex]} \BibitemShut
  {NoStop}%
\bibitem [{\citenamefont {Capozzi}\ \emph {et~al.}(2019)\citenamefont
  {Capozzi}, \citenamefont {Li}, \citenamefont {Zhu},\ and\ \citenamefont
  {Beacom}}]{Capozzi:2018dat}%
  \BibitemOpen
  \bibfield  {author} {\bibinfo {author} {\bibfnamefont {F.}~\bibnamefont
  {Capozzi}}, \bibinfo {author} {\bibfnamefont {S.~W.}\ \bibnamefont {Li}},
  \bibinfo {author} {\bibfnamefont {G.}~\bibnamefont {Zhu}}, \ and\ \bibinfo
  {author} {\bibfnamefont {J.~F.}\ \bibnamefont {Beacom}},\ }\href {\doibase
  10.1103/PhysRevLett.123.131803} {\bibfield  {journal} {\bibinfo  {journal}
  {Phys. Rev. Lett.}\ }\textbf {\bibinfo {volume} {123}},\ \bibinfo {pages}
  {131803} (\bibinfo {year} {2019})},\ \Eprint
  {http://arxiv.org/abs/1808.08232} {arXiv:1808.08232 [hep-ph]} \BibitemShut
  {NoStop}%
\bibitem [{\citenamefont {Abi}\ \emph {et~al.}(2020{\natexlab{b}})\citenamefont
  {Abi} \emph {et~al.}}]{Abi:2020evt}%
  \BibitemOpen
  \bibfield  {author} {\bibinfo {author} {\bibfnamefont {B.}~\bibnamefont
  {Abi}} \emph {et~al.} (\bibinfo {collaboration} {DUNE}),\ }\href@noop {} {\
  (\bibinfo {year} {2020}{\natexlab{b}})},\ \Eprint
  {http://arxiv.org/abs/2002.03005} {arXiv:2002.03005 [hep-ex]} \BibitemShut
  {NoStop}%
\bibitem [{\citenamefont {Abed~Abud}\ \emph {et~al.}(2021)\citenamefont
  {Abed~Abud} \emph {et~al.}}]{AbedAbud:2021hpb}%
  \BibitemOpen
  \bibfield  {author} {\bibinfo {author} {\bibfnamefont {A.}~\bibnamefont
  {Abed~Abud}} \emph {et~al.},\ }\href@noop {} {\  (\bibinfo {year} {2021})},\
  \Eprint {http://arxiv.org/abs/2103.13910} {arXiv:2103.13910
  [physics.ins-det]} \BibitemShut {NoStop}%
\bibitem [{\citenamefont {Capozzi}\ \emph {et~al.}(2017)\citenamefont {Capozzi}
  \emph {et~al.}}]{Capozzi:2017ipn}%
  \BibitemOpen
  \bibfield  {author} {\bibinfo {author} {\bibfnamefont {F.}~\bibnamefont
  {Capozzi}} \emph {et~al.},\ }\href {\doibase 10.1103/PhysRevD.95.096014}
  {\bibfield  {journal} {\bibinfo  {journal} {Phys. Rev. D}\ }\textbf {\bibinfo
  {volume} {95}},\ \bibinfo {pages} {096014} (\bibinfo {year} {2017})},\
  \bibinfo {note} {[Addendum: Phys.Rev.D 101, 116013 (2020)]},\ \Eprint
  {http://arxiv.org/abs/2003.08511} {arXiv:2003.08511 [hep-ph]} \BibitemShut
  {NoStop}%
\bibitem [{\citenamefont {de~Salas}\ \emph {et~al.}(2021)\citenamefont
  {de~Salas}, \citenamefont {Forero}, \citenamefont {Gariazzo}, \citenamefont
  {Mart\'\i{}nez-Mirav\'e}, \citenamefont {Mena}, \citenamefont {Ternes},
  \citenamefont {T\'ortola},\ and\ \citenamefont {Valle}}]{deSalas:2020pgw}%
  \BibitemOpen
  \bibfield  {author} {\bibinfo {author} {\bibfnamefont {P.~F.}\ \bibnamefont
  {de~Salas}}, \bibinfo {author} {\bibfnamefont {D.~V.}\ \bibnamefont
  {Forero}}, \bibinfo {author} {\bibfnamefont {S.}~\bibnamefont {Gariazzo}},
  \bibinfo {author} {\bibfnamefont {P.}~\bibnamefont {Mart\'\i{}nez-Mirav\'e}},
  \bibinfo {author} {\bibfnamefont {O.}~\bibnamefont {Mena}}, \bibinfo {author}
  {\bibfnamefont {C.~A.}\ \bibnamefont {Ternes}}, \bibinfo {author}
  {\bibfnamefont {M.}~\bibnamefont {T\'ortola}}, \ and\ \bibinfo {author}
  {\bibfnamefont {J.~W.~F.}\ \bibnamefont {Valle}},\ }\href {\doibase
  10.1007/JHEP02(2021)071} {\bibfield  {journal} {\bibinfo  {journal} {JHEP}\
  }\textbf {\bibinfo {volume} {02}},\ \bibinfo {pages} {071} (\bibinfo {year}
  {2021})},\ \Eprint {http://arxiv.org/abs/2006.11237} {arXiv:2006.11237
  [hep-ph]} \BibitemShut {NoStop}%
\bibitem [{\citenamefont {Esteban}\ \emph {et~al.}(2020)\citenamefont
  {Esteban}, \citenamefont {Gonzalez-Garcia}, \citenamefont {Maltoni},
  \citenamefont {Schwetz},\ and\ \citenamefont {Zhou}}]{Esteban:2020cvm}%
  \BibitemOpen
  \bibfield  {author} {\bibinfo {author} {\bibfnamefont {I.}~\bibnamefont
  {Esteban}}, \bibinfo {author} {\bibfnamefont {M.~C.}\ \bibnamefont
  {Gonzalez-Garcia}}, \bibinfo {author} {\bibfnamefont {M.}~\bibnamefont
  {Maltoni}}, \bibinfo {author} {\bibfnamefont {T.}~\bibnamefont {Schwetz}}, \
  and\ \bibinfo {author} {\bibfnamefont {A.}~\bibnamefont {Zhou}},\ }\href
  {\doibase 10.1007/JHEP09(2020)178} {\bibfield  {journal} {\bibinfo  {journal}
  {JHEP}\ }\textbf {\bibinfo {volume} {09}},\ \bibinfo {pages} {178} (\bibinfo
  {year} {2020})},\ \Eprint {http://arxiv.org/abs/2007.14792} {arXiv:2007.14792
  [hep-ph]} \BibitemShut {NoStop}%
\bibitem [{\citenamefont {Abe}\ \emph {et~al.}(2021)\citenamefont {Abe} \emph
  {et~al.}}]{Abe:2021gky}%
  \BibitemOpen
  \bibfield  {author} {\bibinfo {author} {\bibfnamefont {K.}~\bibnamefont
  {Abe}} \emph {et~al.} (\bibinfo {collaboration} {T2K}),\ }\href {\doibase
  10.1103/PhysRevD.103.112008} {\bibfield  {journal} {\bibinfo  {journal}
  {Phys. Rev. D}\ }\textbf {\bibinfo {volume} {103}},\ \bibinfo {pages}
  {112008} (\bibinfo {year} {2021})},\ \Eprint
  {http://arxiv.org/abs/2101.03779} {arXiv:2101.03779 [hep-ex]} \BibitemShut
  {NoStop}%
\bibitem [{\citenamefont {Abe}\ \emph {et~al.}(2018{\natexlab{a}})\citenamefont
  {Abe} \emph {et~al.}}]{PhysRevD.97.072001}%
  \BibitemOpen
  \bibfield  {author} {\bibinfo {author} {\bibfnamefont {K.}~\bibnamefont
  {Abe}} \emph {et~al.} (\bibinfo {collaboration} {Super-Kamiokande}),\ }\href
  {\doibase 10.1103/PhysRevD.97.072001} {\bibfield  {journal} {\bibinfo
  {journal} {Phys. Rev.}\ }\textbf {\bibinfo {volume} {D97}},\ \bibinfo {pages}
  {072001} (\bibinfo {year} {2018}{\natexlab{a}})},\ \Eprint
  {http://arxiv.org/abs/1710.09126} {arXiv:1710.09126 [hep-ex]} \BibitemShut
  {NoStop}%
\bibitem [{\citenamefont {Acero}\ \emph {et~al.}(2019)\citenamefont {Acero}
  \emph {et~al.}}]{PhysRevLett.123.151803}%
  \BibitemOpen
  \bibfield  {author} {\bibinfo {author} {\bibfnamefont {M.~A.}\ \bibnamefont
  {Acero}} \emph {et~al.} (\bibinfo {collaboration} {NOvA}),\ }\href {\doibase
  10.1103/PhysRevLett.123.151803} {\bibfield  {journal} {\bibinfo  {journal}
  {Phys. Rev. Lett.}\ }\textbf {\bibinfo {volume} {123}},\ \bibinfo {pages}
  {151803} (\bibinfo {year} {2019})},\ \Eprint
  {http://arxiv.org/abs/1906.04907} {arXiv:1906.04907 [hep-ex]} \BibitemShut
  {NoStop}%
\bibitem [{\citenamefont {Abe}\ \emph {et~al.}(2020)\citenamefont {Abe} \emph
  {et~al.}}]{Abe:2019vii}%
  \BibitemOpen
  \bibfield  {author} {\bibinfo {author} {\bibfnamefont {K.}~\bibnamefont
  {Abe}} \emph {et~al.} (\bibinfo {collaboration} {T2K}),\ }\href {\doibase
  10.1038/s41586-020-2177-0} {\bibfield  {journal} {\bibinfo  {journal}
  {Nature}\ }\textbf {\bibinfo {volume} {580}},\ \bibinfo {pages} {339}
  (\bibinfo {year} {2020})},\ \Eprint {http://arxiv.org/abs/1910.03887}
  {arXiv:1910.03887 [hep-ex]} \BibitemShut {NoStop}%
\bibitem [{\citenamefont {Nunokawa}\ \emph {et~al.}(2008)\citenamefont
  {Nunokawa}, \citenamefont {Parke},\ and\ \citenamefont
  {Valle}}]{Nunokawa:2007qh}%
  \BibitemOpen
  \bibfield  {author} {\bibinfo {author} {\bibfnamefont {H.}~\bibnamefont
  {Nunokawa}}, \bibinfo {author} {\bibfnamefont {S.~J.}\ \bibnamefont {Parke}},
  \ and\ \bibinfo {author} {\bibfnamefont {J.~W.}\ \bibnamefont {Valle}},\
  }\href {\doibase 10.1016/j.ppnp.2007.10.001} {\bibfield  {journal} {\bibinfo
  {journal} {Prog. Part. Nucl. Phys.}\ }\textbf {\bibinfo {volume} {60}},\
  \bibinfo {pages} {338} (\bibinfo {year} {2008})},\ \Eprint
  {http://arxiv.org/abs/0710.0554} {arXiv:0710.0554 [hep-ph]} \BibitemShut
  {NoStop}%
\bibitem [{\citenamefont {Wolfenstein}(1978)}]{Wolfenstein:1977ue}%
  \BibitemOpen
  \bibfield  {author} {\bibinfo {author} {\bibfnamefont {L.}~\bibnamefont
  {Wolfenstein}},\ }\href {\doibase 10.1103/PhysRevD.17.2369} {\bibfield
  {journal} {\bibinfo  {journal} {Phys. Rev.}\ }\textbf {\bibinfo {volume}
  {D17}},\ \bibinfo {pages} {2369} (\bibinfo {year} {1978})}\BibitemShut
  {NoStop}%
\bibitem [{\citenamefont {Mikheev}\ and\ \citenamefont
  {Smirnov}(1985)}]{Mikheev:1986gs}%
  \BibitemOpen
  \bibfield  {author} {\bibinfo {author} {\bibfnamefont {S.}~\bibnamefont
  {Mikheev}}\ and\ \bibinfo {author} {\bibfnamefont {A.~Y.}\ \bibnamefont
  {Smirnov}},\ }\href@noop {} {\bibfield  {journal} {\bibinfo  {journal} {Sov.
  J. Nucl. Phys.}\ }\textbf {\bibinfo {volume} {42}},\ \bibinfo {pages} {913}
  (\bibinfo {year} {1985})}\BibitemShut {NoStop}%
\bibitem [{\citenamefont {Kopp}(2008)}]{Kopp:2006wp}%
  \BibitemOpen
  \bibfield  {author} {\bibinfo {author} {\bibfnamefont {J.}~\bibnamefont
  {Kopp}},\ }\href {\doibase 10.1142/S0129183108012303} {\bibfield  {journal}
  {\bibinfo  {journal} {Int. J. Mod. Phys. C}\ }\textbf {\bibinfo {volume}
  {19}},\ \bibinfo {pages} {523} (\bibinfo {year} {2008})},\ \Eprint
  {http://arxiv.org/abs/physics/0610206} {arXiv:physics/0610206} \BibitemShut
  {NoStop}%
\bibitem [{\citenamefont {Abi}\ \emph {et~al.}(2020{\natexlab{c}})\citenamefont
  {Abi} \emph {et~al.}}]{Abi:2020qib}%
  \BibitemOpen
  \bibfield  {author} {\bibinfo {author} {\bibfnamefont {B.}~\bibnamefont
  {Abi}} \emph {et~al.} (\bibinfo {collaboration} {DUNE}),\ }\href@noop {}
  {\bibfield  {journal} {\bibinfo  {journal} {Eur. Phys. J.}\ }\textbf
  {\bibinfo {volume} {C80}},\ \bibinfo {pages} {978} (\bibinfo {year}
  {2020}{\natexlab{c}})},\ \Eprint {http://arxiv.org/abs/2006.16043}
  {arXiv:2006.16043 [hep-ex]} \BibitemShut {NoStop}%
\bibitem [{\citenamefont {{NOvA Collaboration}}(2019)}]{CAFAna}%
  \BibitemOpen
  \bibinfo {editor} {\bibnamefont {{NOvA Collaboration}}},\ ed.,\ \enquote
  {\bibinfo {title} {{NOvA-ART}},}\ \ (\bibinfo  {publisher} {Redmine},\
  \bibinfo {year} {2019})\ Chap.\ \bibinfo {chapter} {{CAFAna overview}},\
  \bibinfo {note}
  {\url{https://cdcvs.fnal.gov/redmine/projects/novaart/wiki/CAFAna_overview}}\BibitemShut
  {NoStop}%
\bibitem [{\citenamefont {Aliaga}\ \emph {et~al.}(2016)\citenamefont {Aliaga}
  \emph {et~al.}}]{Aliaga:2016oaz}%
  \BibitemOpen
  \bibfield  {author} {\bibinfo {author} {\bibfnamefont {L.}~\bibnamefont
  {Aliaga}} \emph {et~al.} (\bibinfo {collaboration} {MINERvA}),\ }\href
  {\doibase 10.1103/PhysRevD.94.092005, 10.1103/PhysRevD.95.039903} {\bibfield
  {journal} {\bibinfo  {journal} {Phys. Rev.}\ }\textbf {\bibinfo {volume}
  {D94}},\ \bibinfo {pages} {092005} (\bibinfo {year} {2016})},\ \bibinfo
  {note} {[Addendum: Phys. Rev. D95, no.3, 039903 (2017)]},\ \Eprint
  {http://arxiv.org/abs/1607.00704} {arXiv:1607.00704 [hep-ex]} \BibitemShut
  {NoStop}%
\bibitem [{\citenamefont {Andreopoulos}\ \emph {et~al.}(2010)\citenamefont
  {Andreopoulos} \emph {et~al.}}]{Andreopoulos:2009rq}%
  \BibitemOpen
  \bibfield  {author} {\bibinfo {author} {\bibfnamefont {C.}~\bibnamefont
  {Andreopoulos}} \emph {et~al.},\ }\href {\doibase 10.1016/j.nima.2009.12.009}
  {\bibfield  {journal} {\bibinfo  {journal} {Nucl. Instrum. Meth.}\ }\textbf
  {\bibinfo {volume} {A614}},\ \bibinfo {pages} {87} (\bibinfo {year}
  {2010})},\ \Eprint {http://arxiv.org/abs/0905.2517} {arXiv:0905.2517
  [hep-ph]} \BibitemShut {NoStop}%
\bibitem [{\citenamefont {Andreopoulos}\ \emph {et~al.}(2015)\citenamefont
  {Andreopoulos} \emph {et~al.}}]{Andreopoulos:2015wxa}%
  \BibitemOpen
  \bibfield  {author} {\bibinfo {author} {\bibfnamefont {C.}~\bibnamefont
  {Andreopoulos}} \emph {et~al.},\ }\href@noop {} {\  (\bibinfo {year}
  {2015})},\ \Eprint {http://arxiv.org/abs/1510.05494} {arXiv:1510.05494
  [hep-ph]} \BibitemShut {NoStop}%
\bibitem [{\citenamefont {Stowell}\ \emph {et~al.}(2017)\citenamefont {Stowell}
  \emph {et~al.}}]{Stowell:2016jfr}%
  \BibitemOpen
  \bibfield  {author} {\bibinfo {author} {\bibfnamefont {P.}~\bibnamefont
  {Stowell}} \emph {et~al.},\ }\href {\doibase 10.1088/1748-0221/12/01/P01016}
  {\bibfield  {journal} {\bibinfo  {journal} {JINST}\ }\textbf {\bibinfo
  {volume} {12}},\ \bibinfo {pages} {P01016} (\bibinfo {year} {2017})},\
  \Eprint {http://arxiv.org/abs/1612.07393} {arXiv:1612.07393 [hep-ex]}
  \BibitemShut {NoStop}%
\bibitem [{\citenamefont {Bodek}\ and\ \citenamefont
  {Ritchie}(1981)}]{BodekRitchie}%
  \BibitemOpen
  \bibfield  {author} {\bibinfo {author} {\bibfnamefont {A.}~\bibnamefont
  {Bodek}}\ and\ \bibinfo {author} {\bibfnamefont {J.~L.}\ \bibnamefont
  {Ritchie}},\ }\href {\doibase 10.1103/PhysRevD.23.1070} {\bibfield  {journal}
  {\bibinfo  {journal} {Phys. Rev. D}\ }\textbf {\bibinfo {volume} {23}},\
  \bibinfo {pages} {1070} (\bibinfo {year} {1981})}\BibitemShut {NoStop}%
\bibitem [{\citenamefont {Llewellyn~Smith}(1972)}]{llewelyn-smith}%
  \BibitemOpen
  \bibfield  {author} {\bibinfo {author} {\bibfnamefont {C.}~\bibnamefont
  {Llewellyn~Smith}},\ }\href {\doibase 10.1016/0370-1573(72)90010-5}
  {\bibfield  {journal} {\bibinfo  {journal} {Phys. Rept.}\ }\textbf {\bibinfo
  {volume} {3}},\ \bibinfo {pages} {261} (\bibinfo {year} {1972})}\BibitemShut
  {NoStop}%
\bibitem [{\citenamefont {Bradford}\ \emph {et~al.}(2006)\citenamefont
  {Bradford}, \citenamefont {Bodek}, \citenamefont {Budd},\ and\ \citenamefont
  {Arrington}}]{bbba05}%
  \BibitemOpen
  \bibfield  {author} {\bibinfo {author} {\bibfnamefont {R.}~\bibnamefont
  {Bradford}}, \bibinfo {author} {\bibfnamefont {A.}~\bibnamefont {Bodek}},
  \bibinfo {author} {\bibfnamefont {H.}~\bibnamefont {Budd}}, \ and\ \bibinfo
  {author} {\bibfnamefont {J.}~\bibnamefont {Arrington}},\ }\href {\doibase
  10.1016/j.nuclphysbps.2006.08.028} {\bibfield  {journal} {\bibinfo  {journal}
  {Nuclear Physics B - Proceedings Supplements}\ }\textbf {\bibinfo {volume}
  {159}},\ \bibinfo {pages} {127 } (\bibinfo {year} {2006})}\BibitemShut
  {NoStop}%
\bibitem [{\citenamefont {Abe}\ \emph {et~al.}(2018{\natexlab{b}})\citenamefont
  {Abe} \emph {et~al.}}]{Abe:2018wpn}%
  \BibitemOpen
  \bibfield  {author} {\bibinfo {author} {\bibfnamefont {K.}~\bibnamefont
  {Abe}} \emph {et~al.} (\bibinfo {collaboration} {T2K}),\ }\href {\doibase
  10.1103/PhysRevLett.121.171802} {\bibfield  {journal} {\bibinfo  {journal}
  {Phys. Rev. Lett.}\ }\textbf {\bibinfo {volume} {121}},\ \bibinfo {pages}
  {171802} (\bibinfo {year} {2018}{\natexlab{b}})},\ \Eprint
  {http://arxiv.org/abs/1807.07891} {arXiv:1807.07891 [hep-ex]} \BibitemShut
  {NoStop}%
\bibitem [{\citenamefont {Nieves}\ \emph {et~al.}(2011)\citenamefont {Nieves},
  \citenamefont {Ruiz~Simo},\ and\ \citenamefont {Vicente~Vacas}}]{nieves1}%
  \BibitemOpen
  \bibfield  {author} {\bibinfo {author} {\bibfnamefont {J.}~\bibnamefont
  {Nieves}}, \bibinfo {author} {\bibfnamefont {I.}~\bibnamefont {Ruiz~Simo}}, \
  and\ \bibinfo {author} {\bibfnamefont {M.~J.}\ \bibnamefont
  {Vicente~Vacas}},\ }\href {\doibase 10.1103/PhysRevC.83.045501} {\bibfield
  {journal} {\bibinfo  {journal} {Phys. Rev.}\ }\textbf {\bibinfo {volume}
  {C83}},\ \bibinfo {pages} {045501} (\bibinfo {year} {2011})},\ \Eprint
  {http://arxiv.org/abs/1102.2777} {arXiv:1102.2777 [hep-ph]} \BibitemShut
  {NoStop}%
\bibitem [{\citenamefont {Gran}\ \emph {et~al.}(2013)\citenamefont {Gran},
  \citenamefont {Nieves}, \citenamefont {Sanchez},\ and\ \citenamefont
  {Vicente~Vacas}}]{nieves2}%
  \BibitemOpen
  \bibfield  {author} {\bibinfo {author} {\bibfnamefont {R.}~\bibnamefont
  {Gran}}, \bibinfo {author} {\bibfnamefont {J.}~\bibnamefont {Nieves}},
  \bibinfo {author} {\bibfnamefont {F.}~\bibnamefont {Sanchez}}, \ and\
  \bibinfo {author} {\bibfnamefont {M.~J.}\ \bibnamefont {Vicente~Vacas}},\
  }\href {\doibase 10.1103/PhysRevD.88.113007} {\bibfield  {journal} {\bibinfo
  {journal} {Phys. Rev.}\ }\textbf {\bibinfo {volume} {D88}},\ \bibinfo {pages}
  {113007} (\bibinfo {year} {2013})},\ \Eprint {http://arxiv.org/abs/1307.8105}
  {arXiv:1307.8105 [hep-ph]} \BibitemShut {NoStop}%
\bibitem [{\citenamefont {Schwehr}\ \emph {et~al.}(2016)\citenamefont
  {Schwehr}, \citenamefont {Cherdack},\ and\ \citenamefont
  {Gran}}]{Schwehr:2016pvn}%
  \BibitemOpen
  \bibfield  {author} {\bibinfo {author} {\bibfnamefont {J.}~\bibnamefont
  {Schwehr}}, \bibinfo {author} {\bibfnamefont {D.}~\bibnamefont {Cherdack}}, \
  and\ \bibinfo {author} {\bibfnamefont {R.}~\bibnamefont {Gran}},\ }\href@noop
  {} {\  (\bibinfo {year} {2016})},\ \Eprint {http://arxiv.org/abs/1601.02038}
  {arXiv:1601.02038 [hep-ph]} \BibitemShut {NoStop}%
\bibitem [{\citenamefont {Rodrigues}\ \emph
  {et~al.}(2016{\natexlab{a}})\citenamefont {Rodrigues} \emph
  {et~al.}}]{Rodrigues:2015hik}%
  \BibitemOpen
  \bibfield  {author} {\bibinfo {author} {\bibfnamefont {P.~A.}\ \bibnamefont
  {Rodrigues}} \emph {et~al.} (\bibinfo {collaboration} {MINERvA}),\ }\href
  {\doibase 10.1103/PhysRevLett.116.071802} {\bibfield  {journal} {\bibinfo
  {journal} {Phys. Rev. Lett.}\ }\textbf {\bibinfo {volume} {116}},\ \bibinfo
  {pages} {071802} (\bibinfo {year} {2016}{\natexlab{a}})},\ \Eprint
  {http://arxiv.org/abs/1511.05944} {arXiv:1511.05944 [hep-ex]} \BibitemShut
  {NoStop}%
\bibitem [{\citenamefont {Acero}\ \emph {et~al.}(2018)\citenamefont {Acero}
  \emph {et~al.}}]{NOvA:2018gge}%
  \BibitemOpen
  \bibfield  {author} {\bibinfo {author} {\bibfnamefont {M.~A.}\ \bibnamefont
  {Acero}} \emph {et~al.} (\bibinfo {collaboration} {NOvA}),\ }\href {\doibase
  10.1103/PhysRevD.98.032012} {\bibfield  {journal} {\bibinfo  {journal} {Phys.
  Rev.}\ }\textbf {\bibinfo {volume} {D98}},\ \bibinfo {pages} {032012}
  (\bibinfo {year} {2018})},\ \Eprint {http://arxiv.org/abs/1806.00096}
  {arXiv:1806.00096 [hep-ex]} \BibitemShut {NoStop}%
\bibitem [{\citenamefont {Colle}\ \emph {et~al.}(2015)\citenamefont {Colle}
  \emph {et~al.}}]{Colle:2015ena}%
  \BibitemOpen
  \bibfield  {author} {\bibinfo {author} {\bibfnamefont {C.}~\bibnamefont
  {Colle}} \emph {et~al.},\ }\href {\doibase 10.1103/PhysRevC.92.024604}
  {\bibfield  {journal} {\bibinfo  {journal} {Phys. Rev.}\ }\textbf {\bibinfo
  {volume} {C92}},\ \bibinfo {pages} {024604} (\bibinfo {year} {2015})},\
  \Eprint {http://arxiv.org/abs/1503.06050} {arXiv:1503.06050 [nucl-th]}
  \BibitemShut {NoStop}%
\bibitem [{\citenamefont {Rein}\ and\ \citenamefont
  {Sehgal}(1981)}]{Rein:1980wg}%
  \BibitemOpen
  \bibfield  {author} {\bibinfo {author} {\bibfnamefont {D.}~\bibnamefont
  {Rein}}\ and\ \bibinfo {author} {\bibfnamefont {L.~M.}\ \bibnamefont
  {Sehgal}},\ }\href {\doibase 10.1016/0003-4916(81)90242-6} {\bibfield
  {journal} {\bibinfo  {journal} {Annals Phys.}\ }\textbf {\bibinfo {volume}
  {133}},\ \bibinfo {pages} {79} (\bibinfo {year} {1981})}\BibitemShut
  {NoStop}%
\bibitem [{\citenamefont {Wilkinson}\ \emph {et~al.}(2014)\citenamefont
  {Wilkinson}, \citenamefont {Rodrigues}, \citenamefont {Cartwright},
  \citenamefont {Thompson},\ and\ \citenamefont
  {McFarland}}]{Wilkinson:2014yfa}%
  \BibitemOpen
  \bibfield  {author} {\bibinfo {author} {\bibfnamefont {C.}~\bibnamefont
  {Wilkinson}}, \bibinfo {author} {\bibfnamefont {P.}~\bibnamefont
  {Rodrigues}}, \bibinfo {author} {\bibfnamefont {S.}~\bibnamefont
  {Cartwright}}, \bibinfo {author} {\bibfnamefont {L.}~\bibnamefont
  {Thompson}}, \ and\ \bibinfo {author} {\bibfnamefont {K.}~\bibnamefont
  {McFarland}},\ }\href {\doibase 10.1103/PhysRevD.90.112017} {\bibfield
  {journal} {\bibinfo  {journal} {Phys. Rev.}\ }\textbf {\bibinfo {volume}
  {D90}},\ \bibinfo {pages} {112017} (\bibinfo {year} {2014})},\ \Eprint
  {http://arxiv.org/abs/1411.4482} {arXiv:1411.4482 [hep-ex]} \BibitemShut
  {NoStop}%
\bibitem [{\citenamefont {Rodrigues}\ \emph
  {et~al.}(2016{\natexlab{b}})\citenamefont {Rodrigues}, \citenamefont
  {Wilkinson},\ and\ \citenamefont {McFarland}}]{Rodrigues:2016xjj}%
  \BibitemOpen
  \bibfield  {author} {\bibinfo {author} {\bibfnamefont {P.}~\bibnamefont
  {Rodrigues}}, \bibinfo {author} {\bibfnamefont {C.}~\bibnamefont
  {Wilkinson}}, \ and\ \bibinfo {author} {\bibfnamefont {K.}~\bibnamefont
  {McFarland}},\ }\href {\doibase 10.1140/epjc/s10052-016-4314-3} {\bibfield
  {journal} {\bibinfo  {journal} {Eur. Phys. J.}\ }\textbf {\bibinfo {volume}
  {C76}},\ \bibinfo {pages} {474} (\bibinfo {year} {2016}{\natexlab{b}})},\
  \Eprint {http://arxiv.org/abs/1601.01888} {arXiv:1601.01888 [hep-ex]}
  \BibitemShut {NoStop}%
\bibitem [{\citenamefont {Bodek}\ and\ \citenamefont
  {Yang}(2003)}]{Bodek:2002ps}%
  \BibitemOpen
  \bibfield  {author} {\bibinfo {author} {\bibfnamefont {A.}~\bibnamefont
  {Bodek}}\ and\ \bibinfo {author} {\bibfnamefont {U.}~\bibnamefont {Yang}},\
  }\href {\doibase 10.1088/0954-3899/29/8/369} {\bibfield  {journal} {\bibinfo
  {journal} {J. Phys. G}\ }\textbf {\bibinfo {volume} {29}},\ \bibinfo {pages}
  {1899} (\bibinfo {year} {2003})},\ \Eprint
  {http://arxiv.org/abs/hep-ex/0210024} {arXiv:hep-ex/0210024} \BibitemShut
  {NoStop}%
\bibitem [{\citenamefont {Glück}\ \emph {et~al.}(1998)\citenamefont {Glück},
  \citenamefont {Reya},\ and\ \citenamefont {Vogt}}]{Gluck:1998xa}%
  \BibitemOpen
  \bibfield  {author} {\bibinfo {author} {\bibfnamefont {M.}~\bibnamefont
  {Glück}}, \bibinfo {author} {\bibfnamefont {E.}~\bibnamefont {Reya}}, \ and\
  \bibinfo {author} {\bibfnamefont {A.}~\bibnamefont {Vogt}},\ }\href {\doibase
  10.1007/s100520050289} {\bibfield  {journal} {\bibinfo  {journal} {Eur. Phys.
  J. C}\ }\textbf {\bibinfo {volume} {5}},\ \bibinfo {pages} {461} (\bibinfo
  {year} {1998})},\ \Eprint {http://arxiv.org/abs/hep-ph/9806404}
  {arXiv:hep-ph/9806404} \BibitemShut {NoStop}%
\bibitem [{\citenamefont {Yang}\ \emph {et~al.}(2009)\citenamefont {Yang},
  \citenamefont {Andreopoulos}, \citenamefont {Gallagher}, \citenamefont
  {Hoffmann},\ and\ \citenamefont {Kehayias}}]{Yang:2009zx}%
  \BibitemOpen
  \bibfield  {author} {\bibinfo {author} {\bibfnamefont {T.}~\bibnamefont
  {Yang}}, \bibinfo {author} {\bibfnamefont {C.}~\bibnamefont {Andreopoulos}},
  \bibinfo {author} {\bibfnamefont {H.}~\bibnamefont {Gallagher}}, \bibinfo
  {author} {\bibfnamefont {K.}~\bibnamefont {Hoffmann}}, \ and\ \bibinfo
  {author} {\bibfnamefont {P.}~\bibnamefont {Kehayias}},\ }\href {\doibase
  10.1140/epjc/s10052-009-1094-z} {\bibfield  {journal} {\bibinfo  {journal}
  {Eur. Phys. J.}\ }\textbf {\bibinfo {volume} {C63}},\ \bibinfo {pages} {1}
  (\bibinfo {year} {2009})},\ \Eprint {http://arxiv.org/abs/0904.4043}
  {arXiv:0904.4043 [hep-ph]} \BibitemShut {NoStop}%
\bibitem [{\citenamefont {Koba}\ \emph {et~al.}(1972)\citenamefont {Koba},
  \citenamefont {Nielsen},\ and\ \citenamefont {Olesen}}]{Koba:1972ng}%
  \BibitemOpen
  \bibfield  {author} {\bibinfo {author} {\bibfnamefont {Z.}~\bibnamefont
  {Koba}}, \bibinfo {author} {\bibfnamefont {H.~B.}\ \bibnamefont {Nielsen}}, \
  and\ \bibinfo {author} {\bibfnamefont {P.}~\bibnamefont {Olesen}},\ }\href
  {\doibase 10.1016/0550-3213(72)90551-2} {\bibfield  {journal} {\bibinfo
  {journal} {Nucl. Phys.}\ }\textbf {\bibinfo {volume} {B40}},\ \bibinfo
  {pages} {317} (\bibinfo {year} {1972})}\BibitemShut {NoStop}%
\bibitem [{\citenamefont {Sjostrand}\ \emph {et~al.}(2006)\citenamefont
  {Sjostrand}, \citenamefont {Mrenna},\ and\ \citenamefont
  {Skands}}]{Sjostrand:2006za}%
  \BibitemOpen
  \bibfield  {author} {\bibinfo {author} {\bibfnamefont {T.}~\bibnamefont
  {Sjostrand}}, \bibinfo {author} {\bibfnamefont {S.}~\bibnamefont {Mrenna}}, \
  and\ \bibinfo {author} {\bibfnamefont {P.~Z.}\ \bibnamefont {Skands}},\
  }\href {\doibase 10.1088/1126-6708/2006/05/026} {\bibfield  {journal}
  {\bibinfo  {journal} {JHEP}\ }\textbf {\bibinfo {volume} {05}},\ \bibinfo
  {pages} {026} (\bibinfo {year} {2006})},\ \Eprint
  {http://arxiv.org/abs/hep-ph/0603175} {arXiv:hep-ph/0603175 [hep-ph]}
  \BibitemShut {NoStop}%
\bibitem [{\citenamefont {Sanchez}(2018)}]{nova_2018}%
  \BibitemOpen
  \bibfield  {author} {\bibinfo {author} {\bibfnamefont {M.}~\bibnamefont
  {Sanchez}},\ }in\ \href {\doibase 10.5281/zenodo.1286758} {\emph {\bibinfo
  {booktitle} {XXVIII International Conference on Neutrino Physics and
  Astrophysics (Neutrino 2018)}}}\ (\bibinfo  {publisher} {Zenodo},\ \bibinfo
  {year} {2018})\ \bibinfo {note}
  {\url{https://zenodo.org/record/1286758}}\BibitemShut {NoStop}%
\bibitem [{\citenamefont {Dytman}\ and\ \citenamefont
  {Meyer}(2011)}]{Dytman:2011zz}%
  \BibitemOpen
  \bibfield  {author} {\bibinfo {author} {\bibfnamefont {S.}~\bibnamefont
  {Dytman}}\ and\ \bibinfo {author} {\bibfnamefont {A.}~\bibnamefont {Meyer}},\
  }\href {\doibase 10.1063/1.3661588} {\bibfield  {journal} {\bibinfo
  {journal} {AIP Conf. Proc.}\ }\textbf {\bibinfo {volume} {1405}},\ \bibinfo
  {pages} {213} (\bibinfo {year} {2011})}\BibitemShut {NoStop}%
\bibitem [{\citenamefont {Dytman}(2015)}]{Dytman:2015taa}%
  \BibitemOpen
  \bibfield  {author} {\bibinfo {author} {\bibfnamefont {S.}~\bibnamefont
  {Dytman}},\ }\bibfield  {booktitle} {\emph {\bibinfo {booktitle}
  {{Proceedings, Workshop on Neutrino Interactions, Systematic uncertainties
  and near detector physics: Session of CETUP* 2014: Lead/Dead Wood, South
  Dakota, USA, July 22-31, 2014}}},\ }\href {\doibase 10.1063/1.4931864}
  {\bibfield  {journal} {\bibinfo  {journal} {AIP Conf. Proc.}\ }\textbf
  {\bibinfo {volume} {1680}},\ \bibinfo {pages} {020005} (\bibinfo {year}
  {2015})}\BibitemShut {NoStop}%
\bibitem [{\citenamefont {Dytman}(2009)}]{intranuke_2009}%
  \BibitemOpen
  \bibfield  {author} {\bibinfo {author} {\bibfnamefont {S.}~\bibnamefont
  {Dytman}},\ }\href@noop {} {\bibfield  {journal} {\bibinfo  {journal} {Acta
  Phys. Polon.}\ }\textbf {\bibinfo {volume} {B40}},\ \bibinfo {pages} {2445}
  (\bibinfo {year} {2009})}\BibitemShut {NoStop}%
\bibitem [{\citenamefont {Day}\ and\ \citenamefont
  {McFarland}(2012)}]{Day:2012gb}%
  \BibitemOpen
  \bibfield  {author} {\bibinfo {author} {\bibfnamefont {M.}~\bibnamefont
  {Day}}\ and\ \bibinfo {author} {\bibfnamefont {K.~S.}\ \bibnamefont
  {McFarland}},\ }\href {\doibase 10.1103/PhysRevD.86.053003} {\bibfield
  {journal} {\bibinfo  {journal} {Phys. Rev.}\ }\textbf {\bibinfo {volume}
  {D86}},\ \bibinfo {pages} {053003} (\bibinfo {year} {2012})},\ \Eprint
  {http://arxiv.org/abs/1206.6745} {arXiv:1206.6745 [hep-ph]} \BibitemShut
  {NoStop}%
\bibitem [{\citenamefont {Amsler}\ \emph {et~al.}(2015)\citenamefont {Amsler}
  \emph {et~al.}}]{argoncube_loi}%
  \BibitemOpen
  \bibfield  {author} {\bibinfo {author} {\bibfnamefont {C.}~\bibnamefont
  {Amsler}} \emph {et~al.} (\bibinfo {collaboration} {ArgonCube}),\ }\href
  {https://cds.cern.ch/record/1993255} {\emph {\bibinfo {title} {{ArgonCube: a
  novel, fully-modular approach for the realization of large-mass liquid argon
  TPC neutrino detectors}}}},\ \bibinfo {type} {Tech. Rep.}\ \bibinfo {number}
  {CERN-SPSC-2015-009. SPSC-I-243}\ (\bibinfo  {institution} {CERN},\ \bibinfo
  {address} {Geneva},\ \bibinfo {year} {2015})\BibitemShut {NoStop}%
\bibitem [{\citenamefont {Dwyer}\ \emph {et~al.}(2018)\citenamefont {Dwyer}
  \emph {et~al.}}]{Dwyer:2018phu}%
  \BibitemOpen
  \bibfield  {author} {\bibinfo {author} {\bibfnamefont {D.~A.}\ \bibnamefont
  {Dwyer}} \emph {et~al.},\ }\href {\doibase 10.1088/1748-0221/13/10/P10007}
  {\bibfield  {journal} {\bibinfo  {journal} {JINST}\ }\textbf {\bibinfo
  {volume} {13}},\ \bibinfo {pages} {P10007} (\bibinfo {year} {2018})},\
  \Eprint {http://arxiv.org/abs/1808.02969} {arXiv:1808.02969
  [physics.ins-det]} \BibitemShut {NoStop}%
\bibitem [{\citenamefont {Auger}\ \emph {et~al.}(2018)\citenamefont {Auger},
  \citenamefont {Chen}, \citenamefont {Ereditato}, \citenamefont {Goeldi},
  \citenamefont {Kreslo}, \citenamefont {Lorca}, \citenamefont {Luethi},
  \citenamefont {Mettler}, \citenamefont {Sinclair},\ and\ \citenamefont
  {Weber}}]{arclight}%
  \BibitemOpen
  \bibfield  {author} {\bibinfo {author} {\bibfnamefont {M.}~\bibnamefont
  {Auger}}, \bibinfo {author} {\bibfnamefont {Y.}~\bibnamefont {Chen}},
  \bibinfo {author} {\bibfnamefont {A.}~\bibnamefont {Ereditato}}, \bibinfo
  {author} {\bibfnamefont {D.}~\bibnamefont {Goeldi}}, \bibinfo {author}
  {\bibfnamefont {I.}~\bibnamefont {Kreslo}}, \bibinfo {author} {\bibfnamefont
  {D.}~\bibnamefont {Lorca}}, \bibinfo {author} {\bibfnamefont
  {M.}~\bibnamefont {Luethi}}, \bibinfo {author} {\bibfnamefont
  {T.}~\bibnamefont {Mettler}}, \bibinfo {author} {\bibfnamefont {J.~R.}\
  \bibnamefont {Sinclair}}, \ and\ \bibinfo {author} {\bibfnamefont {M.~S.}\
  \bibnamefont {Weber}},\ }\href {\doibase 10.3390/instruments2010003}
  {\bibfield  {journal} {\bibinfo  {journal} {Instruments}\ }\textbf {\bibinfo
  {volume} {2}},\ \bibinfo {pages} {3} (\bibinfo {year} {2018})},\ \Eprint
  {http://arxiv.org/abs/1711.11409} {arXiv:1711.11409 [physics.ins-det]}
  \BibitemShut {NoStop}%
\bibitem [{\citenamefont {Agostinelli}\ \emph {et~al.}(2003)\citenamefont
  {Agostinelli} \emph {et~al.}}]{Agostinelli:2002hh}%
  \BibitemOpen
  \bibfield  {author} {\bibinfo {author} {\bibfnamefont {S.}~\bibnamefont
  {Agostinelli}} \emph {et~al.} (\bibinfo {collaboration} {GEANT4}),\ }\href
  {\doibase 10.1016/S0168-9002(03)01368-8} {\bibfield  {journal} {\bibinfo
  {journal} {Nucl.\ Instrum.\ Meth.\ A}\ }\textbf {\bibinfo {volume} {506}},\
  \bibinfo {pages} {250} (\bibinfo {year} {2003})}\BibitemShut {NoStop}%
\bibitem [{\citenamefont {Abi}\ \emph {et~al.}(2018)\citenamefont {Abi} \emph
  {et~al.}}]{Abi:2018dnh}%
  \BibitemOpen
  \bibfield  {author} {\bibinfo {author} {\bibfnamefont {B.}~\bibnamefont
  {Abi}} \emph {et~al.} (\bibinfo {collaboration} {DUNE}),\ }\href@noop {} {\
  (\bibinfo {year} {2018})},\ \Eprint {http://arxiv.org/abs/1807.10334}
  {arXiv:1807.10334 [physics.ins-det]} \BibitemShut {NoStop}%
\bibitem [{\citenamefont {Abi}\ \emph {et~al.}(2020{\natexlab{d}})\citenamefont
  {Abi} \emph {et~al.}}]{Abi:2020loh}%
  \BibitemOpen
  \bibfield  {author} {\bibinfo {author} {\bibfnamefont {B.}~\bibnamefont
  {Abi}} \emph {et~al.} (\bibinfo {collaboration} {DUNE}),\ }\href {\doibase
  10.1088/1748-0221/15/08/T08010} {\bibfield  {journal} {\bibinfo  {journal}
  {JINST}\ }\textbf {\bibinfo {volume} {15}},\ \bibinfo {pages} {T08010}
  (\bibinfo {year} {2020}{\natexlab{d}})},\ \Eprint
  {http://arxiv.org/abs/2002.03010} {arXiv:2002.03010 [physics.ins-det]}
  \BibitemShut {NoStop}%
\bibitem [{\citenamefont {Marshall}\ and\ \citenamefont
  {Thomson}(2015)}]{Marshall:2015rfa}%
  \BibitemOpen
  \bibfield  {author} {\bibinfo {author} {\bibfnamefont {J.~S.}\ \bibnamefont
  {Marshall}}\ and\ \bibinfo {author} {\bibfnamefont {M.~A.}\ \bibnamefont
  {Thomson}},\ }\href {\doibase 10.1140/epjc/s10052-015-3659-3} {\bibfield
  {journal} {\bibinfo  {journal} {Eur. Phys. J.}\ }\textbf {\bibinfo {volume}
  {C75}},\ \bibinfo {pages} {439} (\bibinfo {year} {2015})},\ \Eprint
  {http://arxiv.org/abs/1506.05348} {arXiv:1506.05348 [physics.data-an]}
  \BibitemShut {NoStop}%
\bibitem [{\citenamefont {Acciarri}\ \emph {et~al.}(2018)\citenamefont
  {Acciarri} \emph {et~al.}}]{Acciarri:2017hat}%
  \BibitemOpen
  \bibfield  {author} {\bibinfo {author} {\bibfnamefont {R.}~\bibnamefont
  {Acciarri}} \emph {et~al.} (\bibinfo {collaboration} {MicroBooNE}),\ }\href
  {\doibase 10.1140/epjc/s10052-017-5481-6} {\bibfield  {journal} {\bibinfo
  {journal} {Eur. Phys. J.}\ }\textbf {\bibinfo {volume} {C78}},\ \bibinfo
  {pages} {82} (\bibinfo {year} {2018})},\ \Eprint
  {http://arxiv.org/abs/1708.03135} {arXiv:1708.03135 [hep-ex]} \BibitemShut
  {NoStop}%
\bibitem [{\citenamefont {Abi}\ \emph {et~al.}(2020{\natexlab{e}})\citenamefont
  {Abi} \emph {et~al.}}]{cvn_paper}%
  \BibitemOpen
  \bibfield  {author} {\bibinfo {author} {\bibfnamefont {B.}~\bibnamefont
  {Abi}} \emph {et~al.} (\bibinfo {collaboration} {DUNE}),\ }\href@noop {}
  {\bibfield  {journal} {\bibinfo  {journal} {Phys. Rev.}\ }\textbf {\bibinfo
  {volume} {D102}},\ \bibinfo {pages} {092003} (\bibinfo {year}
  {2020}{\natexlab{e}})},\ \Eprint {http://arxiv.org/abs/2006.15052}
  {arXiv:2006.15052 [physics.ins-det]} \BibitemShut {NoStop}%
\bibitem [{\citenamefont {James}(1994)}]{James:1994vla}%
  \BibitemOpen
  \bibfield  {author} {\bibinfo {author} {\bibfnamefont {F.}~\bibnamefont
  {James}},\ }\href@noop {} {\enquote {\bibinfo {title} {{MINUIT Function
  Minimization and Error Analysis: Reference Manual Version 94.1}},}\ }
  (\bibinfo {year} {1994}),\ \bibinfo {note} {{CERN-D-506,
  CERN-D506}}\BibitemShut {NoStop}%
\bibitem [{\citenamefont {Esteban}\ \emph {et~al.}(2019)\citenamefont
  {Esteban}, \citenamefont {Gonzalez-Garcia}, \citenamefont
  {Hernandez-Cabezudo}, \citenamefont {Maltoni},\ and\ \citenamefont
  {Schwetz}}]{Esteban:2018azc}%
  \BibitemOpen
  \bibfield  {author} {\bibinfo {author} {\bibfnamefont {I.}~\bibnamefont
  {Esteban}}, \bibinfo {author} {\bibfnamefont {M.~C.}\ \bibnamefont
  {Gonzalez-Garcia}}, \bibinfo {author} {\bibfnamefont {A.}~\bibnamefont
  {Hernandez-Cabezudo}}, \bibinfo {author} {\bibfnamefont {M.}~\bibnamefont
  {Maltoni}}, \ and\ \bibinfo {author} {\bibfnamefont {T.}~\bibnamefont
  {Schwetz}},\ }\href {\doibase 10.1007/JHEP01(2019)106} {\bibfield  {journal}
  {\bibinfo  {journal} {JHEP}\ }\textbf {\bibinfo {volume} {01}},\ \bibinfo
  {pages} {106} (\bibinfo {year} {2019})},\ \Eprint
  {http://arxiv.org/abs/1811.05487} {arXiv:1811.05487 [hep-ph]} \BibitemShut
  {NoStop}%
\bibitem [{\citenamefont {Esteban}\ \emph {et~al.}(2018)\citenamefont
  {Esteban}, \citenamefont {Gonzalez-Garcia}, \citenamefont
  {Hernandez-Cabezudo}, \citenamefont {Maltoni},\ and\ \citenamefont
  {Schwetz}}]{nufitweb}%
  \BibitemOpen
  \bibfield  {author} {\bibinfo {author} {\bibfnamefont {I.}~\bibnamefont
  {Esteban}}, \bibinfo {author} {\bibfnamefont {M.~C.}\ \bibnamefont
  {Gonzalez-Garcia}}, \bibinfo {author} {\bibfnamefont {A.}~\bibnamefont
  {Hernandez-Cabezudo}}, \bibinfo {author} {\bibfnamefont {M.}~\bibnamefont
  {Maltoni}}, \ and\ \bibinfo {author} {\bibfnamefont {T.}~\bibnamefont
  {Schwetz}},\ }\href {{http://www.nu-fit.org/}} {\enquote {\bibinfo {title}
  {Nufit4.0},}\ } (\bibinfo {year} {2018})\BibitemShut {NoStop}%
\bibitem [{\citenamefont {Roe}(2017)}]{Roe:2017zdw}%
  \BibitemOpen
  \bibfield  {author} {\bibinfo {author} {\bibfnamefont {B.}~\bibnamefont
  {Roe}},\ }\href {\doibase 10.1103/PhysRevD.95.113004} {\bibfield  {journal}
  {\bibinfo  {journal} {Phys. Rev.}\ }\textbf {\bibinfo {volume} {D95}},\
  \bibinfo {pages} {113004} (\bibinfo {year} {2017})},\ \Eprint
  {http://arxiv.org/abs/1707.02322} {arXiv:1707.02322 [hep-ex]} \BibitemShut
  {NoStop}%
\bibitem [{\citenamefont {Abrah\~ao}\ \emph {et~al.}(2021)\citenamefont
  {Abrah\~ao} \emph {et~al.}}]{Abrahao:2020ztg}%
  \BibitemOpen
  \bibfield  {author} {\bibinfo {author} {\bibfnamefont {T.}~\bibnamefont
  {Abrah\~ao}} \emph {et~al.} (\bibinfo {collaboration} {Double Chooz}),\
  }\href {\doibase 10.1007/JHEP01(2021)190} {\bibfield  {journal} {\bibinfo
  {journal} {JHEP}\ }\textbf {\bibinfo {volume} {01}},\ \bibinfo {pages} {190}
  (\bibinfo {year} {2021})},\ \Eprint {http://arxiv.org/abs/2007.13431}
  {arXiv:2007.13431 [hep-ex]} \BibitemShut {NoStop}%
\bibitem [{\citenamefont {Adey}\ \emph {et~al.}(2018)\citenamefont {Adey} \emph
  {et~al.}}]{Adey:2018zwh}%
  \BibitemOpen
  \bibfield  {author} {\bibinfo {author} {\bibfnamefont {D.}~\bibnamefont
  {Adey}} \emph {et~al.} (\bibinfo {collaboration} {Daya Bay}),\ }\href@noop {}
  {\bibfield  {journal} {\bibinfo  {journal} {Phys. Rev. Lett.}\ }\textbf
  {\bibinfo {volume} {121}},\ \bibinfo {pages} {241805} (\bibinfo {year}
  {2018})},\ \Eprint {http://arxiv.org/abs/1809.02261} {arXiv:1809.02261
  [hep-ex]} \BibitemShut {NoStop}%
\bibitem [{\citenamefont {Bak}\ \emph {et~al.}(2018)\citenamefont {Bak} \emph
  {et~al.}}]{Bak:2018ydk}%
  \BibitemOpen
  \bibfield  {author} {\bibinfo {author} {\bibfnamefont {G.}~\bibnamefont
  {Bak}} \emph {et~al.} (\bibinfo {collaboration} {RENO}),\ }\href@noop {}
  {\bibfield  {journal} {\bibinfo  {journal} {Phys. Rev. Lett.}\ }\textbf
  {\bibinfo {volume} {121}},\ \bibinfo {pages} {201801} (\bibinfo {year}
  {2018})},\ \Eprint {http://arxiv.org/abs/1806.00248} {arXiv:1806.00248
  [hep-ex]} \BibitemShut {NoStop}%
\bibitem [{\citenamefont {Tanabashi}\ \emph {et~al.}(2018)\citenamefont
  {Tanabashi} \emph {et~al.}}]{Tanabashi:2018oca}%
  \BibitemOpen
  \bibfield  {author} {\bibinfo {author} {\bibfnamefont {M.}~\bibnamefont
  {Tanabashi}} \emph {et~al.} (\bibinfo {collaboration} {Particle Data
  Group}),\ }\href {\doibase 10.1103/PhysRevD.98.030001} {\bibfield  {journal}
  {\bibinfo  {journal} {Phys. Rev.}\ }\textbf {\bibinfo {volume} {D98}},\
  \bibinfo {pages} {030001} (\bibinfo {year} {2018})}\BibitemShut {NoStop}%
\bibitem [{\citenamefont {Cowan}\ \emph {et~al.}(2011)\citenamefont {Cowan},
  \citenamefont {Cranmer}, \citenamefont {Gross},\ and\ \citenamefont
  {Vitells}}]{Cowan:2010js}%
  \BibitemOpen
  \bibfield  {author} {\bibinfo {author} {\bibfnamefont {G.}~\bibnamefont
  {Cowan}}, \bibinfo {author} {\bibfnamefont {K.}~\bibnamefont {Cranmer}},
  \bibinfo {author} {\bibfnamefont {E.}~\bibnamefont {Gross}}, \ and\ \bibinfo
  {author} {\bibfnamefont {O.}~\bibnamefont {Vitells}},\ }\href {\doibase
  10.1140/epjc/s10052-011-1554-0, 10.1140/epjc/s10052-013-2501-z} {\bibfield
  {journal} {\bibinfo  {journal} {Eur. Phys. J.}\ }\textbf {\bibinfo {volume}
  {C71}},\ \bibinfo {pages} {1554} (\bibinfo {year} {2011})},\ \bibinfo {note}
  {[Erratum: Eur. Phys. J. C73, 2501(2013)]},\ \Eprint
  {http://arxiv.org/abs/1007.1727} {arXiv:1007.1727 [physics.data-an]}
  \BibitemShut {NoStop}%
\bibitem [{\citenamefont {Wilks}(1938)}]{wilks}%
  \BibitemOpen
  \bibfield  {author} {\bibinfo {author} {\bibfnamefont {S.~S.}\ \bibnamefont
  {Wilks}},\ }\href {\doibase 10.1214/aoms/1177732360} {\bibfield  {journal}
  {\bibinfo  {journal} {Ann. Math. Stat.}\ }\textbf {\bibinfo {volume} {9}},\
  \bibinfo {pages} {60 } (\bibinfo {year} {1938})}\BibitemShut {NoStop}%
\bibitem [{\citenamefont {Feldman}\ and\ \citenamefont
  {Cousins}(1998)}]{Feldman:1997qc}%
  \BibitemOpen
  \bibfield  {author} {\bibinfo {author} {\bibfnamefont {G.~J.}\ \bibnamefont
  {Feldman}}\ and\ \bibinfo {author} {\bibfnamefont {R.~D.}\ \bibnamefont
  {Cousins}},\ }\href {\doibase 10.1103/PhysRevD.57.3873} {\bibfield  {journal}
  {\bibinfo  {journal} {Phys. Rev. D}\ }\textbf {\bibinfo {volume} {57}},\
  \bibinfo {pages} {3873} (\bibinfo {year} {1998})},\ \Eprint
  {http://arxiv.org/abs/physics/9711021} {arXiv:physics/9711021} \BibitemShut
  {NoStop}%
\bibitem [{\citenamefont {Rice}(2006)}]{rice2006mathematical}%
  \BibitemOpen
  \bibfield  {author} {\bibinfo {author} {\bibfnamefont {J.~A.}\ \bibnamefont
  {Rice}},\ }\href@noop {} {\emph {\bibinfo {title} {Mathematical Statistics
  and Data Analysis.}}},\ \bibinfo {edition} {3rd}\ ed.\ (\bibinfo  {publisher}
  {Belmont, CA: Duxbury Press.},\ \bibinfo {year} {2006})\BibitemShut {NoStop}%
\bibitem [{\citenamefont {Ciuffoli}\ \emph {et~al.}(2014)\citenamefont
  {Ciuffoli}, \citenamefont {Evslin},\ and\ \citenamefont
  {Zhang}}]{Ciuffoli:2013rza}%
  \BibitemOpen
  \bibfield  {author} {\bibinfo {author} {\bibfnamefont {E.}~\bibnamefont
  {Ciuffoli}}, \bibinfo {author} {\bibfnamefont {J.}~\bibnamefont {Evslin}}, \
  and\ \bibinfo {author} {\bibfnamefont {X.}~\bibnamefont {Zhang}},\ }\href
  {\doibase 10.1007/JHEP01(2014)095} {\bibfield  {journal} {\bibinfo  {journal}
  {JHEP}\ }\textbf {\bibinfo {volume} {01}},\ \bibinfo {pages} {095} (\bibinfo
  {year} {2014})},\ \Eprint {http://arxiv.org/abs/1305.5150} {arXiv:1305.5150
  [hep-ph]} \BibitemShut {NoStop}%
\bibitem [{\citenamefont {Qian}\ \emph {et~al.}(2012)\citenamefont {Qian} \emph
  {et~al.}}]{Qian:2012zn}%
  \BibitemOpen
  \bibfield  {author} {\bibinfo {author} {\bibfnamefont {X.}~\bibnamefont
  {Qian}} \emph {et~al.},\ }\href {\doibase 10.1103/PhysRevD.86.113011}
  {\bibfield  {journal} {\bibinfo  {journal} {Phys. Rev.}\ }\textbf {\bibinfo
  {volume} {D86}},\ \bibinfo {pages} {113011} (\bibinfo {year} {2012})},\
  \Eprint {http://arxiv.org/abs/1210.3651} {arXiv:1210.3651 [hep-ph]}
  \BibitemShut {NoStop}%
\bibitem [{\citenamefont {Blennow}\ \emph {et~al.}(2014)\citenamefont
  {Blennow}, \citenamefont {Coloma}, \citenamefont {Huber},\ and\ \citenamefont
  {Schwetz}}]{Blennow:2013oma}%
  \BibitemOpen
  \bibfield  {author} {\bibinfo {author} {\bibfnamefont {M.}~\bibnamefont
  {Blennow}}, \bibinfo {author} {\bibfnamefont {P.}~\bibnamefont {Coloma}},
  \bibinfo {author} {\bibfnamefont {P.}~\bibnamefont {Huber}}, \ and\ \bibinfo
  {author} {\bibfnamefont {T.}~\bibnamefont {Schwetz}},\ }\href {\doibase
  10.1007/JHEP03(2014)028} {\bibfield  {journal} {\bibinfo  {journal} {JHEP}\
  }\textbf {\bibinfo {volume} {1403}},\ \bibinfo {pages} {028} (\bibinfo {year}
  {2014})},\ \Eprint {http://arxiv.org/abs/1311.1822} {arXiv:1311.1822
  [hep-ph]} \BibitemShut {NoStop}%
\end{thebibliography}%
